\documentclass{aastex6}

\begin{document}
\title{A Submillimeter Perspective on the GOODS Fields (SUPER GOODS)---I.
An Utradeep SCUBA-2 Survey of the GOODS-N\altaffilmark{1,2,3,4}
}

\author{
L.~L.~Cowie\altaffilmark{5},
A.~J.~Barger\altaffilmark{5,6,7}, 
L.-Y.~Hsu\altaffilmark{5},
Chian-Chou~Chen\altaffilmark{8},
F.~N.~Owen\altaffilmark{9},
W.-H.~Wang\altaffilmark{10}
}

\altaffiltext{1}{The James Clerk Maxwell Telescope is operated by the 
East Asian Observatory on behalf of The National Astronomical
Observatory of Japan, Academia Sinica Institute of Astronomy and
Astrophysics, the Korea Astronomy and Space Science Institute,
the National Astronomical Observatories of China and the Chinese
Academy of Sciences (Grant No. XDB09000000), with additional
funding support from the Science and Technology Facilities Council
of the United Kingdom and participating universities in the United
Kingdom and Canada.}
\altaffiltext{2}{The National Radio Astronomy Observatory is a facility of 
the National Science Foundation operated under cooperative agreement by 
Associated Universities, Inc.}
\altaffiltext{3}{The Submillimeter Array is a joint project between the
Smithsonian Astrophysical Observatory and the Academia Sinica Institute
of Astronomy and Astrophysics and is funded by the Smithsonian Institution
and the Academia Sinica.}
\altaffiltext{4}{The W.~M.~Keck Observatory is operated as a scientific
partnership among the the California Institute of Technology, the University
of California, and NASA, and was made possible by the generous financial
support of the W.~M.~Keck Foundation.}
\altaffiltext{5}{Institute for Astronomy, University of Hawaii,
2680 Woodlawn Drive, Honolulu, HI 96822, USA}
\altaffiltext{6}{Department of Astronomy, University of Wisconsin-Madison,
475 N. Charter Street, Madison, WI 53706, USA}
\altaffiltext{7}{Department of Physics and Astronomy, University of Hawaii,
2505 Correa Road, Honolulu, HI 96822, USA}
\altaffiltext{8}{Centre for Extragalactic Astronomy, Department of Physics, 
Durham University, South Road, Durham DH1 3LE, UK}
\altaffiltext{9}{National Radio Astronomy Observatory, P.O. Box O, 
Socorro, NM 87801, USA}
\altaffiltext{10}{Academia Sinica Institute of Astronomy and Astrophysics, 
P.O. Box 23-141, Taipei 10617, Taiwan}

\slugcomment{Accepted to The Astrophysical Journal}

\begin{abstract}
In this first paper in the SUPER GOODS series on powerfully star-forming 
galaxies in the two GOODS fields, we present a deep SCUBA-2
survey of the GOODS-N at both 850\,$\mu$m and 450\,$\mu$m
(central rms noise of 0.28\,mJy and 2.6\,mJy, respectively).
In the central region the 850\,$\mu$m observations 
cover the GOODS-N to near the confusion limit of $\sim1.65$~mJy,
while over a wider 450\,arcmin$^2$ region---well complemented by 
{\em Herschel\/} far-infrared imaging---they have a median 
$4\,\sigma$ limit of 3.5\,mJy.
We present $\ge4\,\sigma$ catalogs of 186 850\,$\mu$m and 31
450\,$\mu$m selected sources.
We use interferometric observations from the SMA and the VLA to obtain precise 
positions for 114 SCUBA-2 sources (28 from the SMA, all of which
are also VLA sources).
We present new spectroscopic redshifts and include all existing
spectroscopic or photometric redshifts. 
We also compare redshifts estimated using the 20\,cm to 850\,$\mu$m and the 
250\,$\mu$m to 850\,$\mu$m flux ratios. We show that the
redshift distribution increases with increasing flux, and we parameterize the dependence.
We compute the star formation history and the star formation rate (SFR) 
density distribution functions in various redshift intervals, finding that they reach a 
peak at $z=2-3$ before dropping to higher redshifts. We show that the number density
per unit volume of SFR\,$\gtrsim500~M_\odot~{\rm yr}^{-1}$ galaxies measured from the
SCUBA-2 sample does not change much relative to that of lower SFR galaxies
from UV selected samples over $z=2-5$, suggesting that, apart from changes in the normalization, 
the shape in the number density as a function of SFR is invariant over this redshift interval.
\end{abstract}


\section{Introduction}
\label{secintro}

The distant, dusty, ultraluminous galaxies first discovered
with the Submillimeter Common-User Bolometer Array (SCUBA; Holland et al.\ 1999) 
on the 15~m James Clerk Maxwell Telescope (JCMT)
(Smail et al.\ 1997; Barger et al.\ 1998; Hughes et al.\ 1998; Eales et al.\ 1999)
are some of the most powerfully star-forming galaxies in the universe
(see Casey et al.\ 2014 for a recent review).
Mapping these galaxies is key to understanding how the
most massive galaxies formed and how the shape of the star formation rate (SFR)
density distribution function 
evolves with cosmic time, especially since a significant number of these 
galaxies are not detected in UV/optical selections.
Understanding these galaxies requires wide and deep far-infrared (FIR) and submillimeter
surveys with the best possible ancillary data, since such data are essential for 
determining the redshifts and bolometric luminosities of individual galaxies. 
In this series of papers, we describe the results of our ultradeep submillimeter
survey with the revolutionary SCUBA-2 camera (Holland et al.\ 2013) on the JCMT 
of the most heavily studied regions on the sky:  the two GOODS fields.

Although ultraluminous infrared galaxies (ULIRGs; $L_{\rm IR}>10^{12}~L_\odot$) 
contribute only a tiny fraction of the luminosity density in the present-day universe, about
2$\%$ at $z=0.3$ (Le Floc'h et al.\ 2005), 
by $z\sim2$, the high-redshift analogs of ULIRGs---dusty, ultraluminous galaxies with 
SFRs in excess of $500\,M_\odot$~yr$^{-1}$ that are detected at submillimeter 
wavelengths---contain a significant fraction
of the total star formation out to at least a redshift of 5
(Barger et al.\ 2000, 2012, 2014;
Chapman et al.\ 2005; Wardlow et al.\ 2011; Casey et al.\ 2013; Swinbank et al.\ 2014). 
At low redshifts ($z\sim1$), the evolution of dusty star-forming 
galaxies can be studied with observations from the {\em Herschel\/} satellite alone
(e.g., Elbaz et al.\ 2011; Gruppioni et al.\ 2013; Magnelli et al.\ 2013). 
However, at higher redshifts, the {\em Herschel\/} bands that correspond to the peak 
wavelength in the FIR spectral energy distribution (SED) near 100~$\mu$m
become less sensitive, and by $z \gtrsim 5$, these bands sample wavelengths
shorter than the peak wavelength, making longer wavelength observations essential.

At the faintest fluxes, such observations can be obtained from direct
Atacama Large Millimeter/submillimeter Array (ALMA) surveys
(e.g., Aravena et al.\ 2016; Bouwens et al.\ 2016; Fujimoto et al.\ 2016;
Gonz{\'a}lez-L\'opez et al.\ 2016; Oteo et al.\ 2016; Yamaguchi et al.\ 2016; Dunlop et al.\ 2017), 
or by targeting massive cluster lensing fields with single-dish submillimeter telescopes
(e.g., Blain et al.\ 1999; Cowie et al.\ 2002; Knudsen et al.\ 2008; 
Chen et al.\ 2013a,b; Hsu et al.\ 2016).
However, at brighter fluxes ($\gtrsim2$\,mJy at 850\,$\mu$m), 
the surface density of sources is low, 
and the small interferometric fields of view and small highly magnified cluster regions make 
such surveys inefficient. Instead, surveys at these brighter fluxes 
are best carried out with wide-field imagers on single-dish submillimeter telescopes. 

SCUBA-2 is currently the most powerful means to carry out deep, wide-field surveys 
at submillimeter wavelengths (e.g., Casey et al.\ 2013; Chen et al.\ 2013b; 
Geach et al.\ 2013, 2017; Barger et al.\ 2014; Hsu et al.\ 2016). Many of the highest 
redshift and most luminous galaxies found in these surveys are not detected 
in the rest-frame UV/optical, despite extremely deep {\em HST\/} observations
(e.g., HDF850.1 in the Hubble Deep Field-North (HDF-N), originally detected 
in the SCUBA map by Hughes et al.\ 1998; the galaxy's redshift of $z=5.18$ was
eventually measured from CO observations by Walter et al.\ 2012).
Even when these sources are also detected in the UV/optical, determining their
extinction corrections is extremely difficult (e.g., Reddy et al.\ 2012).
Barger et al.\ (2014) showed that submillimeter selection finds galaxies that 
are substantially disjoint from UV selection, even when extinction corrections 
are taken into account.
Thus, the dusty, powerfully star-forming galaxy population has to be directly detected 
and characterized independently of rest-frame UV/optical surveys.

In this first paper in the SUbmillimeter PERspective on the GOODS fields
(SUPER GOODS) series, we present our SCUBA-2 observations of 
the GOODS-N/CANDELS/{\em Chandra\/} Deep 
Field-North (CDF-N). The natural limit of single-dish submillimeter
observations is the depth at which confusion becomes important. (This refers 
to the blending of sources and/or the noise being dominated by unresolved 
contributions from fainter sources.)
For the JCMT, the confusion limit at 850\,$\mu$m is $\sim1.65$\,mJy
(see Section~\ref{secscub}),
which corresponds to a SFR of $\sim 220\,M_\odot$~yr$^{-1}$ for a 
Kroupa (2001) initial mass function (IMF) (see Section~\ref{seddisc}).
Since at this flux we can probe the overlap region between submillimeter 
selected and UV/optical selected galaxy samples,
we observed the multiwavelength GOODS-N field to near the confusion 
limit for a $4\,\sigma$ detection (i.e., an rms level of $\sim 0.4$~mJy). 

Barger et al.\ (2014) showed that accurate (subarcsecond) positions for the 
bulk of the SCUBA-2 sources in the GOODS-N can be found using the deep 
Karl G. Jansky Very Large Array (VLA) 20\,cm (1.4~GHz) observations
of F. Owen (2017, in preparation). However,
to identify counterparts for sources with ambiguous or missing radio counterparts,
we have also obtained submillimeter interferometry with the Submillimeter Array 
(SMA; Ho et al.\ 2004). The SMA observations provide accurate
positions for 28 of the 186 850\,$\mu$m SCUBA-2 sources. 
A further 86 have unique VLA counterparts.

The combination of our uniquely uniform and deep SCUBA-2 map with
{\em Chandra\/}, {\em Herschel\/}, {\em HST\/}, {\em Spitzer\/}, 
VLA, and ground-based data provides a rare opportunity to explore in detail 
the properties of dusty, powerfully star-forming galaxies in the distant universe.
(Unfortunately, because of its high northern declination, the GOODS-N
cannot be observed with ALMA.) 
In Section~\ref{secdata}, we present our SCUBA-2 data set and its reduction. 
We describe the identification of multiwavelength counterparts using our VLA 
and SMA data, and we present the associated
redshifts, some of which come from new Keck spectroscopy.
We provide $\ge4\,\sigma$ 850\,$\mu$m and 450\,$\mu$m catalogs with 
the counterpart and redshift information, and we compare
these source lists with previous work. We also examine how successful the 
use of $K_s-4.5\,\mu$m colors is in picking out the galaxies in our 
850\,$\mu$m catalog.
In Section~\ref{disc}, we construct SEDs, determine the conversion of
submillimeter flux to 
SFR, and estimate redshifts using 20\,cm to 850\,$\mu$m and 250\,$\mu$m
to 850\,$\mu$m flux ratios. 
We then compute the contributions of the SCUBA-2 sample to
cosmic star formation and compare with the
total star formation history from the extinction-corrected UV.
Finally, we determine the evolution of the SFR density distribution 
functions with redshift and compare them with those in the UV.
In Section~\ref{secconcl}, we summarize our results.
We postpone much of the discussion of the X-ray data to Paper~3
of the series, where we combine the present data with the submillimeter 
sample for the CDF-S.

We assume the Wilkinson Microwave
Anisotropy Probe cosmology of $H_0=70.5$\,km\,s$^{-1}$\,Mpc$^{-1}$,
$\Omega_{\rm M}=0.27$, and $\Omega_\Lambda=0.73$ (Larson et al.\ 2011).

\section{850 Micron Sources in the CDF-N}
\label{secdata}
The depth of multiwavelength coverage on the GOODS-N field makes 
it one of the best targets for the deepest possible wide-field submillimeter survey.
The GOODS-N has a 2~Ms {\em Chandra\/} X-ray observation 
(Alexander et al.\ 2003), four optical bands ($B$ through $z$) from the 
{\em HST\/} GOODS-N survey using ACS (Giavalisco et al.\ 2004), and two
near-infrared (NIR) bands from the {\em HST\/} CANDELS survey using WFC3 
(Koekemoer et al.\ 2011; Grogin et al.\ 2011), as well as $K_s$ from
Canada-France-Hawaii Telescope (CFHT) observations with WIRCAM 
(Wang et al.\ 2010). The {\em Spitzer\/} mid-infrared (MIR) observations 
using MIPS (PI: M.~Dickinson) and the {\em Herschel\/} FIR observations 
using PACS and SPIRE (Elbaz et al.\ 2011; Lutz et al.\ 2011) are among the deepest 
ever taken and provide sufficient sensitivity to constrain tightly the thermal SEDs. 
There is also an ultra-deep 20~cm image from the upgraded VLA 
(F. Owen 2017, in preparation).
Finally, the field has secure spectroscopic observations for many thousands 
of galaxies (e.g., Cohen et al.\ 2000; Wirth et al.\ 2004, 2015; Cowie et al.\ 2004, 2016; 
Barger et al.\ 2008) and photometric redshifts for many more (e.g., Rafferty et al.\ 2011). 
In this section, we present our SCUBA-2 observations and catalogs and some
new spectroscopic redshifts of the SCUBA-2 sources that we obtained with the Keck~10\,m
telescopes. We also describe our use of the impressive ancillary data sets.

\subsection{SCUBA-2 Observations}
\label{secscub}
We obtained the data from early 2012 through to mid-2016 using the SCUBA-2 
camera. We centered all of the observations on the aimpoint of the {\em Chandra\/}
X-ray observations at right ascension 12:36:49.6,  
declination 62:13:53.0 (J2000.0) (Alexander et al.\ 2003).
In the central region, we used the \textsc{CV Daisy} scan pattern  to obtain 
the maximum depth. The \textsc{CV Daisy} field size ($5\farcm5$ 
radius; out to this radius the noise is less than twice the central value) is well 
matched to the deepest portion of the X-ray image. To cover the outer regions
to find brighter but rarer sources, we used the \textsc{PONG-900} scan pattern
($10\farcm5$ radius; by this radius the noise is twice the central value).
Detailed information about the SCUBA-2 scan patterns 
can be found in Holland et al.\ (2013).

\begin{figure*}
\includegraphics[width=18.4cm]{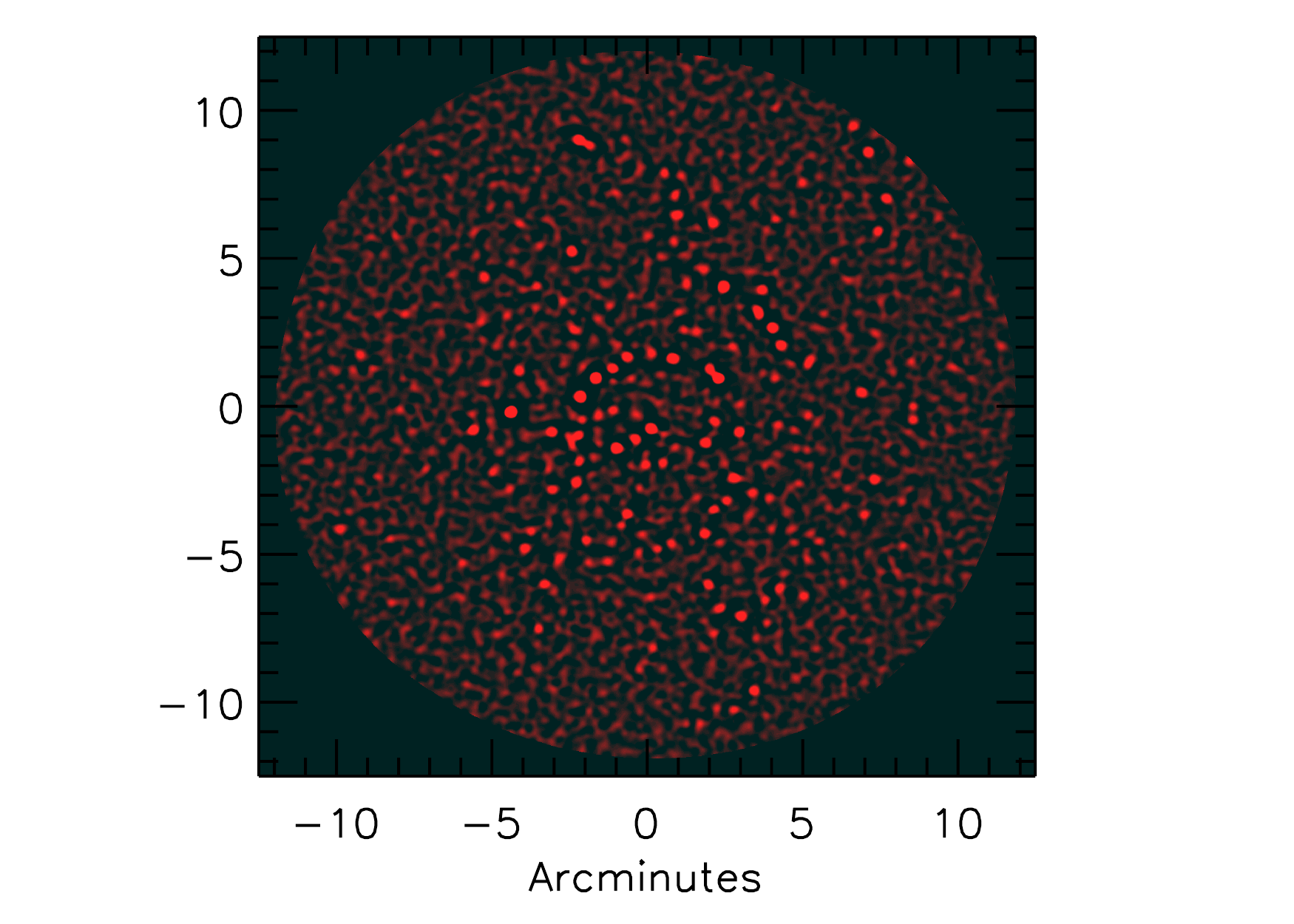}
\includegraphics[width=10cm,angle=0]{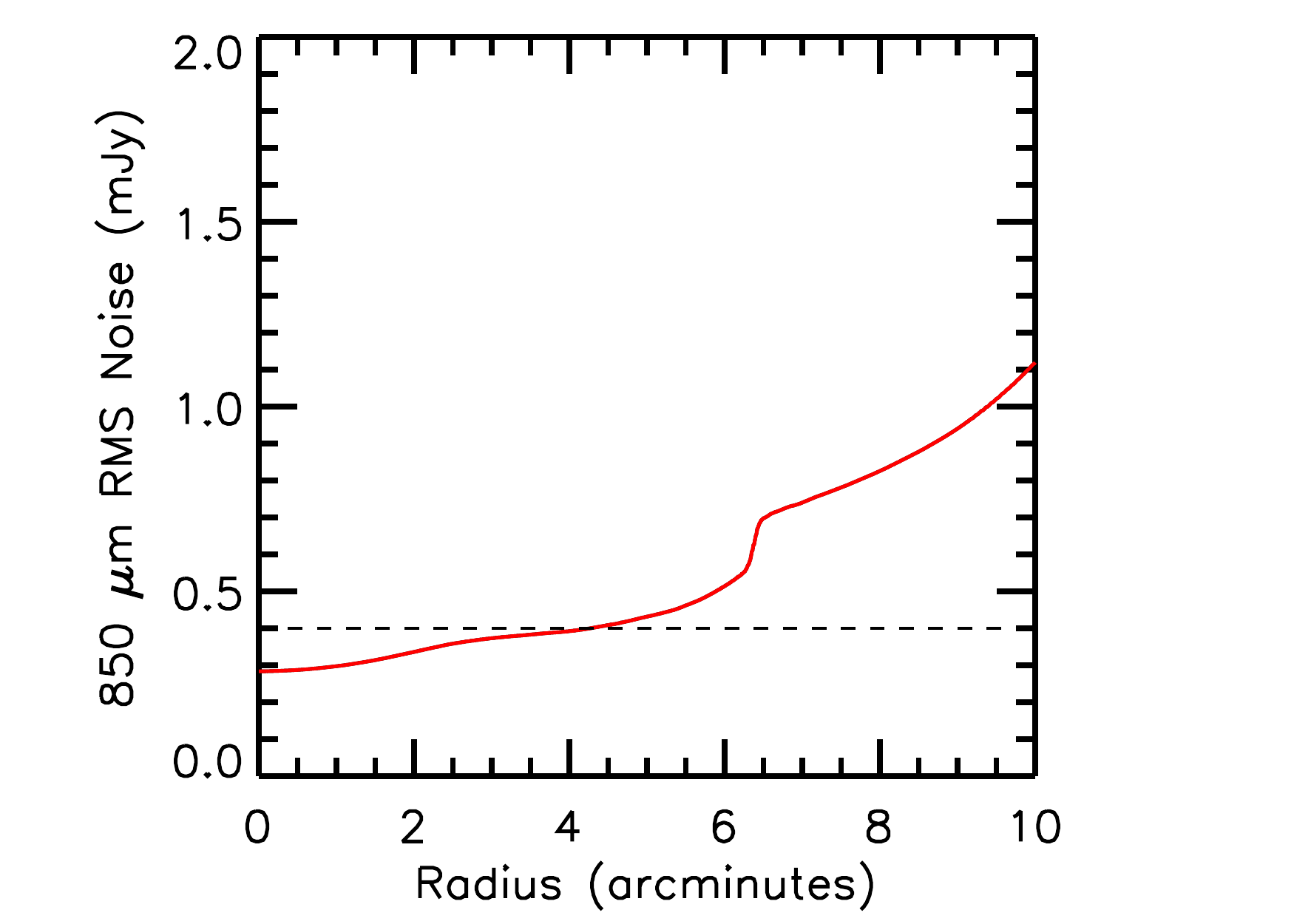}
\hskip -1.5cm\includegraphics[width=10cm,angle=0]{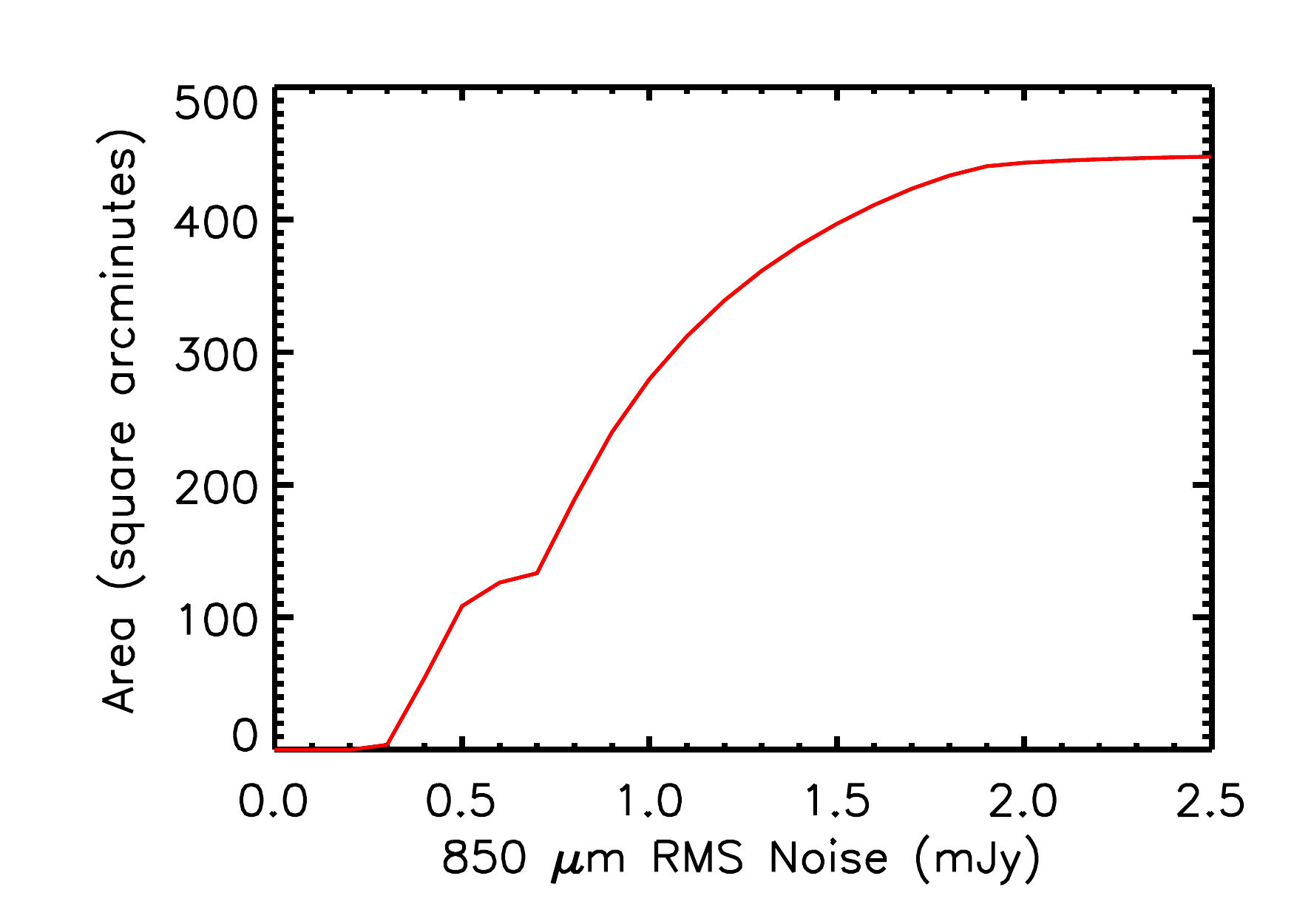}
\caption{
(a) 850\,$\mu$m matched-filter S/N image.
The area of the image is 450~arcmin$^2$.
(b) Azimuthally averaged 850\,$\mu$m rms noise vs. radius.
The more sensitive central region (radius less than $6'$) 
is dominated by the \textsc{CV Daisy}
observations, while the outer region is covered by
the \textsc{PONG-900} observations. The black dashed
line shows the rms noise corresponding to a $4\,\sigma$ detection
threshold of 1.6~mJy, the approximate confusion limit
for the JCMT at 850\,$\mu$m. In (a), the surface
density of sources is higher in the central low S/N
region of the image. 
(c) Cumulative area covered vs. 850\,$\mu$m rms noise.
\label{pretty_850}
}
\end{figure*}

\begin{deluxetable*}{cccccc}
\tablecaption{SCUBA-2 Observations}
\tablehead{Field & R.A. & Decl.& Weather  & Scan & Exposure\\ & (J2000.0) & (J2000.0) & Band & Type & (Hrs)}
\startdata
CDF-N   &12:36:49.6&62:13:53.0&     1  &    \textsc{CV Daisy}  &19.8   \\
      & & & 2	&     \textsc{CV Daisy} & 25.0  \\
      & &  & 3  &    \textsc{CV Daisy}  &4.5        \\
    & & &   1  &    \textsc{PONG-900}  & 14.0      \\
    & & &   2  &    \textsc{PONG-900}  &42.0       \\
    & & &   3  &    \textsc{PONG-900}  &23.3        \\
\enddata
\label{obs}
\end{deluxetable*}

We summarize the details of our observations in Table~\ref{obs}, where we give the
exposure times and weather conditions 
(band~1, $\tau_{\rm 225~GHz}<0.05$; band~2, $0.05<\tau_{\rm 225~GHz}<0.08$;
band~3, $0.08<\tau_{\rm 225~GHz}<0.12$) for each of the scanning patterns 
observed\footnote{The program IDs are M12AH15B, M13AH24A, M13BH16B,
M14AH22B, M14BH03, M15AH04, M15BH03B, M16AH07B.}.
If we exclude the poorer weather band~3 observations, then we obtained 101~hours on the field, 
roughly equally divided between the two scanning modes.
While SCUBA-2 obtains simultaneous 450\,$\mu$m and 850\,$\mu$m observations, 
the predominantly band~2 weather implies that the 450\,$\mu$m observations are 
not extremely deep, and we focus primarily on the 850\,$\mu$m data.

Our reduction procedures follow those described in Chen et al.\ (2013b).
We reduced the data using the Dynamic Iterative Map Maker (DIMM) 
(Jenness et al.\ 2011; Chapin et al.\ 2013).
DIMM performs iterative estimations on the common mode signal, 
the astronomical signal, and the white noise. It also applies
flat-field and extinction corrections and uses a Fourier Transform filter 
to remove low-frequency excess signal relative to the white noise that is not 
able to be removed through common mode subtraction (Chapin et al.\ 2013).
We used the standard configuration faint source file \textit{dimmconfig\_blank\_field.lis}. 
We ran DIMM on each bolometer subarray individually to avoid data splitting, 
and we used the \textsc{MOSAIC\_JCMT\_IMAGES} recipe in the Pipeline for 
Combining and Analyzing Reduced Data (PICARD; Jenness et al.\ 2008) to 
form the final images and noise maps. 

We expect nearly all the galaxies to be much smaller than the $\sim$\,14$''$ resolution 
of the JCMT at 850\,$\mu$m and hence to appear as unresolved sources.
We therefore applied a matched-filter to our maps, which
provides a maximum likelihood estimate of the source strength (e.g., Serjeant et al.\ 2003). 
Assuming $S(i,j)$ and $\sigma$(i,j) are the signal and r.m.s noise maps produced by DIMM 
and $PSF(i,j)$ is the signal point spread function, then the filtered signal map $F(i,j)$ is
given by
\begin{equation}
F(i,j) = \frac{\sum_{i,j} [S(i,j) / \sigma(i,j)^2 \times PSF(i,j)]}{\sum_{i,j} [1 / \sigma(i,j)^2 \times PSF(i,j)^2]} \,,
\end{equation} 
and the filtered noise map $N(i,j)$ is given by
\begin{equation}
N(i,j) = \frac{1}{\sqrt{\sum_{i,j} [1 / \sigma(i,j)^2 \times PSF(i,j)^2]}} \,.
\end{equation} 

Ideally, the PSF for the matched-filter algorithm is a 
Gaussian normalized to a peak of unity with FWHM equal to the JCMT beam size 
at a given wavelength (i.e., $7\farcs5$ at 450\,$\mu$m and $14''$ at 850\,$\mu$m). 
However, the map produced from DIMM has low spatial frequency structures 
that need to be removed.  
Thus, before running the matched-filter, we convolved the map with a broad Gaussian 
normalized to a sum of unity, and we subtracted this convolved map from the original map. 
Chen et al.\ (2013b) showed that the source fluxes and the 
signal-to-noise ratio (S/N) are not 
sensitive to the size of the FWHM for reasonable choices. We adopted 
values of $20''$ at 450\,$\mu$m and $30''$ at 850\,$\mu$m. 
The PSF used for the matched-filter must be processed in the same way, 
which results in a Mexican hat-like wavelet with the central Gaussian
surrounded by a negative trough.

\begin{figure*}
\includegraphics[width=18.4cm]{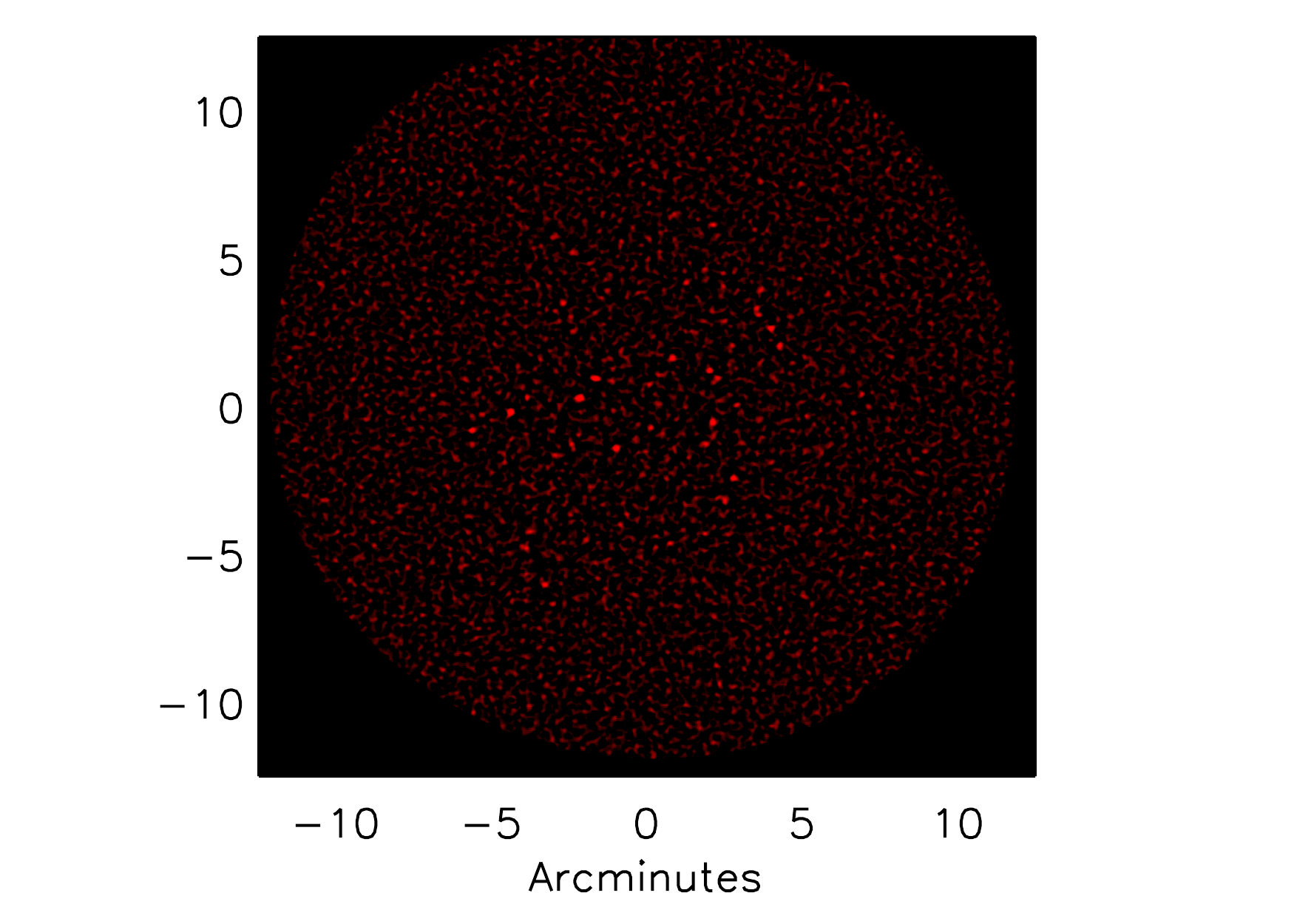}
\includegraphics[width=10cm,angle=0]{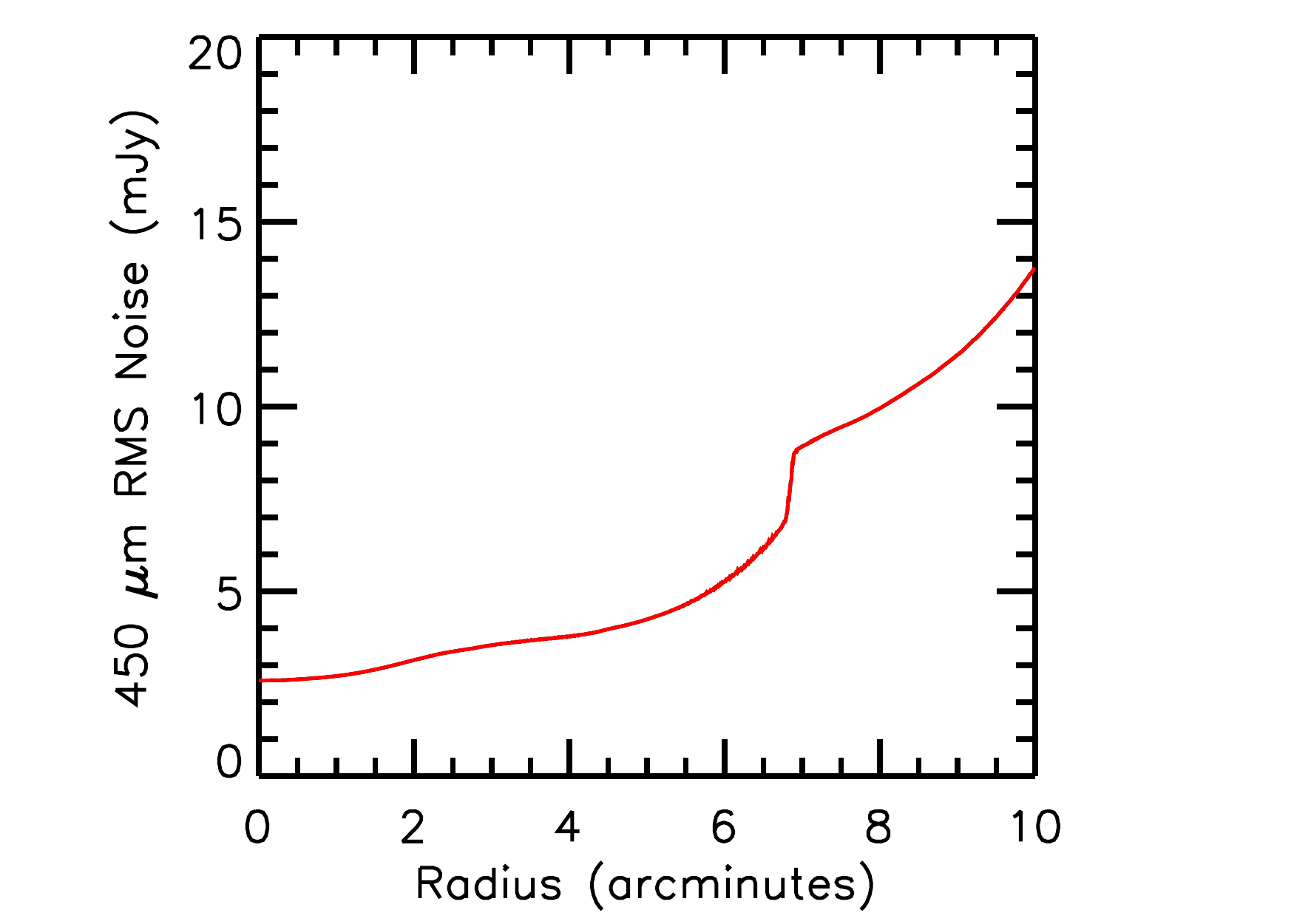}
\hskip -1.5cm\includegraphics[width=10cm,angle=0]{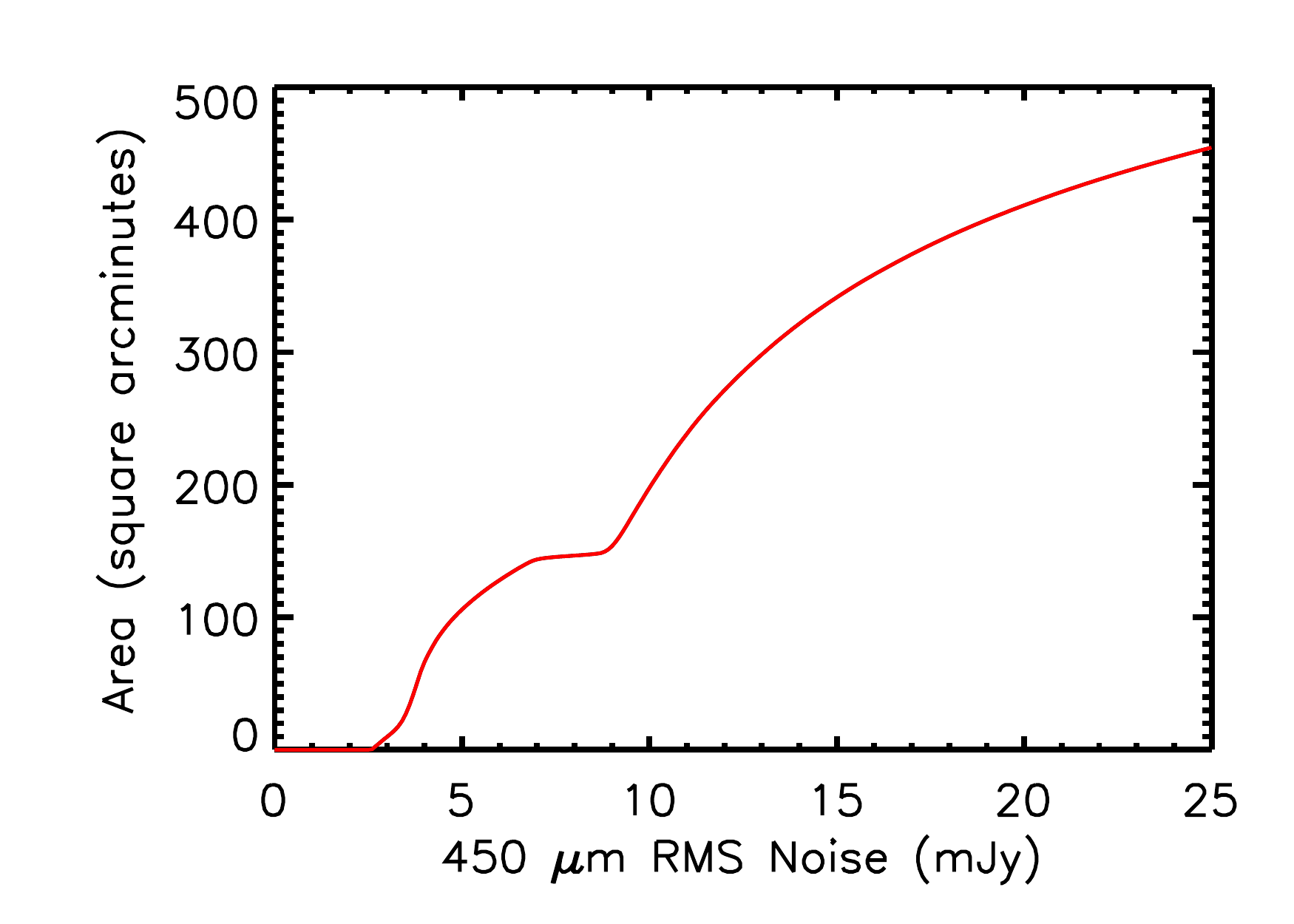}
\caption{(a) 450\,$\mu$m matched-filter S/N image. 
Only the central region covered
by the \textsc{CV Daisy} scans is shown, since the noise outside
this region is high. 
All but one of the $\ge4\,\sigma$ sources (white points) lie within this region.
(b) Azimuthally averaged 450\,$\mu$m rms noise vs. radius.
The more sensitive central region (radius less than $6^{\prime}$) 
is dominated by the \textsc{CV Daisy}
observations, while the outer region is covered by
the \textsc{PONG-900} observations. (c) Cumulative area
covered vs. 450\,$\mu$m rms noise.
\label{noise_radius_450}
}
\end{figure*}

\begin{figure*}
\centerline{\includegraphics[width=9cm]{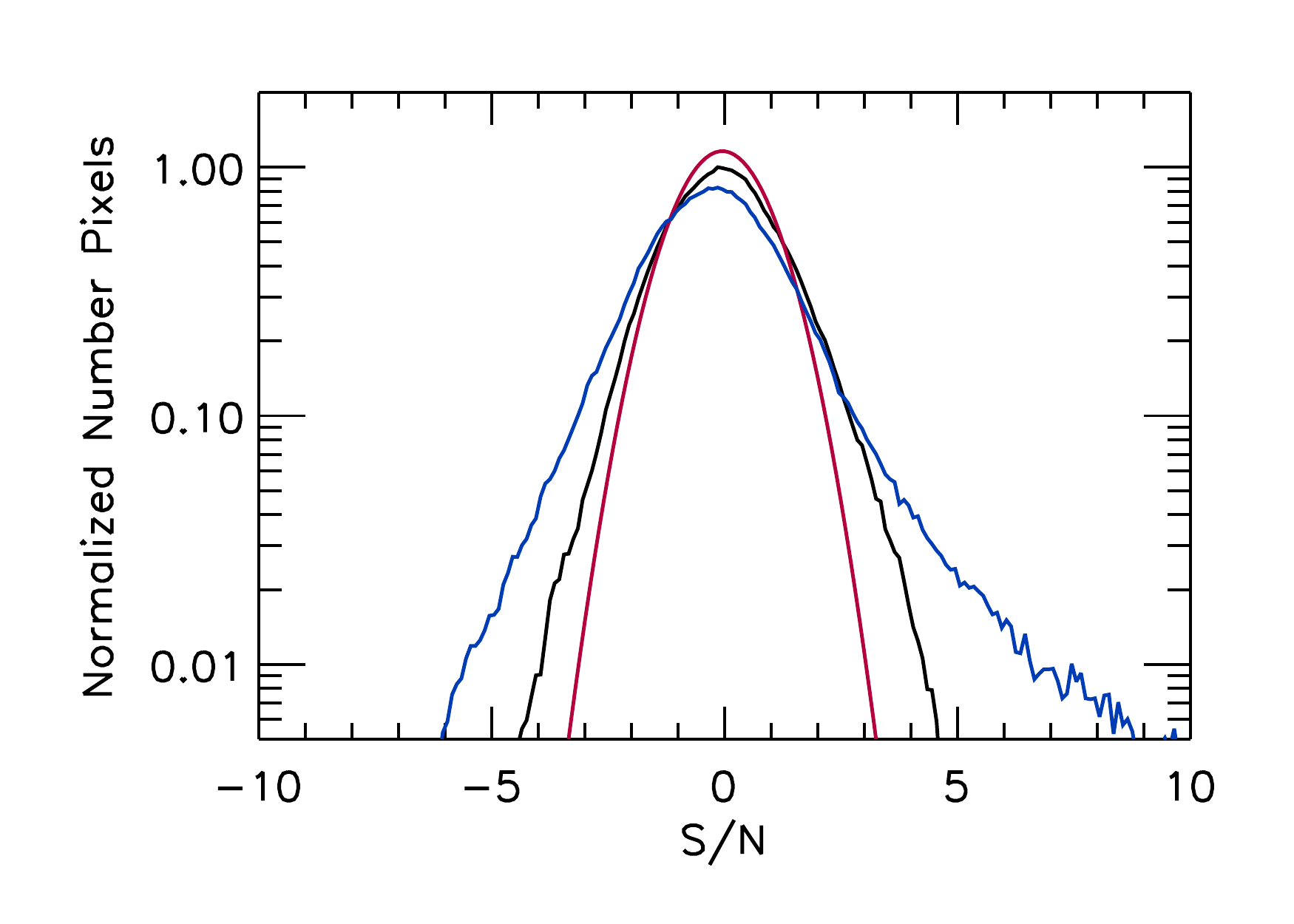}
\includegraphics[width=9cm]{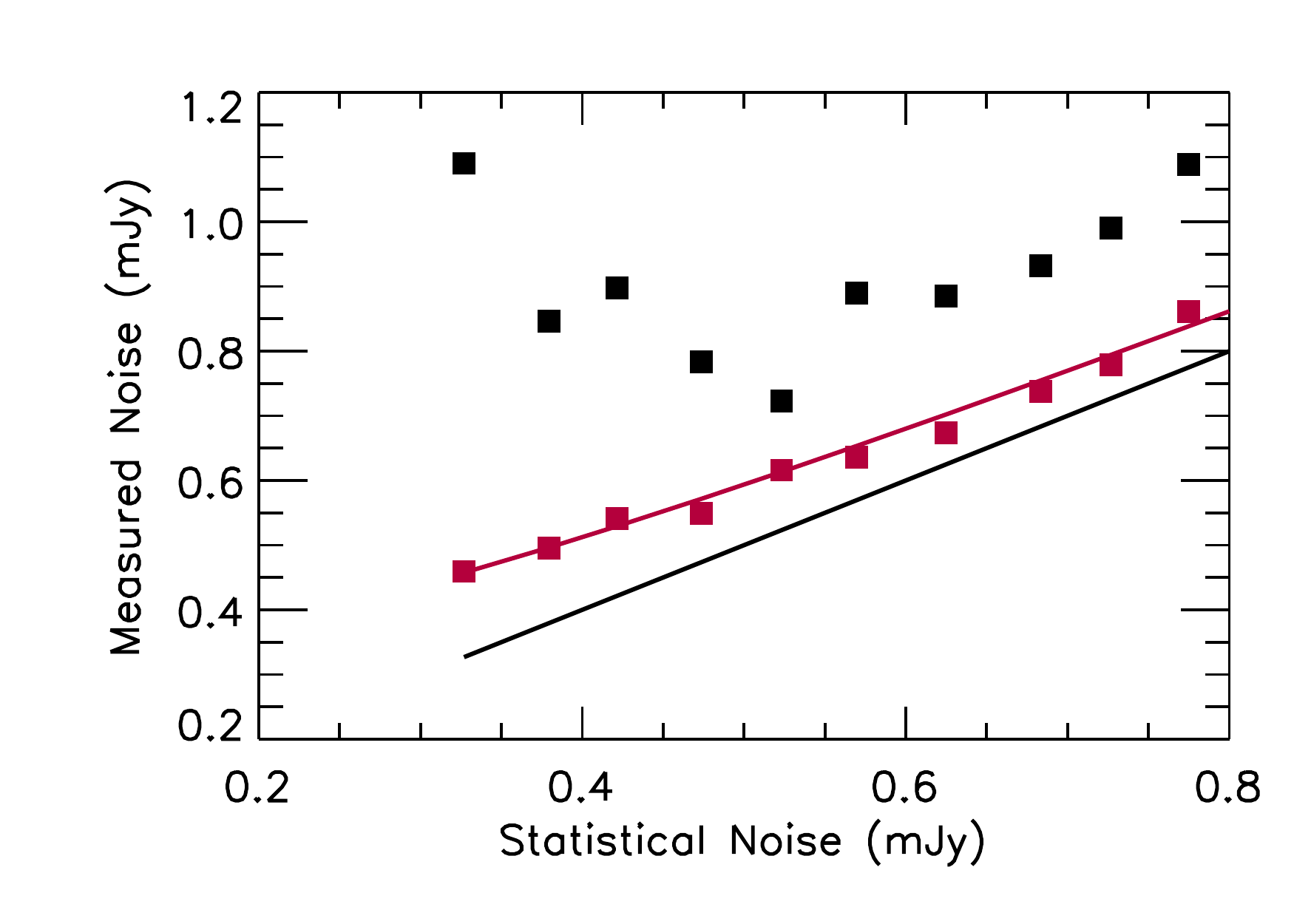}}
\caption{
(a) Distribution of S/N from pixel
to pixel in the central (low noise) region of the image
where the 850\,$\mu$m rms noise is less than 0.6\,mJy.
The blue curve shows the distribution for the raw image,
while the black curve shows the distribution when all
sources brighter than 2\,mJy are removed. The red curve
shows the expected distribution from the statistical noise.
(b) Measured 850\,$\mu$m rms noise in circular annuli vs. 
mean statistical noise. 
The black squares show the values measured in the raw
image. The red squares the values measured in the image after
the sources brighter than 2\,mJy are removed. The black curve shows
the expectation from the statistical noise, and the red curve
the result when a confusion noise of $\sigma_c = 0.32$\,mJy
is added in quadrature. 
\label{confused}
}
\end{figure*}

We calibrated the fluxes using standard Flux Conversion Factors 
(FCFs; 491\,Jy\,pW$^{-1}$ for 450\,$\mu$m and 537\,Jy\,pW$^{-1}$ for 850\,$\mu$m). 
We applied 10\% upward corrections to compensate for the flux lost 
during filtering (Chen et al.\ 2013b). We find that the flux calibrations 
(Dempsey et al.\ 2013) are accurate to better than 10\%
based on the variation in the calibrators and the comparison with
the SMA data.

In Figure~\ref{pretty_850}(a), we show the 850\,$\mu$m image
made from all of the observations, including the band 3 data.
In Figure~\ref{pretty_850}(b), we show the rms noise as a function of the 
offset angle from the centroid position. In Figure~\ref{pretty_850}(c),
we show the cumulative area observed below a given rms noise.
For the 450\,$\mu$m image, we only used the band 1 and band 2 data.
We show this image in Figure~\ref{noise_radius_450}(a). In
Figure~\ref{noise_radius_450}(b), we show rms noise as a function of the 
offset angle from the centroid position. In Figure~\ref{noise_radius_450}(c),
we show the cumulative area observed below a given rms noise.

Following Chen et al.\ (2013b), we generated the source catalogs by 
identifying the peak S/N pixel, subtracting this peak pixel 
and its surrounding areas using the PSF scaled and centered
on the value and position of that pixel, and then searching 
for the next S/N peak. We iterated this process until we 
reached a S/N of 3.5. We then limited the sample 
to the sources with S/N above four. There are 186
detected sources at 850\,$\mu$m, of which 154 have fluxes above 2\,mJy. 
We present the SCUBA-2 catalogs in Section~\ref{catalog};
the 850\,$\mu$m sample can be found in Table~5.

Given the number of resolution elements ($\sim 10^4$) in the SCUBA-2 
image, a normal distribution would give $\sim 0.3$ false sources at
the $4\,\sigma$ level. In order to address systematic effects,
we also analyzed the negative of the image. To remove the
negative troughs produced in the processing, we first cleaned
the image of all the positively detected sources. We then
searched for sources in the negative of the image, finding four sources
with fluxes above 2\,mJy and S/N above 4. This may be compared
with the 156 sources detected in the positive image with the
same thresholds.

The central regions of the 850\,$\mu$m image have very low noise, 
and confusion effects may become significant (e.g., Condon 1992).
(Confusion may be safely ignored in the 450\,$\mu$m image, which is
higher resolution and shallower.)
There are two types of confusion: the first is blending, which is usually 
considered to become important when there is more than one
source per 30 beams, and the second is noise from
sources below the flux limit of interest, which we call $\sigma_c$.
The total noise is the combination of the noise measured from the
SCUBA-2 signal, which we refer to as the 
statistical noise, $\sigma_s$, and $\sigma_c$; it is given by
$\sigma=(\sigma_s^2 + \sigma_c^2)^{0.5}$.

We can estimate both the blending limit and $\sigma_c$ from the fits to 
the 850\,$\mu$m counts of Hsu et al.\ (2016). Based on the counts, we find
more than one source per 30 beams below 1.64\,mJy, which is a lower flux
than that of any of the sources in the present sample. 
Thus, blending is not that important for our sample.
In order to estimate $\sigma_c$, we simulated an image using the
counts and the PSF of the matched-filter image and then measured the
dispersion in the simulated image. 
We found $\sigma_c=0.33$\,mJy for sources
with fluxes less than 2\,mJy. The value of $\sigma_c$
is not sensitive to the low flux cutoff, which we took as 0.005\,mJy.

We can also estimate $\sigma_c$ directly from the image
by comparing the measured dispersion with the statistical noise.
As the statistical noise becomes smaller, we expect to see the
measured noise rise above it and reach a constant value equal
to $\sigma_c$. In Figure~\ref{confused}(a), using only the portion of the 
image with noise less than 0.6\,mJy, we show 
the distribution of the actual S/N relative to
the statistical noise for both the raw 
matched-filter image (blue curve) and 
the matched-filter image with the sources brighter than 2\,mJy removed (black curve).
The red curve shows the expectation from the statistical noise
alone. The figure emphasizes the importance of removing the
brighter sources, which otherwise would dominate the noise, but
even with these removed, the black curve is wider than the red curve
due to the confusion noise. In Figure~\ref{confused}(b), we show
the comparison of the measured noise in radial annuli (the statistical
noise increases with radius) versus the mean statistical noise
in each annulus. The black squares show the values measured from  
the matched-filter 
image, and the red squares the values measured from the matched-filter image with 
the sources brighter than 2\,mJy removed. 
The black squares have a median value  
of $\sim 0.89$\,mJy for the points with statistical noise less tha 0.6\,mJy,
which is very similar to that measured by Geach et al.\ (2017) for
the same quantity. This noise is dominated by the brighter
sources and shows significant variation due to the small
number statistics of these sources. In order to measure
$\sigma_c$ from the faint sources only, we use the measured noise
with the bright sources removed. We can compare this
with the statistical noise (black curve)
to estimate the $\sigma_c$ from sources fainter than 2\,mJy.
The red curve shows the result of adding $\sigma_c=0.33$
in quadrature to the statistical noise; it gives a good approximation
to the values measured from the image with 
the sources brighter than 2\,mJy removed. 
This value of $\sigma_c$ is in good agreement
with the value that we estimated from the counts and is comparable to
the minimum statistical noise in the image. Thus, we conclude
that we are just reaching the confusion noise limit in the image.

We finally tested the completeness of source recovery in the
central region. To do so, we added a small number of sources
(usually 5) of a given flux at random positions in the central low noise portion
of the image ($<0.6$\,mJy). 
We then ran the finding procedure to obtain
the source list including the introduced sources. We considered a source
to be recovered if it was within $4\farcs5$ of the
original position and had a flux that was less than $3\sigma$ 
different from its actual value. We repeated the procedure
100 times for each flux. We summarize the results in Table~2
for sources with input fluxes of 5\,mJy, 3\,mJy, and 2\,mJy,
where we give the input flux, the 
percentage recovered, the mean recovered flux divided by the input flux,
and the dispersion of the recovered sources,  which we show as the 
the mean statistical noise  of the sample together with the additional
$\sigma_c$ required to match the measured dispersion. We omit $\sigma_c$ from the table for the
2\,mJy sources where the dispersion is biased by the omission of the 
downward scattered sources which are not recoverd.
The recovery rate is high ($>90$\% at
3\,mJy), upward boosts are small, and the inferred noise is consistent
with our previous estimates of $\sigma_c$.

\begin{deluxetable*}{ccccc}
\tablecaption{Source Recovery}
\tablehead{Flux & Recovered  & Boost & $\sigma_s$ & $\sigma_c$   \\ (mJy) & Fraction &  & (mJy) & (mJy)}
\startdata
5   & 0.93 &     1.01  &  0.40 & 0.37   \\
3   & 0.91 &     1.01  &  0.40 & 0.33   \\
2   & 0.72 &     1.11  &  0.40 & \nodata   \\
\enddata
\label{recover}
\end{deluxetable*}

\subsection{SMA Observations}
\label{sec_SMA}

We need interferometric follow-up observations
to identify the true optical, NIR, and MIR
counterparts to the SCUBA-2 sources. With such data, we can
find accurate positions and determine whether a
SCUBA-2 source is a blend of emission from multiple galaxies. 
Such multiples are not uncommon in the brighter submillimeter detected
populations (e.g., Wang et al.\ 2011; Barger et al.\ 2012; 
Smol\v{c}i\'{c} et al.\ 2012; Chen et al.\ 2013b; Hodge et al.\ 2013; 
Karim et al.\ 2013; Simpson et al.\ 2015; Miettinen et al.\ 2015).
In Section~\ref{vla}, we discuss how most of the SCUBA-2 sources can 
be identified with ultra-deep VLA 20\,cm imaging. 
However, sometimes there are two or more potential 20\,cm counterparts; 
thus, it is generally best if the positions can be measured from submillimeter 
interferometry. 
We observed nearly all of the brighter SCUBA-2 sources with the SMA. 
Including archival data from other SMA programs, we have identifications for 
33 submillimeter sources with the SMA. These correspond
to 28 of the SCUBA-2 sources. Because of the sensitivity
limits, we did not attempt spectroscopy with the SMA.

We performed the calibration and data inspection using the
IDL-based Caltech package MIR modified for the SMA. We
generated continuum data by averaging the spectral channels
after doing the passband phase calibration. We used both gain 
calibrators to derive gain curves. For consistency checks,
we compared these results with those obtained by adopting 
just one calibrator. We did not find any systematic differences. 
We performed the flux calibrations using data taken under 
conditions (time, hour angle, and elevation) similar to the
conditions of the flux calibrator observations.
The flux calibration error is typically 
within $\sim10$\% with this method.
We exported the calibrated interferometric visibility data
to the package MIRIAD for subsequent imaging and analysis.
We weighted the visibility data inversely proportional to the 
system temperature and Fourier transformed them to form images.
We also applied the ``robust weighting'' of Briggs (1995), with
a robust parameter of 1.0, to obtain a better balance between
beam size and S/N. We CLEANed the images
around detected sources to approximately 1.5 times the
noise level to remove the effects of the sidelobes. 
(The results are not sensitive to choosing a slightly deeper CLEANing
level, such as 1.0 times the noise.) 
The upgrade of the SMA to a 4~GHz bandwidth during the course
of our observing program considerably improved the continuum 
sensitivity and made calibrations with fainter quasars easier.
We typically achieved rms $\sim1.2-1.5$\,mJy in a night with the old
2\,GHz bandwidth and rms $\sim0.7-0.9$\,mJy in a night with the new 4\,GHz 
bandwidth. All of the SMA fluxes and flux errors that we quote are primary-beam 
corrected. We measured source positions and fluxes by fitting the 
images with point-source models using the MIRIAD IMFIT routine.

In Table~3, we provide the identification number (Column~1), the name based 
on the SMA coordinate (Column~2), 
the right ascension and declination measured from the SMA image
(Columns~3 and 4), the 860\,$\mu$m SMA flux and error (Columns~5 and 6),
and the S/N (Column~7) for the 33 4$\,\sigma$ detected SMA sources.
All of the SMA detected sources correspond
to 20\,cm sources (Section~\ref{vla}), and we include the
20\,cm fluxes in the table (Column~8). We also include spectroscopic
redshifts, where available (Column~9; see Section~\ref{redshifts}).
Finally, we give the identification number from Table~5
of the nearest SCUBA-2 850\,$\mu$m source (Column~10).

\begin{figure}
\begin{center}
\centering{\includegraphics[width=4.1in,angle=0]{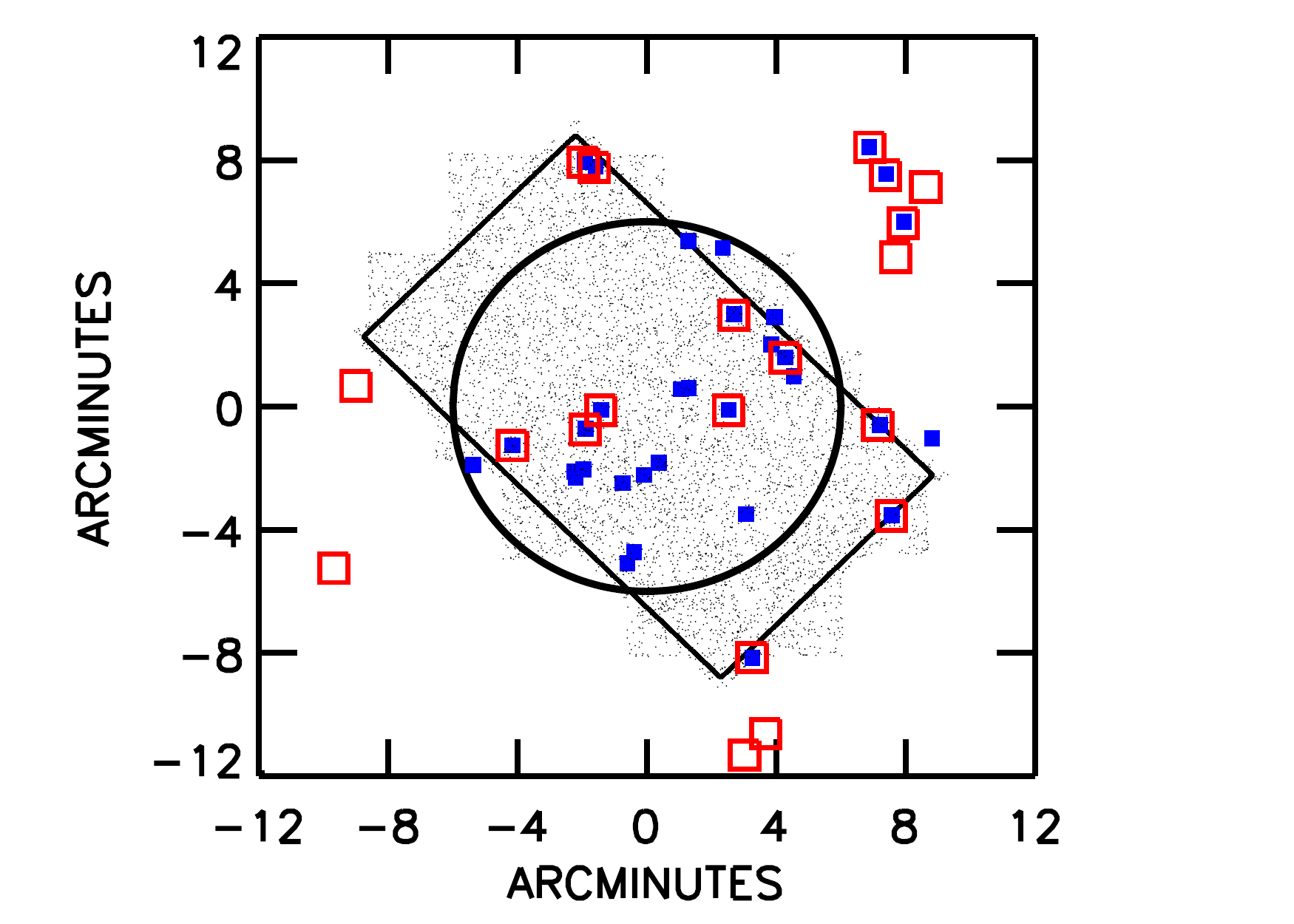}}
\caption{
Positions of the 19 $4\,\sigma$ SCUBA-2 sources with fluxes $>7$\,mJy 
(red open squares) and of the 33 $4\,\sigma$ SMA sources (blue squares).
The large circle shows a $6'$ radius where the 850\,$\mu$m rms noise 
reaches the confusion limit for the JCMT of 2\,mJy ($4\,\sigma$). 
This is also the deepest portion of the {\em Chandra\/} X-ray image.
The shading shows the full region of the {\em HST\/} GOODS-N imaging,
while the rectangle shows the most uniformly covered portion. 
The GOODS-{\em Herschel\/} PACS observations
cover roughly the same area as the {\em HST\/} GOODS-N observations,
while the GOODS-{\em Herschel\/} SPIRE observations cover the full SCUBA-2 field.
\label{scuba2_schematic}
}
\end{center}
\end{figure}

Six of the SMA detected sources (three pairs) are separated by less
than the SCUBA-2 FWHM
(identification numbers (21, 27), (28, 32), and (29, 31) from Table~3, 
with the last being
an extremely close pair with a $2''$ separation). All of
these sources were identified as multiple contributors to single
SCUBA sources in Wang et al.\ (2011) and Barger et al.\ (2012). 
(Note that the (28, 32) pair was considered a triplet with 22 in those
papers, but 22 is farther away and now corresponds to
SCUBA-2 source 77 in Table~5.)
While two of the pairs still correspond to single sources in the present SCUBA-2
data, the (28, 32) pair now corresponds to two separate SCUBA-2 sources,
119 and 102, respectively, in Table~5.
Thus, we are left with only two clear multiples in the SCUBA-2 sources
with SMA observations.

\begin{deluxetable*}{ccccrrrccc}
\renewcommand\baselinestretch{1.0}
\tablewidth{0pt}
\tablecaption{SMA $4\,\sigma$ Sources}
\scriptsize
\tablehead{No & Name & R.A. & Decl.&  Flux  & Error & S/N  & 20cm & z & SCUBA-2 \\ & & J2000.0 & J2000.0 & (mJy) & (mJy) &  & ($\mu$Jy) &  & No.\\ (1) & (2) & (3) & (4) & (5) & (6) & (7) & (8) & (9) & (10) }
\startdata
   1 &  SMA123711622211 &       12       37 11.70 &       62       22 11.9 &  23.9 &  2.50 &  9.5 &   87 &   4.055 &      2\cr
   2 &  SMA123555622239 &       12       35 55.85 &       62       22 39.3 &  17.0 &  1.90 &  8.9 &   56 & \nodata &      4\cr
   3 &  SMA123730621258 &       12       37 30.79 &       62       12 58.9 &  14.9 & 0.90 &  16.5 &   126 & \nodata &      3\cr
   4 &  SMA123546622013 &       12       35 46.65 &       62       20 13.7 &  14.0 &  2.20 &  6.3 &   85 & \nodata &      5\cr
   5 &  SMA123551622147 &       12       35 51.35 &       62       21 47.3 &  13.7 &  2.80 &  4.8 &   51 & \nodata &      1\cr
   6 &  SMA123633621408 &       12       36 33.44 &       62       14 8.69 &  12.0 &  1.40 &  8.5 &   33 &   4.042 &      8\cr
   7 &  SMA123627620605 &       12       36 27.19 &       62 06 5.50 &  11.5 & 0.70 &  16.4 &   34 & \nodata &      7\cr
   8 &  SMA123550621042 &       12       35 50.22 &       62       10 42.4 &  10.1 &  2.70 &  3.7 &   25 & \nodata &     12\cr
   9 &  SMA123708622202 &       12       37 8.825 &       62       22 2.09 &  9.2 &  1.00 &  9.2 &   143 &   4.051 &     10\cr
  10 &  SMA123634621923 &       12       36 34.92 &       62       19 23.7 &  8.9 &  2.10 &  4.2 &   63 & \nodata &     20\cr
  11 &  SMA123539621312 &       12       35 39.49 &       62       13 12.1 &  8.5 & 0.50 &  17.0 &   35 & \nodata &     39\cr
  12 &  SMA123651621225 &       12       36 51.98 &       62       12 25.7 &  7.8 &  1.00 &  7.8 &   12 &   5.183 &     26\cr
  13 &  SMA123628621045 &       12       36 28.84 &       62       10 45.2 &  7.7 & 0.90 &  8.5 &   48 & \nodata &     63\cr
  14 &  SMA123618621550 &       12       36 18.32 &       62       15 50.7 &  7.2 & 0.70 &  10.2 &   163 &   2.000 &     18\cr
  15 &  SMA123707621408 &       12       37 7.210 &       62       14 8.30 &  7.1 &  1.40 &  5.0 &   25 &   2.490 &     15\cr
  16 &  SMA123741621220 &       12       37 41.14 &       62       12 20.4 &  7.1 &  1.80 &  3.9 &   27 & \nodata &     35\cr
  17 &  SMA123631621714 &       12       36 31.92 &       62       17 14.7 &  7.1 & 0.50 &  14.2 &   21 & \nodata &     14\cr
  18 &  SMA123711621331 &       12       37 11.31 &       62       13 31.1 &  6.7 & 0.60 &  11.1 &   123 &   1.995 &     13\cr
  19 &  SMA123644621937 &       12       36 44.08 &       62       19 37.9 &  6.4 &  1.10 &  5.8 &   25 & \nodata &     24\cr
  20 &  SMA123622621616 &       12       36 22.08 &       62       16 16.2 &  5.4 & 0.60 &  9.0 &   26 & \nodata &     42\cr
  21 &  SMA123643621450 &       12       36 43.97 &       62       14 50.7 &  5.3 &  1.10 &  4.8 &   30 &   2.095 &     31\cr
  22 &  SMA123712621212 &       12       37 12.00 &       62       12 12.2 &  5.3 & 0.90 &  5.8 &   33 &   2.914 &     77\cr
  23 &  SMA123701621145 &       12       37 1.600 &       62       11 45.9 &  4.8 & 0.60 &  8.0 &   95 &   1.760 &     25\cr
  24 &  SMA123658620931 &       12       36 58.53 &       62 09 31.6 &  4.6 & 0.60 &  7.6 &   27 & \nodata &     65\cr
  25 &  SMA123655621201 &       12       36 55.92 &       62       12 1.91 &  4.5 & 0.80 &  5.6 &   25 &   2.737 &     80\cr
  26 &  SMA123553621337 &       12       35 53.23 &       62       13 37.9 &  4.3 & 0.80 &  5.3 &   41 &   2.098 &     19\cr
  27 &  SMA123646621448 &       12       36 46.09 &       62       14 48.5 &  4.2 & 0.80 &  5.2 &   101 & \nodata &     31\cr
  28 &  SMA123714621208 &       12       37 14.26 &       62       12 8.10 &  4.1 & 0.70 &  5.8 &   25 &   3.157 &    119\cr
  29 &  SMA123621621708 &       12       36 21.29 &       62       17 8.08 &  3.5 & 0.70 &  5.0 &   164 & \nodata &     51\cr
  30 &  SMA123616621513 &       12       36 16.10 &       62       15 13.7 &  3.4 & 0.60 &  5.6 &   36 &   2.578 &     43\cr
  31 &  SMA123621621709 &       12       36 21.10 &       62       17 9.59 &  3.4 & 0.60 &  5.6 &   44 &   1.992 &     51\cr
  32 &  SMA123714621156 &       12       37 14.03 &       62       11 56.4 &  3.2 & 0.90 &  3.5 &   22 & \nodata &    102\cr
  33 &  SMA123700620909 &       12       37 0.300 &       62 09 9.89 &  3.1 & 0.60 &  5.1 &   297 & \nodata &    112\cr
\enddata
\label{sma_table}
\end{deluxetable*}

In Figure~\ref{scuba2_schematic}, we compare the 
positions of the SMA detected sources (blue squares) with those of
the brighter SCUBA-2 sources (red open squares).
For the latter, we only show fluxes greater than 7\,mJy
that would be detected above the $4\,\sigma$ level throughout
the entire 450\,arcmin$^2$ field. 
All of the bright SCUBA-2 sources in the deep portion of the 
CDF-N (large circle) and the {\em HST\/} GOODS-N 
region (shading) have SMA detections.

\begin{figure}
\centerline{\includegraphics[width=4in,angle=0]{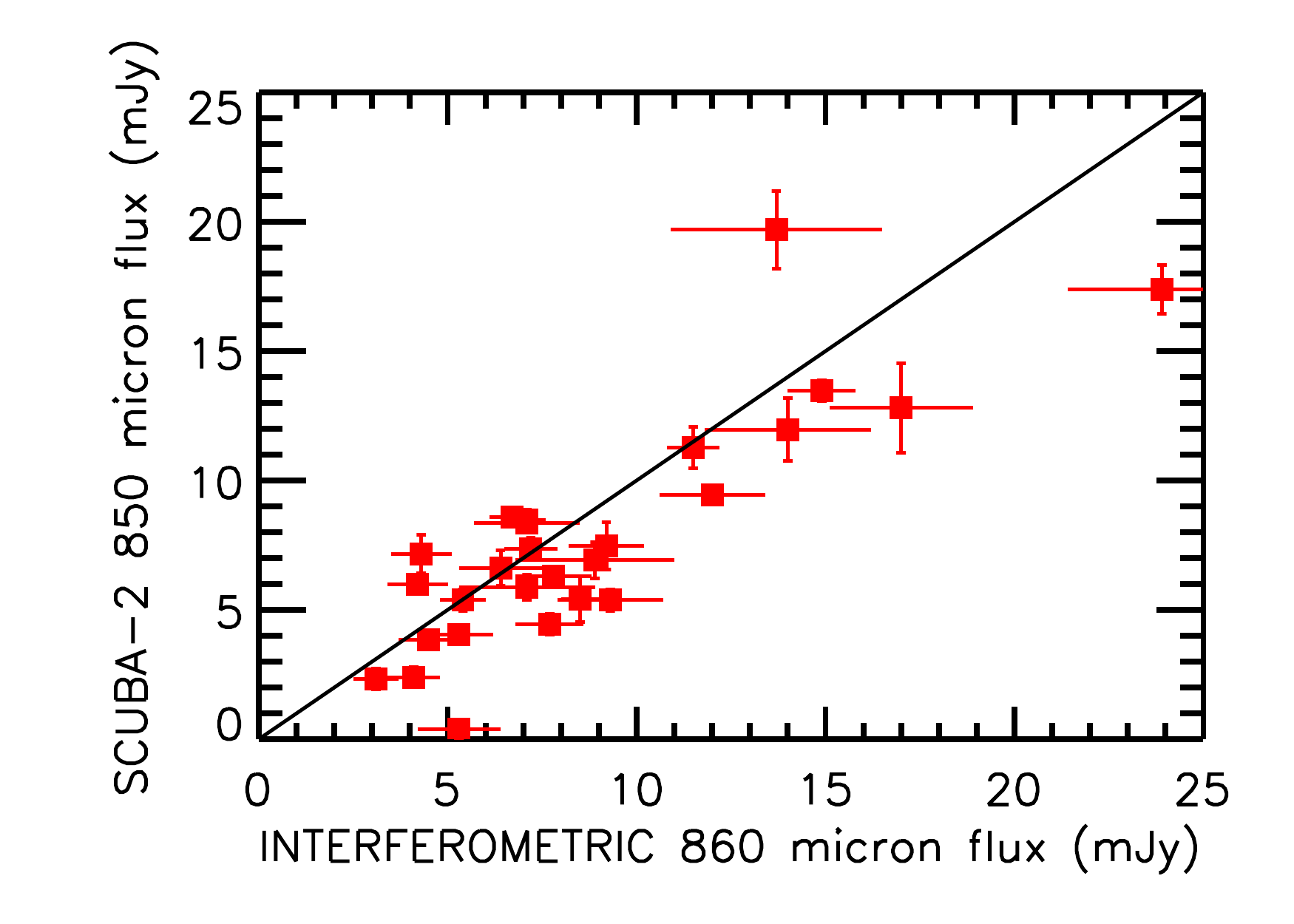}}
\caption{
SCUBA-2 flux vs. SMA flux (red squares).
Only SCUBA-2 sources with a single counterpart in the SMA field
are shown to avoid SCUBA-2 sources with multiple contributing galaxies.
The SCUBA-2 fluxes are measured at the nominal positions of
the SMA sources, which provide the most accurate positions in the
SCUBA-2 image.
Error bars are $1\,\sigma$ in each axis.
\label{flux_cal}}
\end{figure}

Since the measured SMA and SCUBA-2 fluxes are independent, we 
can use them to check the flux calibration. In Figure~\ref{flux_cal},
we plot SCUBA-2 850\,$\mu$m flux versus SMA flux.
The SCUBA-2 fluxes are 8\% lower than the SMA fluxes, which
is within the systematic error in the calibration. 
The point with the higher SMA flux than SCUBA-2 flux
is GN20 (Pope et al.\ 2005), 
where the measured SMA flux is from Younger et al.\ (2008).
Younger et al.\ measured GN20 in the extended configuration of the SMA;
thus, the correction to a total flux is somewhat uncertain. The
remaining sources were observed in the compact mode of the SMA
where the spatial resolution is $2''$ at 345\,GHz. We
do not expect these sources to be resolved, and the measured 
SMA flux should be a good approximation to the total flux.

\subsection{VLA 20~cm Observations}
\label{vla}
The upgrade of the VLA has greatly increased the sensitivity
of the decimetric continuum observations that can be obtained,
making it possible to identify radio counterparts to nearly all of the
brighter SCUBA-2 sources.
Here we use the extremely deep 20\,cm image of the CDF-N reduced by
F.~Owen (2017, in preparation).
The image covers a $40'$ diameter region with an effective resolution of
$1\farcs8$. The absolute radio positions are known to $0\farcs1-0\farcs2$ rms.
The highest sensitivity region, about $9'$ in radius, is closely
matched to the full area of the present SCUBA-2
survey and completely covers the deepest region, providing a relatively uniform 
20\,cm rms of 2.3\,$\mu$Jy for determining the radio fluxes of the
SCUBA-2 sources. There are 787 distinct $\ge5\,\sigma$ radio sources within the
$9'$ radius, excluding sources that appear to be parts of other sources.

\begin{figure*}
\vskip -4.5cm
\centerline{\includegraphics[width=4.5in,angle=90]{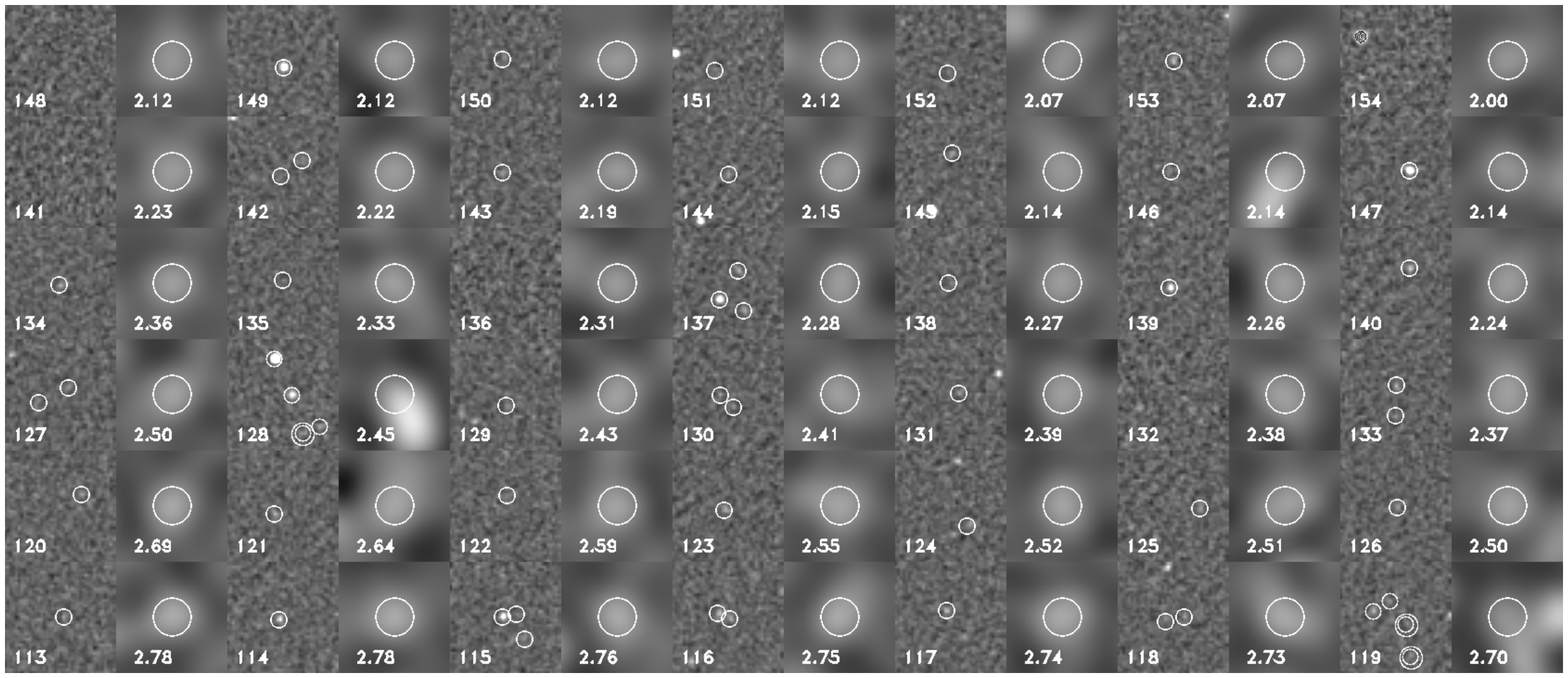}}
\vskip -3.8cm
\centerline{\includegraphics[width=4.5in,angle=90]{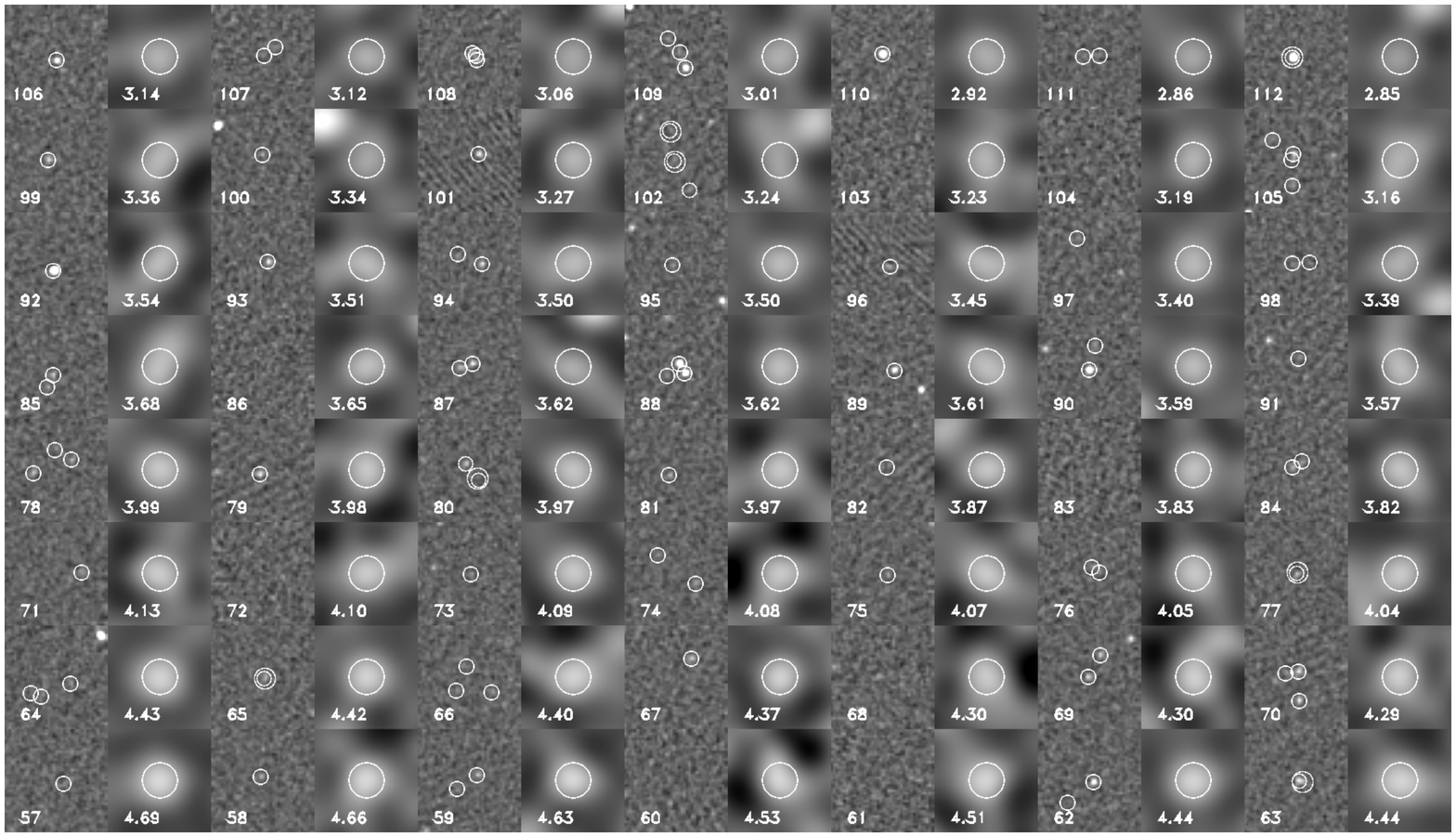}}
\vskip -3.8cm
\centerline{\includegraphics[width=4.5in,angle=90]{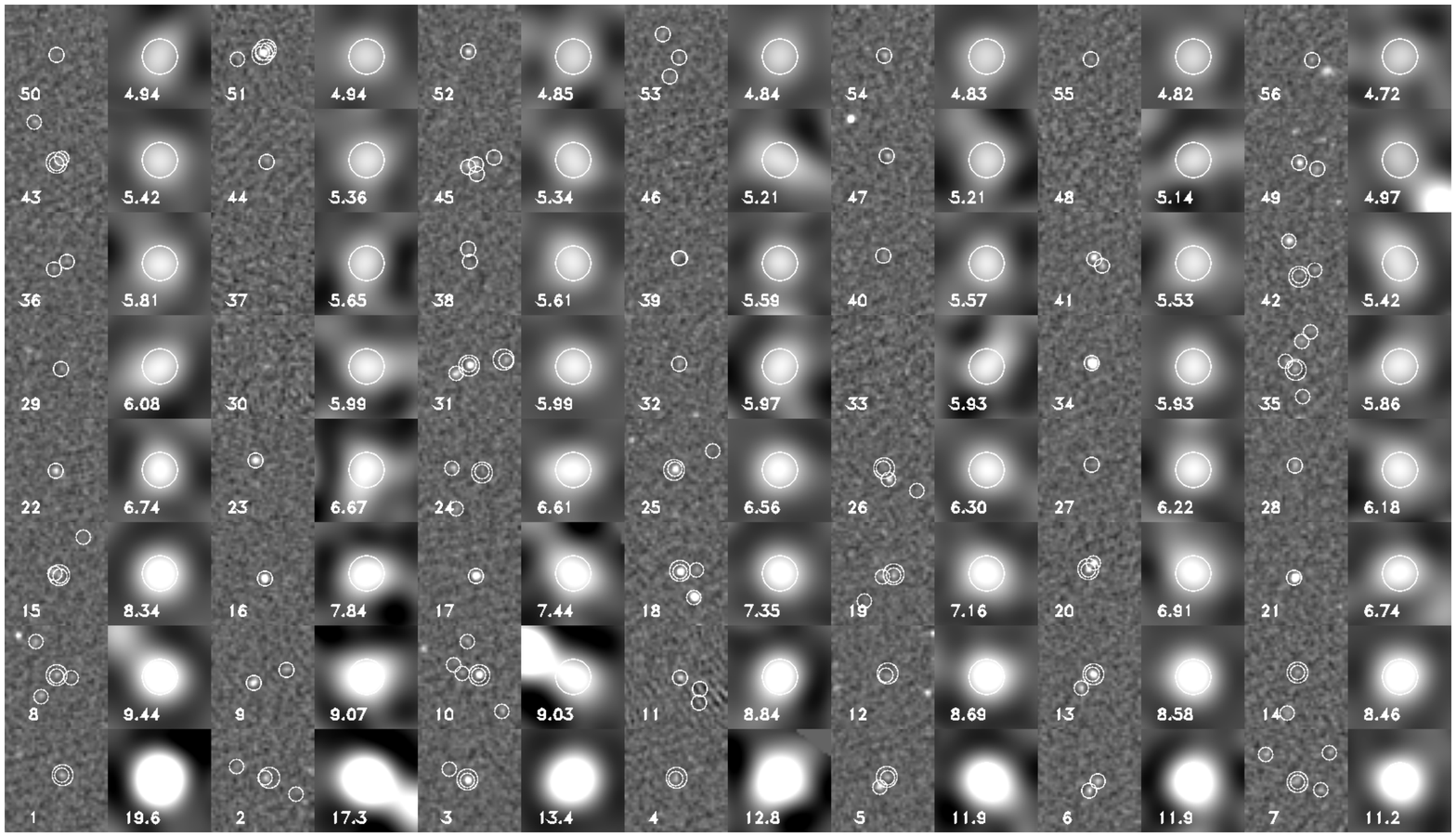}}
\caption{
Thumbnails ($40''$ on a side) of the $\ge4\,\sigma$ SCUBA-2 
sources with 850\,$\mu$m fluxes above 2\,mJy. 
In each case, we show the radio image 
in the left thumbnail and the SCUBA-2 image in the right thumbnail. 
In the left thumbnails, the small circles show
the positions of the $\ge5\,\sigma$ 20\,cm sources, while the larger circles
show the positions of the SMA detections. 
In the right thumbnails, the large circle
shows the $14''$ diameter beam width. The thumbnails
(source number given at the bottom of the left thumbnails) are shown in order 
of decreasing 850\,$\mu$m flux (flux given at the bottom of the right thumbnails), 
starting from the bottom left of the figure.
\label{radio_images}
}
\end{figure*}

\begin{figure*}
\vskip -4cm
\centerline{\includegraphics[width=5.0in,angle=90]{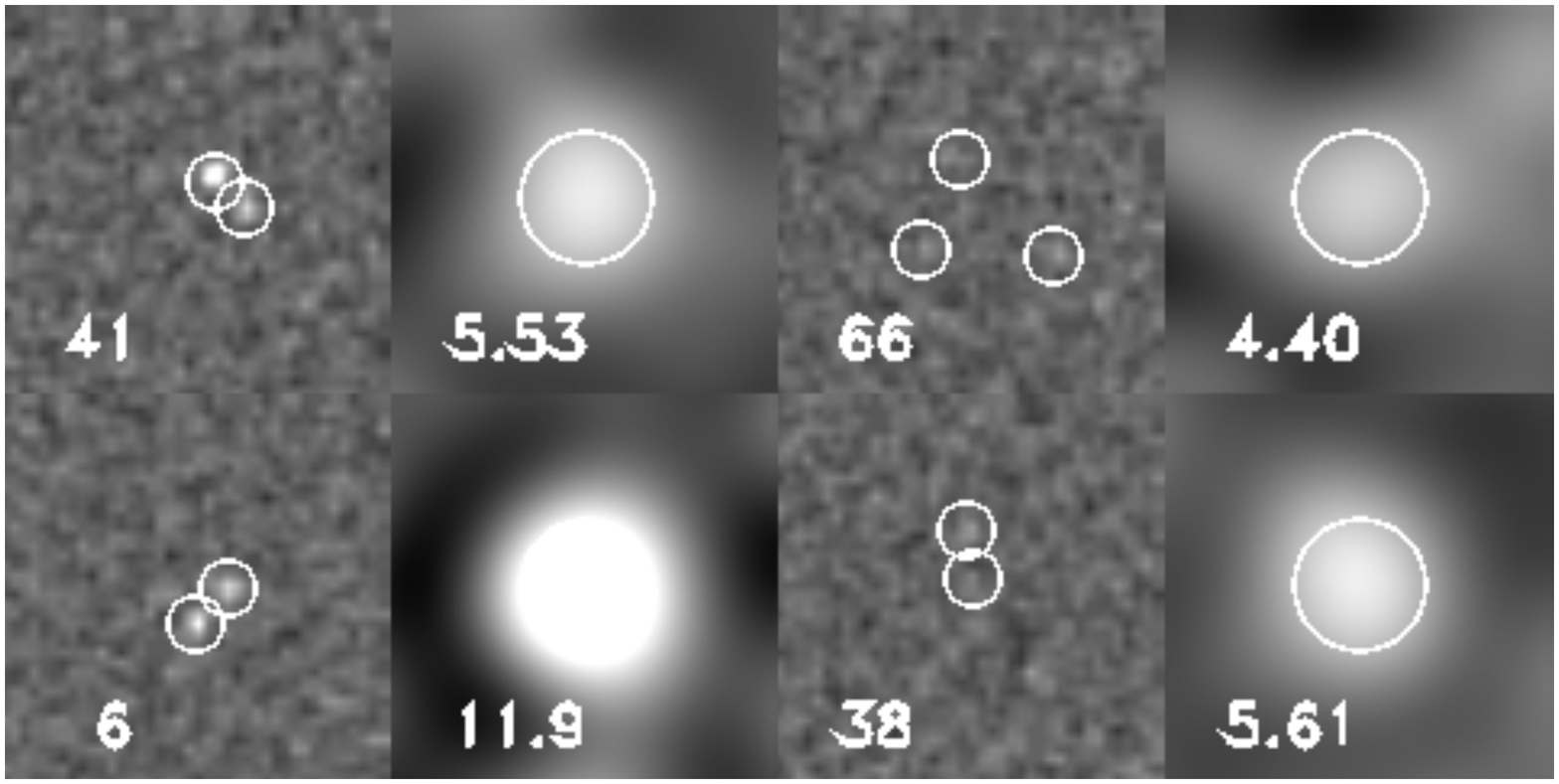}}
\caption{
Blow-ups of the thumbnails in Figure~\ref{radio_images}
corresponding to the four sources with
850\,$\mu$m fluxes above 4\,mJy that have multiple
radio counterparts but no SMA follow-up observations. 
In each case, we show the radio image 
in the left thumbnail and the SCUBA-2 image in the right thumbnail. 
In the left thumbnails, the small circles show
the positions of the $\ge5\,\sigma$ 20\,cm sources.
In the right thumbnails, the large circle
shows the $14''$ diameter beam width. The thumbnails
(source number given at the bottom of the left thumbnails) are shown in order 
of decreasing 850\,$\mu$m flux (flux given at the bottom of the right thumbnails), 
starting from the bottom left of the figure.
\label{thumb_blowup}
}
\end{figure*}

We first inspected each of the SCUBA-2 sources to determine
possible radio counterparts. In Figure~\ref{radio_images},
we show images of each of the SCUBA-2 sources with fluxes above 2\,mJy 
(right thumbnails) ordered by increasing 850\,$\mu$m flux (see Table~5) 
starting from the bottom left of the figure, along with
the corresponding radio images (left thumbnails). 
In the left thumbnails, we mark the sources from the $\ge5\,\sigma$ radio 
catalog with small circles and any SMA detections with larger circles.
As discussed in Section~\ref{sec_SMA} and shown in Table~3,
all of the SMA detections correspond to a radio source.

When an SMA detection determines the radio counterpart, or when
there is a single radio source within a $4\farcs5$ search radius from the 
SCUBA-2 position, then we consider the SCUBA-2 source to
have a unique counterpart identification, and we adopt the radio position 
as the accurate position for the SCUBA-2 source. This localizes
114 of the 186 sources, including 26
of the 29 sources with 850\,$\mu$m fluxes above 6\,mJy. 
Many of the remaining SCUBA-2 sources have multiple potential
radio counterparts and no SMA follow-up observation to decide on the correct one.
(See Figure~\ref{thumb_blowup}, where we show the four such
sources with 850\,$\mu$m fluxes above 4\,mJy.)  
However, 39 of the 154 850\,$\mu$m 
SCUBA-2 sources with fluxes greater than 2\,mJy 
have no potential radio counterparts within our match
radius. These are primarily drawn from the faintest SCUBA-2 sources;
the brightest source without a radio counterpart has a flux
of 6.67\,mJy (source 23 in Table~5 and Figure~\ref{radio_images}).

\begin{figure}
\centerline{\includegraphics[width=4in,angle=90]{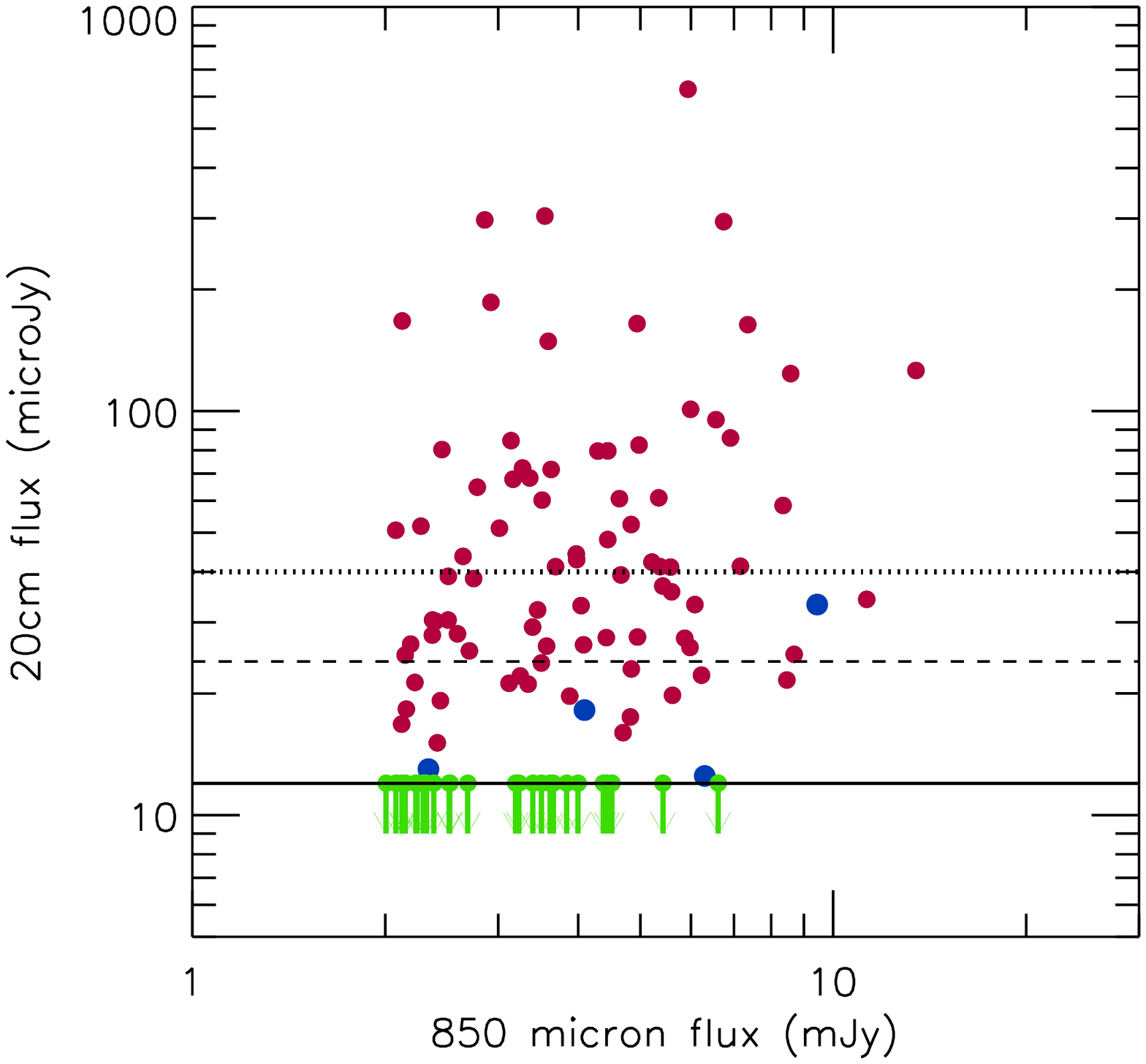}}
\vskip -1.0cm
\caption{
Radio counterparts to the $\ge4\sigma$ SCUBA-2 sources with 850\,$\mu$m
fluxes above 2\,mJy and lying within the $9'$ highest
sensitivity region of the radio image (red circles---sources with an SMA 
identification, or where there is a single radio counterpart within $4\farcs5$ 
of the SCUBA-2 position; green circles with downward pointing 
arrows---sources where there is no radio counterpart within the $4\farcs5$ search radius;
blue circles---sources where there is a spectroscopic redshift $z>3$.
The horizontal lines show the $5\,\sigma$ limits of the F.~Owen (2017, in preparation)
(solid), Morrison et al.\ (2010) (dashed), and Richards et al.\ (2000) (dotted)
20\,cm observations. 
\label{radio_counterpart}
}
\end{figure}

The present radio identification rate (more than 75\% for sources with
850\,$\mu$m flux $>2$\,mJy) represents a substantial improvement over 
previous radio matches, which typically yielded 60--70\% 
matching for sources with 850\,$\mu$m fluxes $>$5--8\,mJy
(e.g., Barger et al.\ 2000; Ivison et al.\ 2002; Chapman et al.\ 2003). 
In Figure~\ref{radio_counterpart}, we illustrate the gain
in counterpart matching as a function of the 20\,cm rms.

The radio blank sources are of particular interest
as being potentially at high redshift. The highest
redshift sources in the sample with known spectroscopic redshifts
$z>3$ (blue circles in Figure~\ref{radio_counterpart}) 
have low radio fluxes relative to their submillimeter fluxes (Carilli \& Yun 1999;
Barger et al.\ 2000), reflecting the opposing $K$-corrections
in the radio and submillimeter. 

\begin{figure}
\centerline{\includegraphics[width=4.0in,angle=0]{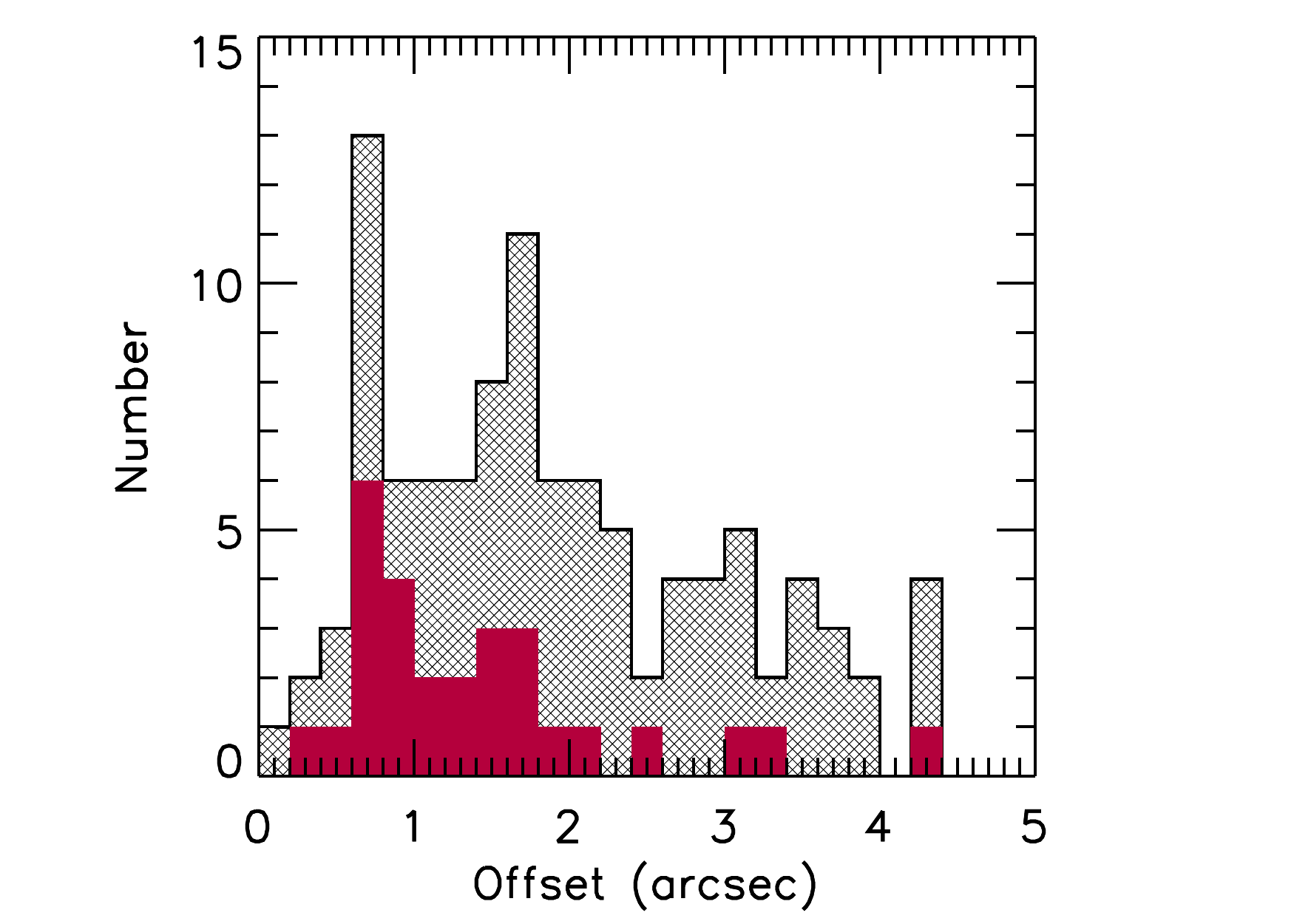}}
\caption{
Offsets from the SCUBA-2 850\,$\mu$m centroid positions
of the individual radio (black cross-hatching; the mean offset is $1\farcs9$) 
and SMA (red; the mean offset is $1\farcs4$
for these higher flux 850\,$\mu$m sources) counterparts. 
\label{offset_hist}
}
\end{figure}

We may use the offsets of the counterparts relative to the SCUBA-2
850\,$\mu$m centroid positions to estimate the
accuracy of the SCUBA-2 positions and to justify 
our choice of a $4\farcs5$ matching radius. We first checked
the absolute astrometric pointing of the SCUBA-2 observations. Comparing
with the SMA detected sources, we find an absolute offset of $0\farcs75$ in 
right ascension and $0\farcs4$ in declination. 
Comparing with the VLA detected sources, we find $0\farcs5$ in right ascension
and $0\farcs3$ in declination. Both offsets are small compared with the positional
uncertainties in the SCUBA-2 sources. In Figure~\ref{offset_hist},
we show the offsets of the individual SMA (red) and
radio (black cross-hatching) counterparts. The mean offsets
are $1\farcs4$ for the (on average, brighter) sample with SMA
counterparts and $1\farcs9$ for the larger and fainter sample with
radio counterparts. Nearly all of the SMA counterparts
lie within $3\farcs5$ of the SCUBA-2 position, and 96\% of the
radio counterparts lie within $4''$. This positional uncertainty emphasizes
the difficulty of working directly with the SCUBA-2 positions.
In our analysis, we will focus on the SCUBA-2 sources with
unique counterparts and accurate positions. However, it should be noted 
that most of the very high S/N SCUBA-2 sources do have
fairly accurate positions from the SCUBA-2 data alone. For example,
for the SCUBA-2 sources with S/N$>10$, the mean offset from
the SMA or radio counterparts is $1\farcs2$, while at S/N$>6$,
it is $1\farcs7$.

\subsection{Optical and NIR Counterparts}
\label{sec_optnir}
Of the 102 SCUBA-2 sources with accurate positions (i.e., with SMA 
measurements and/or single VLA counterparts) and 850\,$\mu$m
fluxes above 2\,mJy (there are 114 total with accurate positions; 
see Section~\ref{vla}), 57 lie in the most uniformly covered portion of the 
{\em HST\/} GOODS-N (rectangle in Figure~\ref{scuba2_schematic}).
Of these, 43 have magnitudes from the {\em HST\/} GOODS-N catalog
(Giavalisco et al.\ 2004; we use the auto magnitudes) using
a $1\farcs5$ matching radius. In the NIR, we use
similarly measured auto magnitudes from the {\em HST\/} CANDELS data 
(Grogin et al.\ 2011; Koekemoer et al.\ 2011).
The CFHT WIRCAM $K_s$ image from Wang et al.\ (2010) covers
the full SCUBA-2 field, so we also use corrected $3''$ diameter aperture 
magnitudes from this image (102 sources; see Table~5).

\begin{figure*}[h]
\includegraphics[width=3.0in,angle=0]{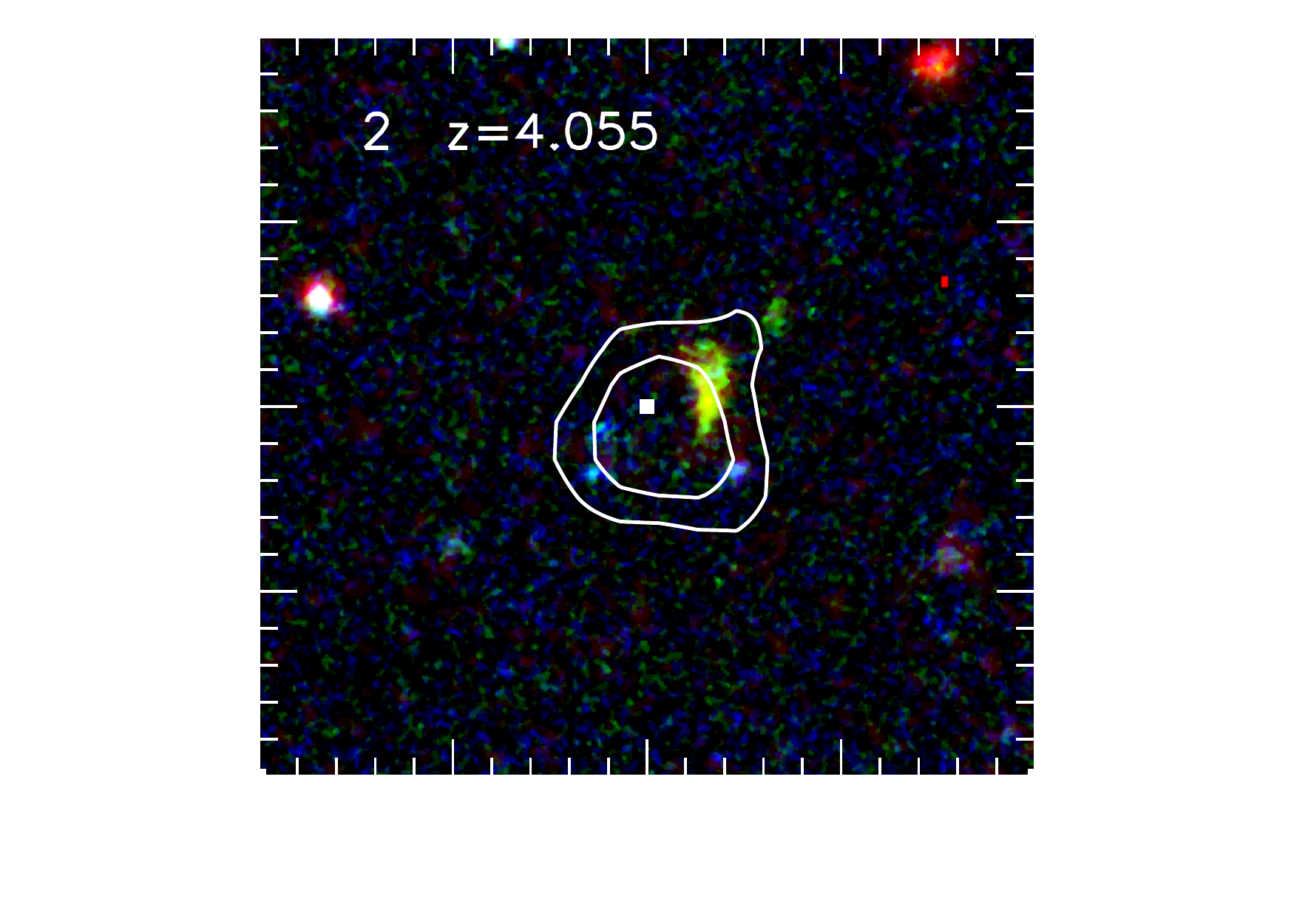}
\hspace{-2.5cm}\includegraphics[width=3.0in,angle=0]{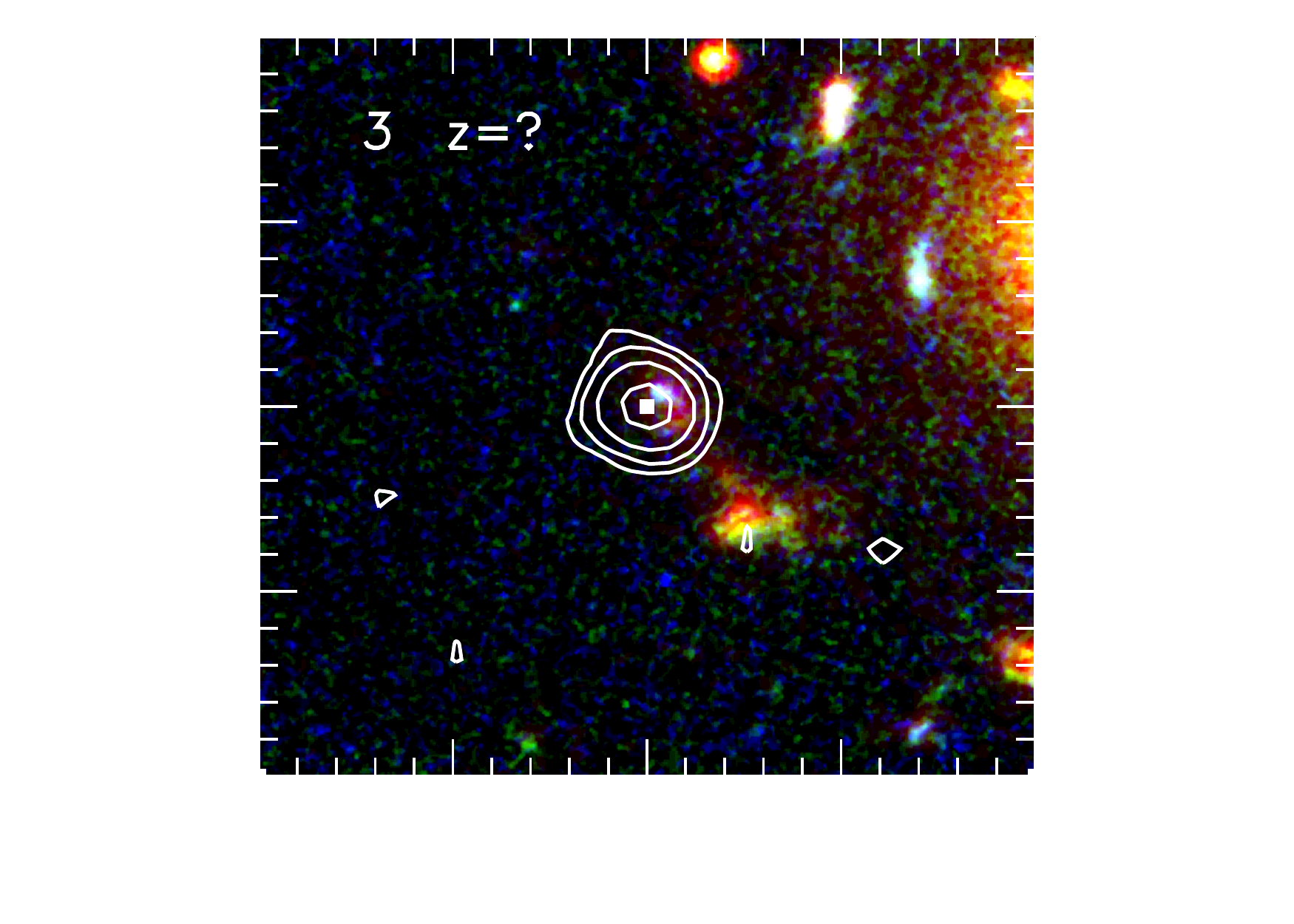}
\hspace{-2.5cm}\includegraphics[width=3.0in,angle=0]{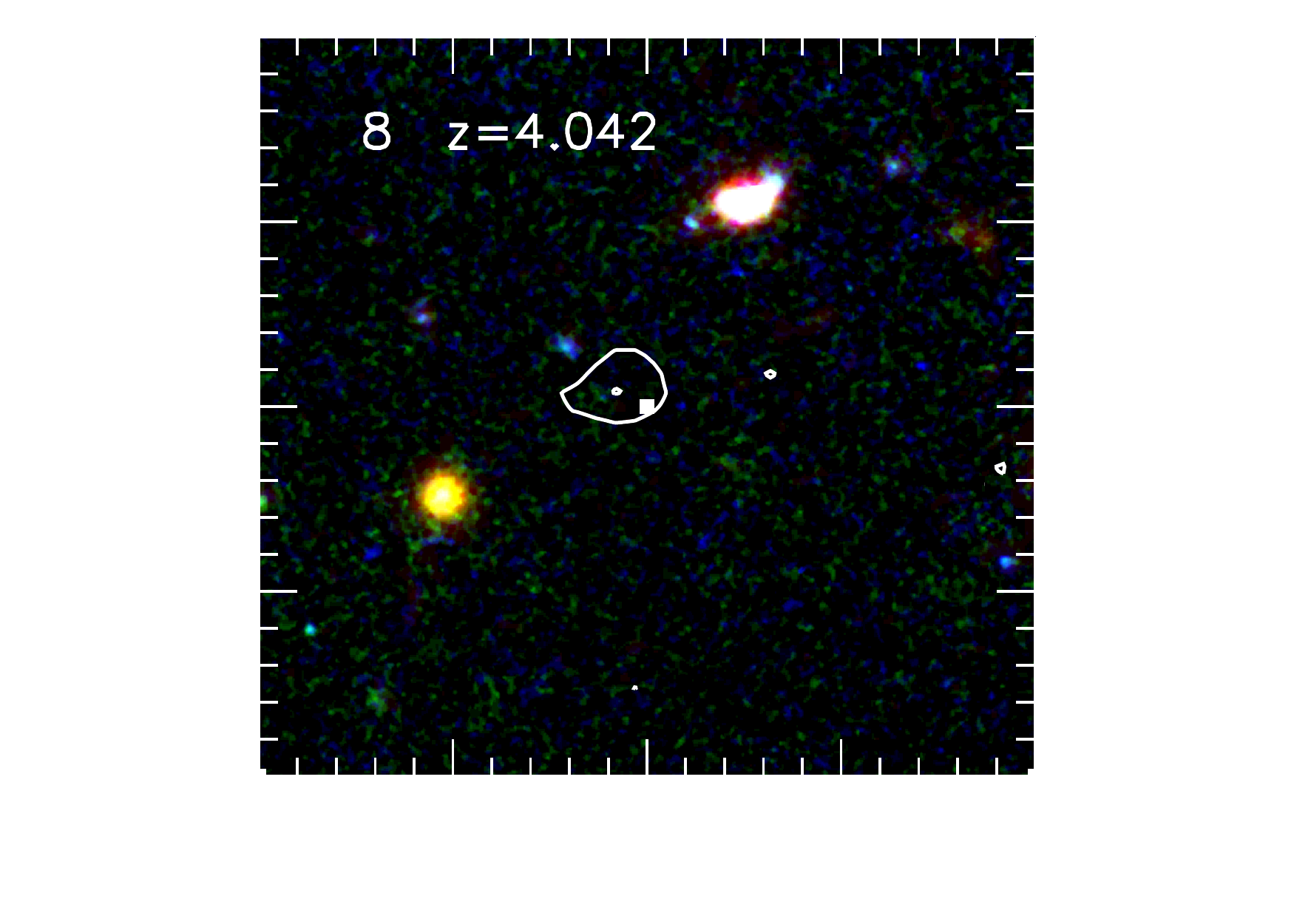}
\vspace{-0.5cm}
\includegraphics[width=3.0in,angle=0]{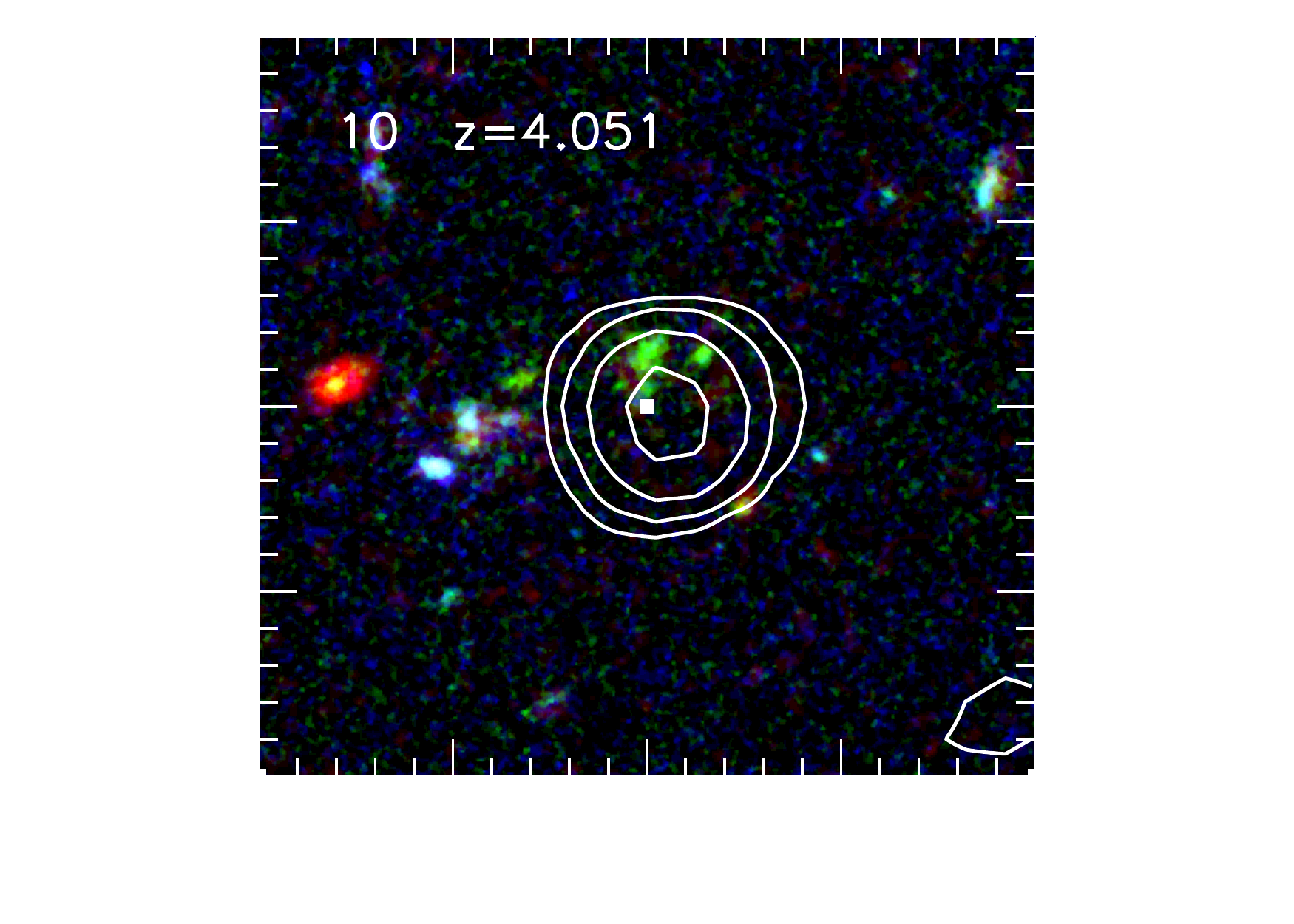}
\hspace{-2.5cm}\includegraphics[width=3.0in,angle=0]{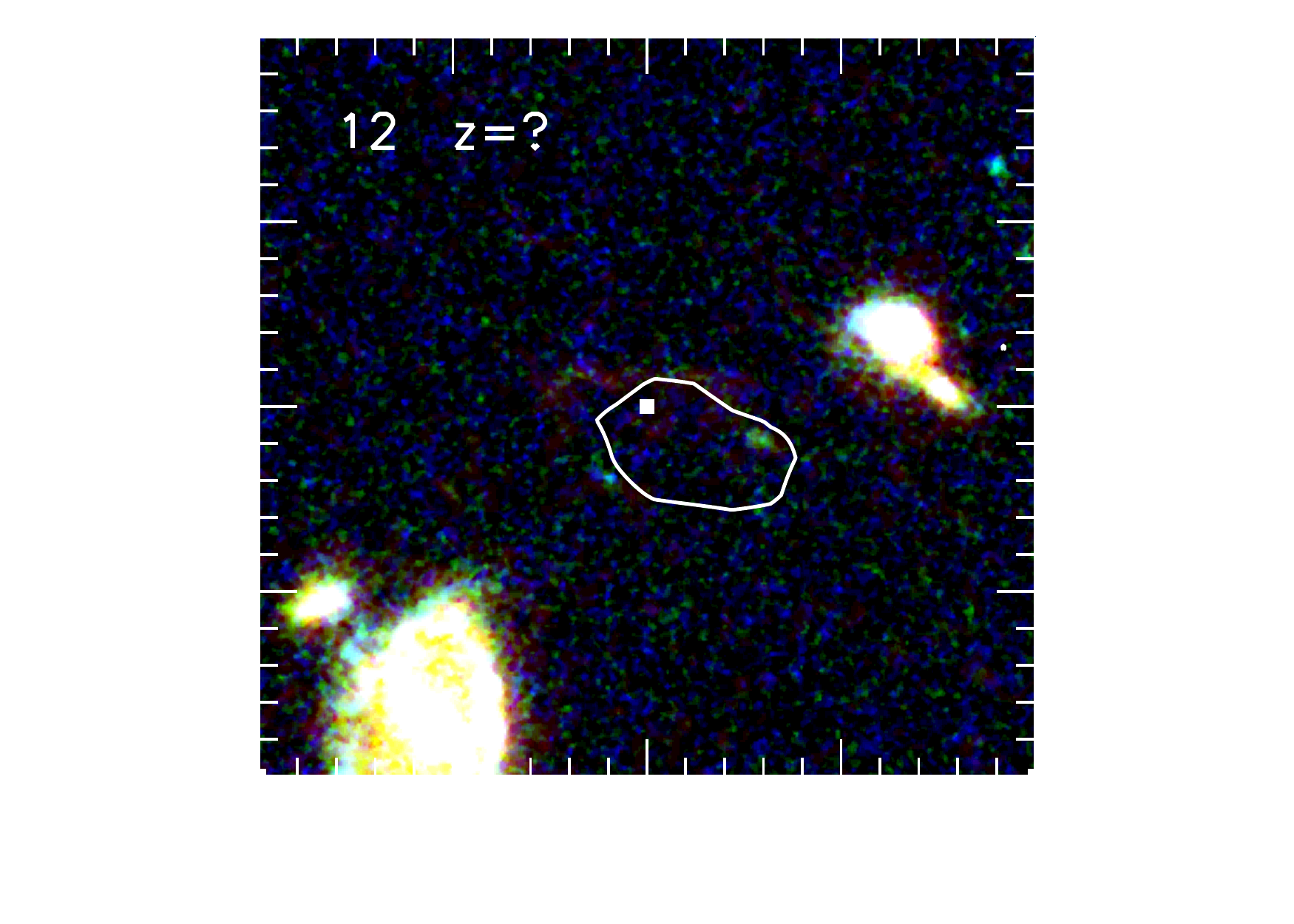}
\hspace{-2.5cm}\includegraphics[width=3.0in,angle=0]{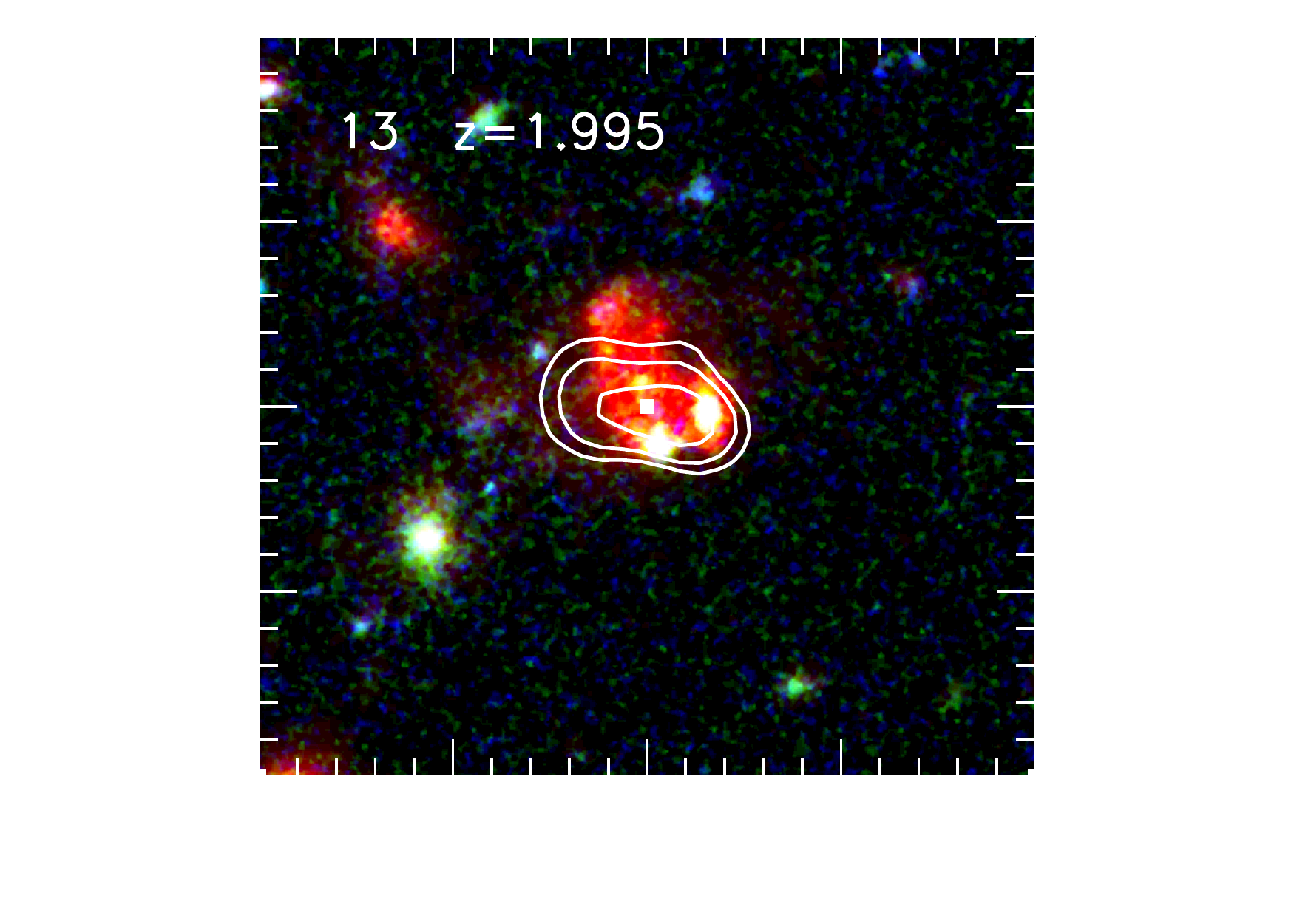}
\vspace{-0.5cm}
\includegraphics[width=3.0in,angle=0]{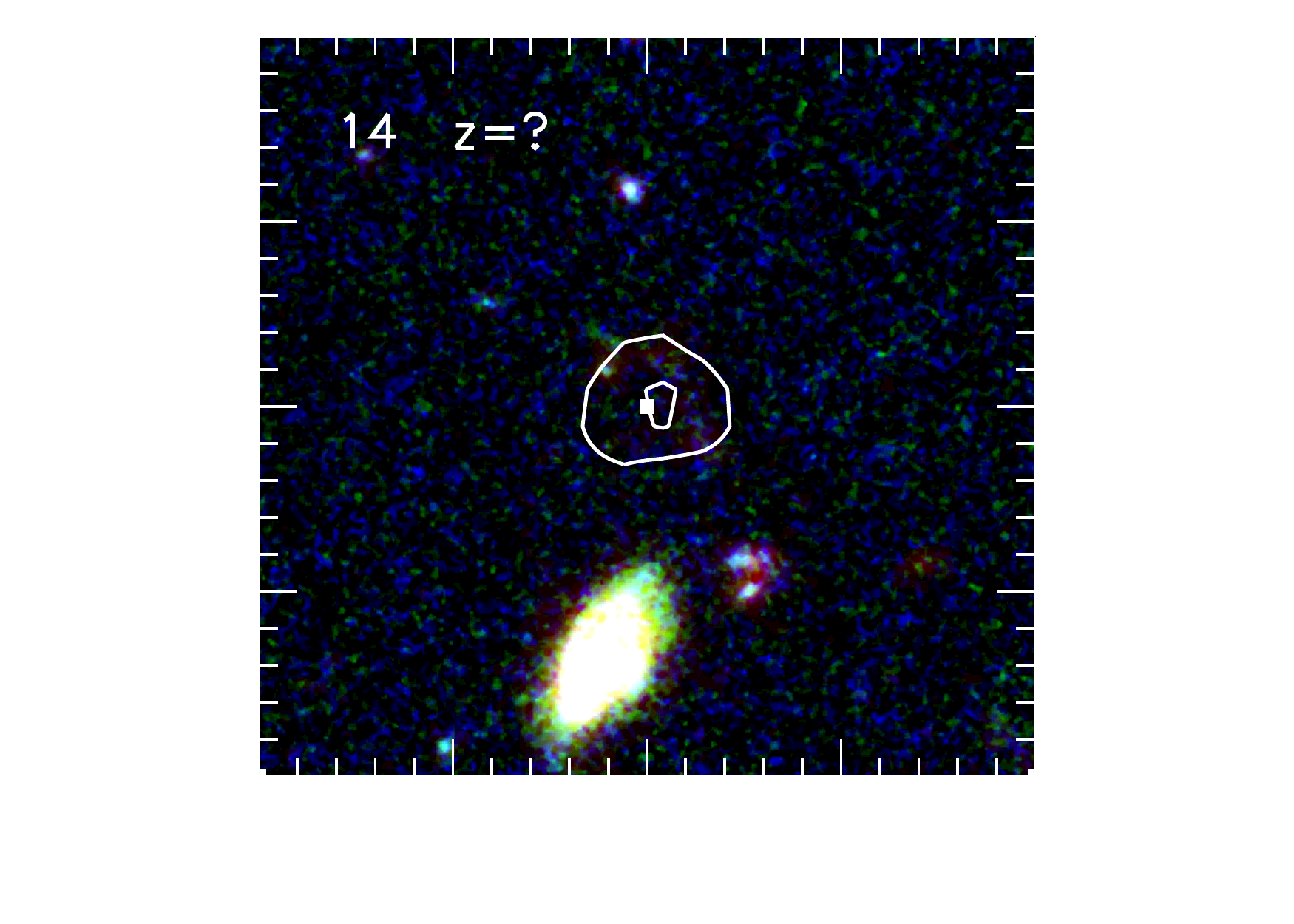}
\hspace{-2.5cm}\includegraphics[width=3.0in,angle=0]{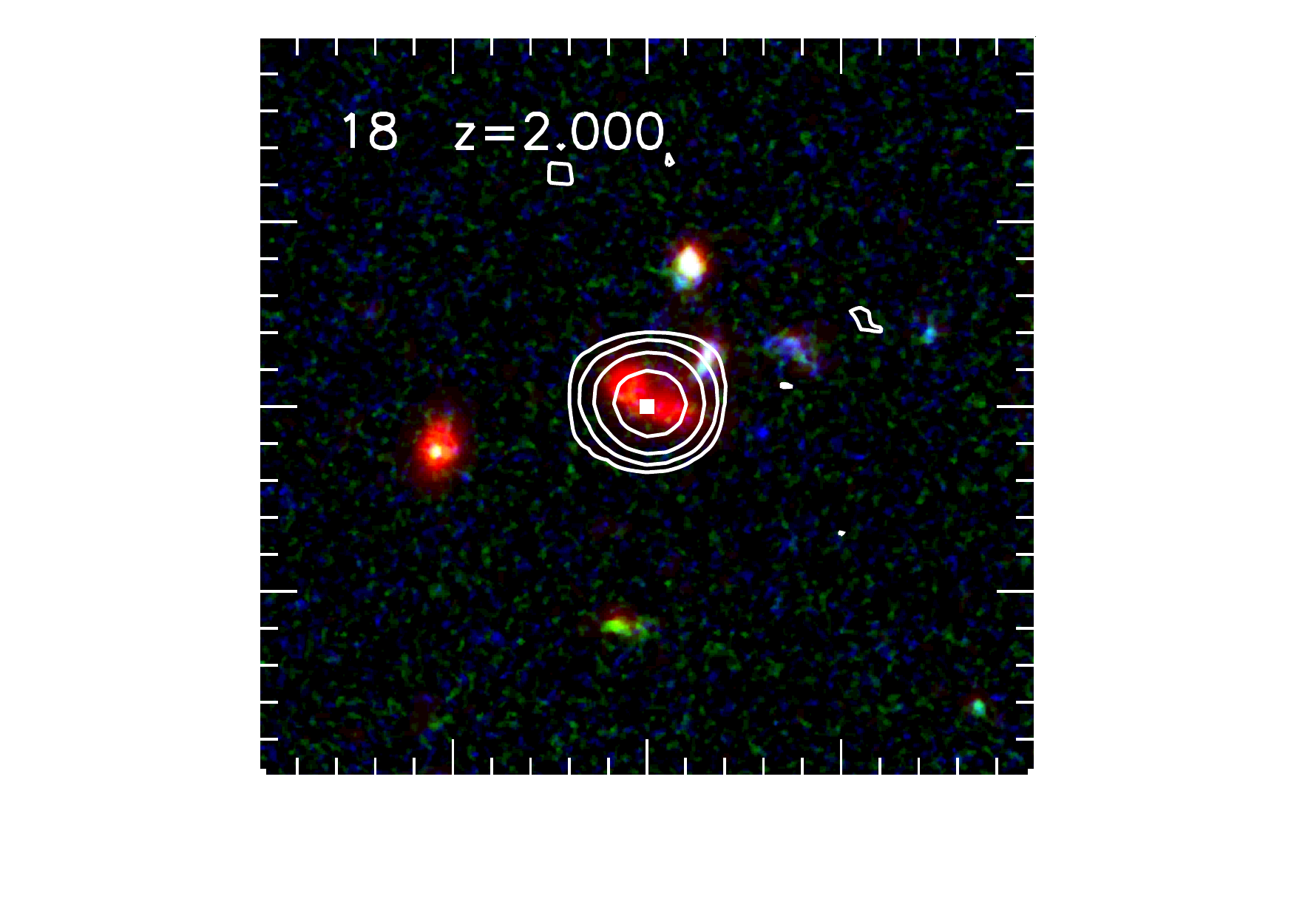}
\hspace{-2.5cm}\includegraphics[width=3.0in,angle=0]{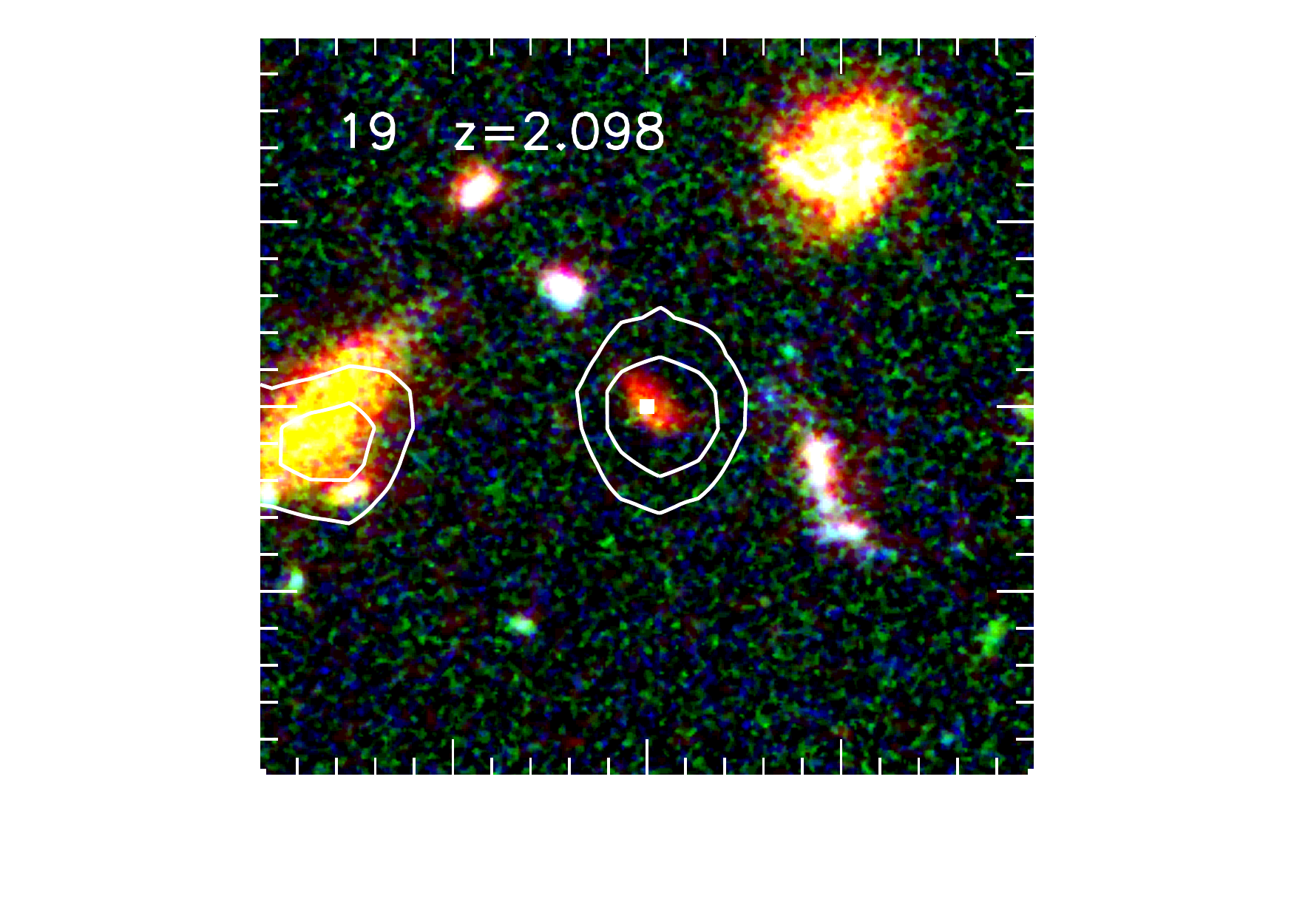}
\vspace{-0.5cm}
\includegraphics[width=3.0in,angle=0]{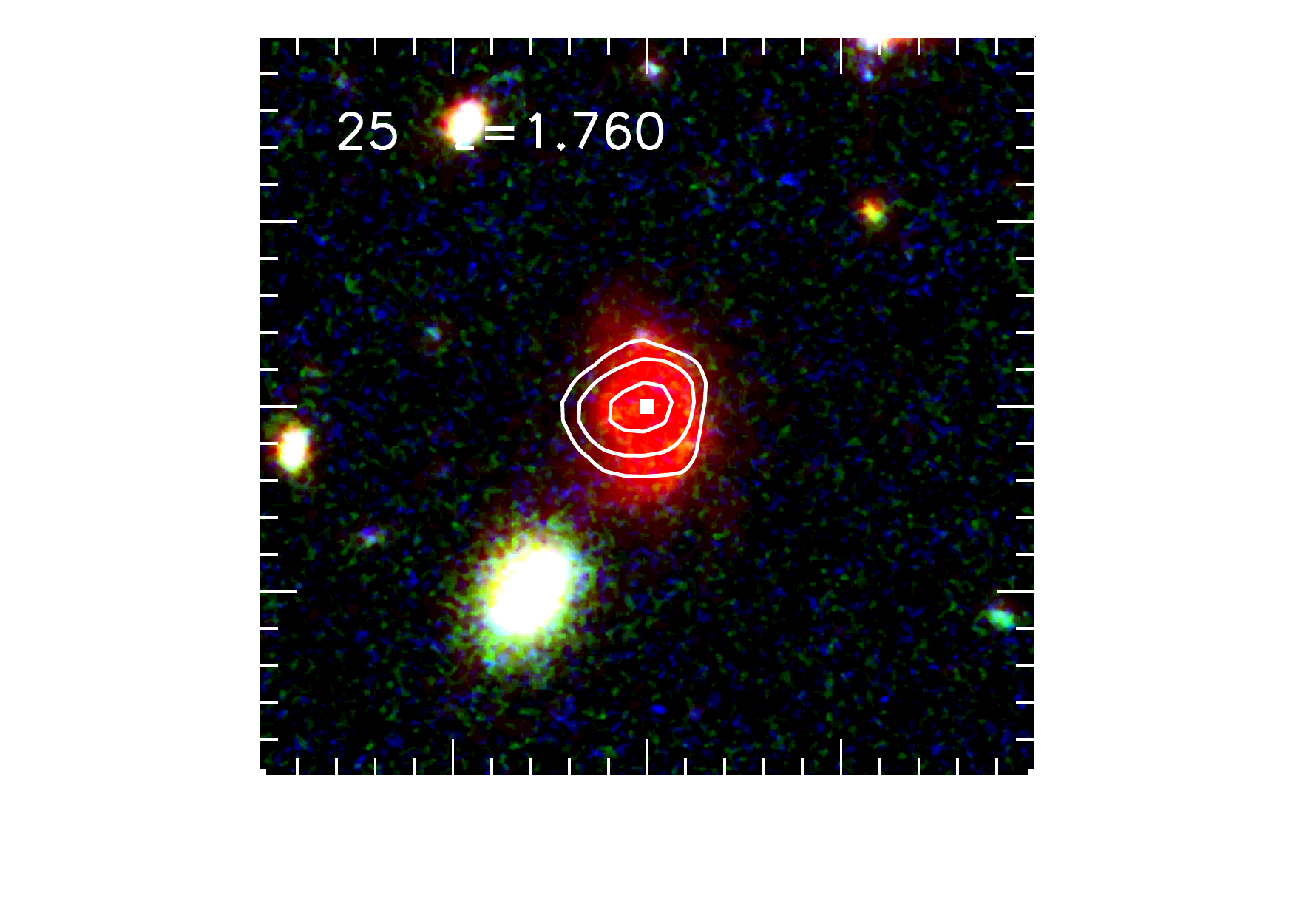}
\hspace{-2.5cm}\includegraphics[width=3.0in,angle=0]{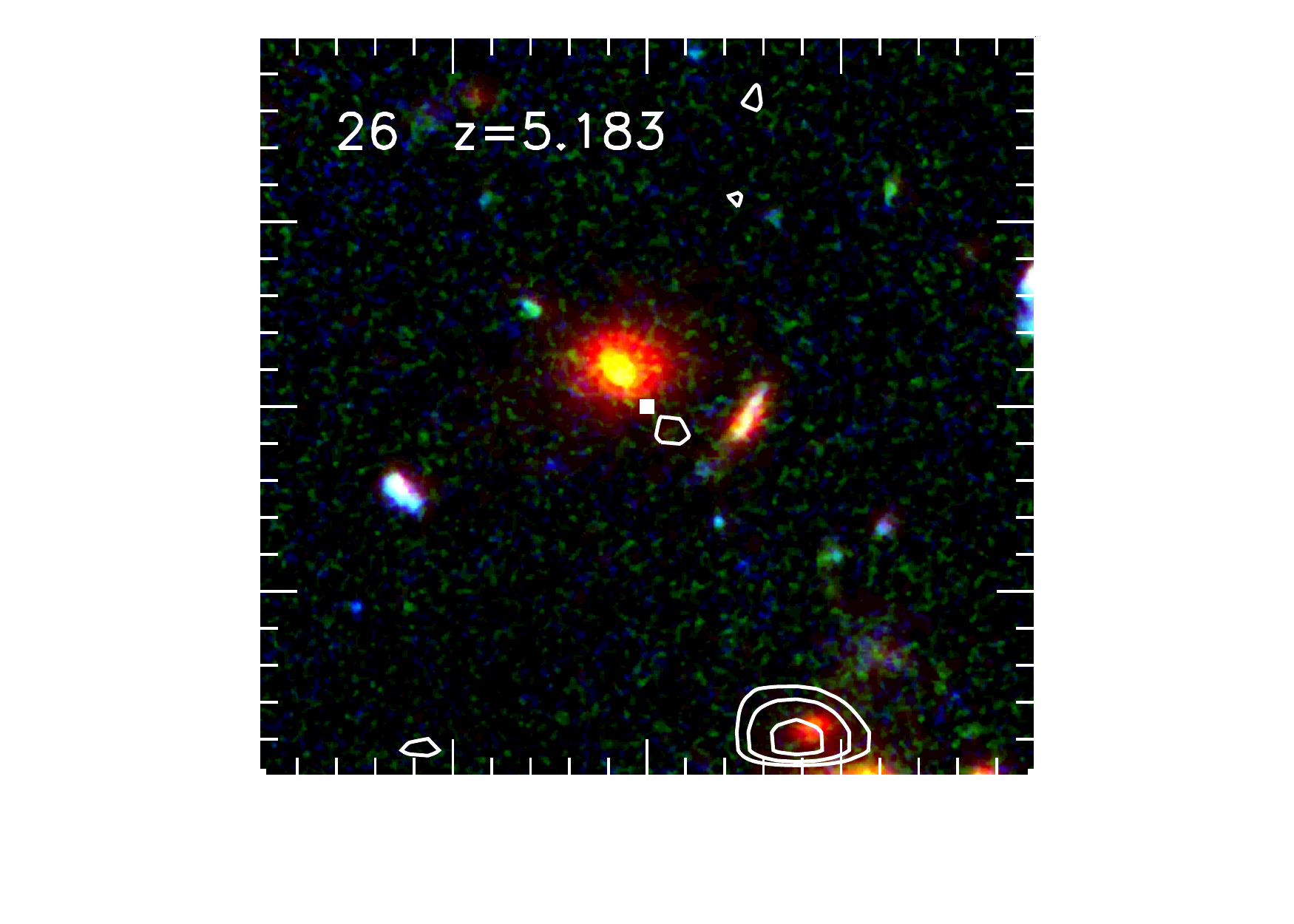}
\hspace{-2.5cm}\includegraphics[width=3.0in,angle=0]{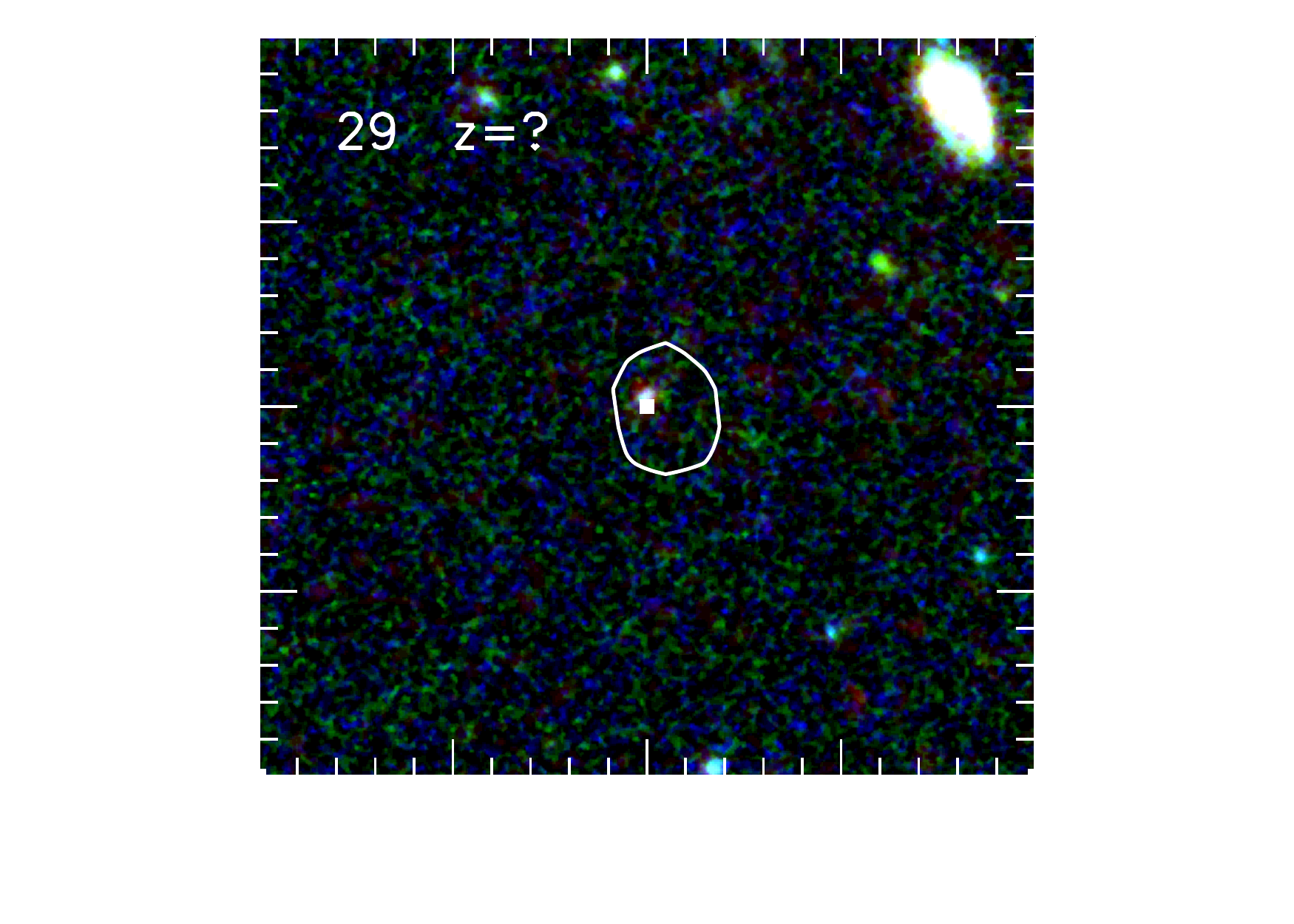}
\caption{
Three-color (F450W, F814W, and F160W) {\em HST\/} images 
of the 12 brightest SCUBA-2 sources in the {\em HST\/} GOODS-N region. 
The thumbnails are $10''$ on a side.
In each case, the white square marks the centroid from
the SMA observation, and the white contours are from the VLA 20~cm 
observation.
The identification numbers refer to Table~5, and the redshifts are spectroscopic
(either optical/NIR or CO). The sources with identification numbers 
2, 10, and 26 are the well-known
galaxies GN20, GN20.2, and HDF850.1, respectively.
\label{nir_radio}}
\end{figure*}

All of the sources with fluxes above 6\,mJy in the {\em HST\/} 
GOODS-N have SMA detections and precise positions.
In Figure~\ref{nir_radio}, we show three-color thumbnails for
these 12 sources,
with blue representing F450W ($B$-band), green F814W ($I$-band), 
and red F160W ($H$-band).
The positions of the SCUBA-2 sources are marked by the
SMA centroids (white squares) and the VLA radio contours (white).
If there is a spectroscopic redshift 
(CO or optical/NIR; see Section~\ref{redshifts}), 
then it is given in the upper left of the thumbnail image next to the
identification number from Table~5.
The images show the very dusty nature of the galaxies. The
lower redshift galaxies are usually only visible in the NIR band,
and the higher redshift galaxies are often absent in all the bands.
Where emission is seen in the higher redshift galaxies, 
it generally lies outside the {\it high surface brightness
regions in the submillimeter and radio images}. For example,
in the sources with identification numbers 2 and 10,  
known as GN20 and GN20.2 (Pope et al.\ 2005), respectively,
the green blobs show material at the redshift of the
galaxy but lying outside the bright radio  emission (Hodge et al.\ 2015). 
The heavy levels of extinction make determining morphologies
extremely difficult, and only one of the SCUBA-2 galaxies shown 
in Figure~\ref{nir_radio} (identification number 13) appears to be a 
clear merger based on both the NIR and radio morphologies.
The sources with identification numbers 3 and 18 could perhaps be 
fainter mergers.

\subsection{IRAC Fluxes}
\label{fmir}
We used the catalog of Wang et al.\ (2010) to obtain the IRAC fluxes.
Wang et al.\ used $K_s$ priors to deblend sources in the IRAC 
3.6\,$\mu$m, 4.5\,$\mu$m, 5.6\,$\mu$m, and 8.0\,$\mu$m
images obtained from the {\em Spitzer\/} GOODS Legacy program 
(PI:  M.~Dickinson).  Wang et al.\ (2010) also included
sources without $K_s$ counterparts in their final catalog.
Of the 102 SCUBA-2 sources with accurate positions and 850\,$\mu$m 
fluxes above 2\,mJy (there are 114 total with accurate positions; 
see Section~\ref{vla}), 71 lie within the deep IRAC
area covered by the Wang et al.\ catalog. Of these, 68 have
4.5\,$\mu$m counterparts within a $1\farcs5$ matching radius.
The remaining 3 sources each lie too close to a very bright neighbor for 
accurate measurements.

\begin{figure}
\includegraphics[width=3.4in,angle=0]{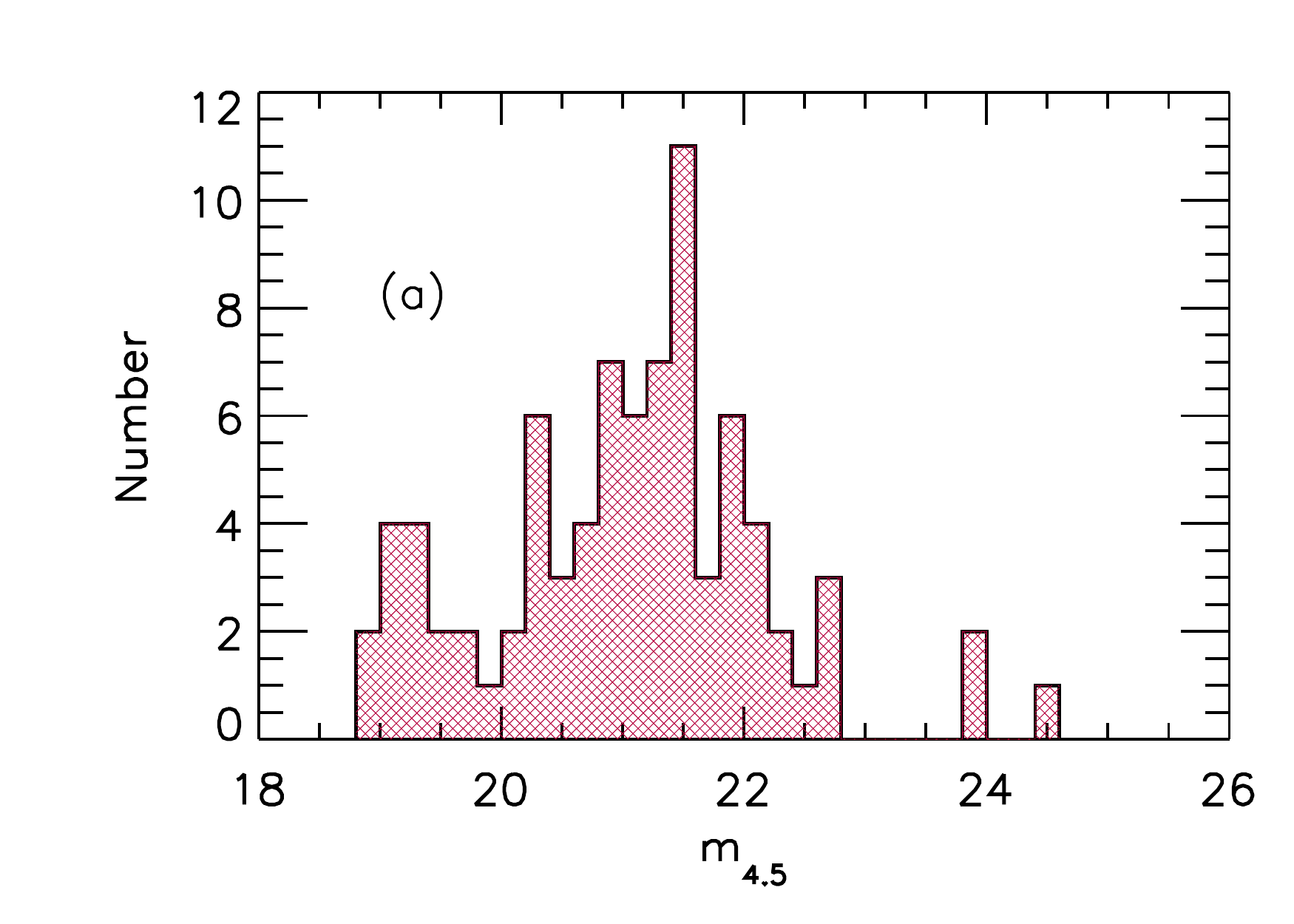}
\includegraphics[width=3.4in,angle=0]{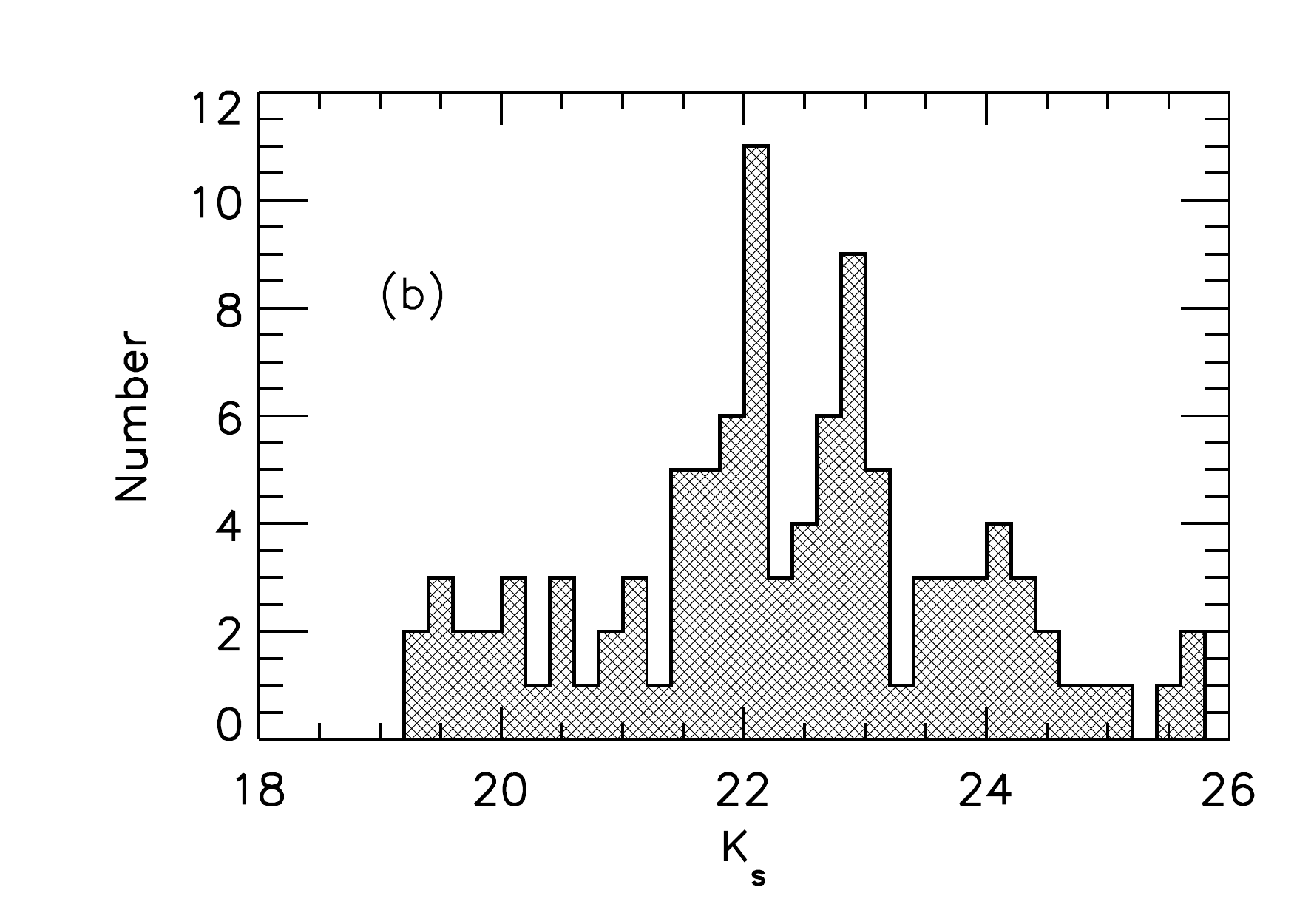}
\caption{Distribution of (a) 4.5\,$\mu$m
and (b) $K_s$ (2.1~$\mu$m) magnitudes. 
All SCUBA-2 sources with accurate positions 
and 850\,$\mu$m fluxes above 2\,mJy
are shown in (b), while only the sources in the deep IRAC area 
are shown in (a). Three sources lying on the edge of brighter 4.5\,$\mu$m 
sources are excluded in (a). Sources with $K_s$ magnitudes 
fainter than 25.5 are shown at that magnitude in (b). 
\label{mag_hist}}
\end{figure}

In Figure~\ref{mag_hist}, we show the distributions of 
(a) the 4.5\,$\mu$m magnitudes (68 sources) and 
(b) the $K_s$ magnitudes (102 sources).
The distribution of the 4.5\,$\mu$m magnitudes
is considerably brighter than that of the $K_s$ magnitudes,
with all but three of the detected sources having a 4.5\,$\mu$m
magnitude brighter than 23. This reflects the well-known
result that bright submillimeter detected galaxies are very red at these 
wavelengths. Indeed, Wang et al.\ (2012) developed a method that
uses red $K_s-4.5\,\mu$m$>1.6$ colors to select high-redshift, dusty galaxies 
(called KIEROs).
Other groups have similarly used red $H-4.5\,\mu$m$>2.25$ colors 
(HIEROS; e.g., Caputi et al.\ 2012; Wang et al.\ 2016)
or red colors in optical-infrared bands (OIRTC; Chen et al.\ 2016).

\begin{figure}
\centerline{\includegraphics[width=4.4in,angle=0]{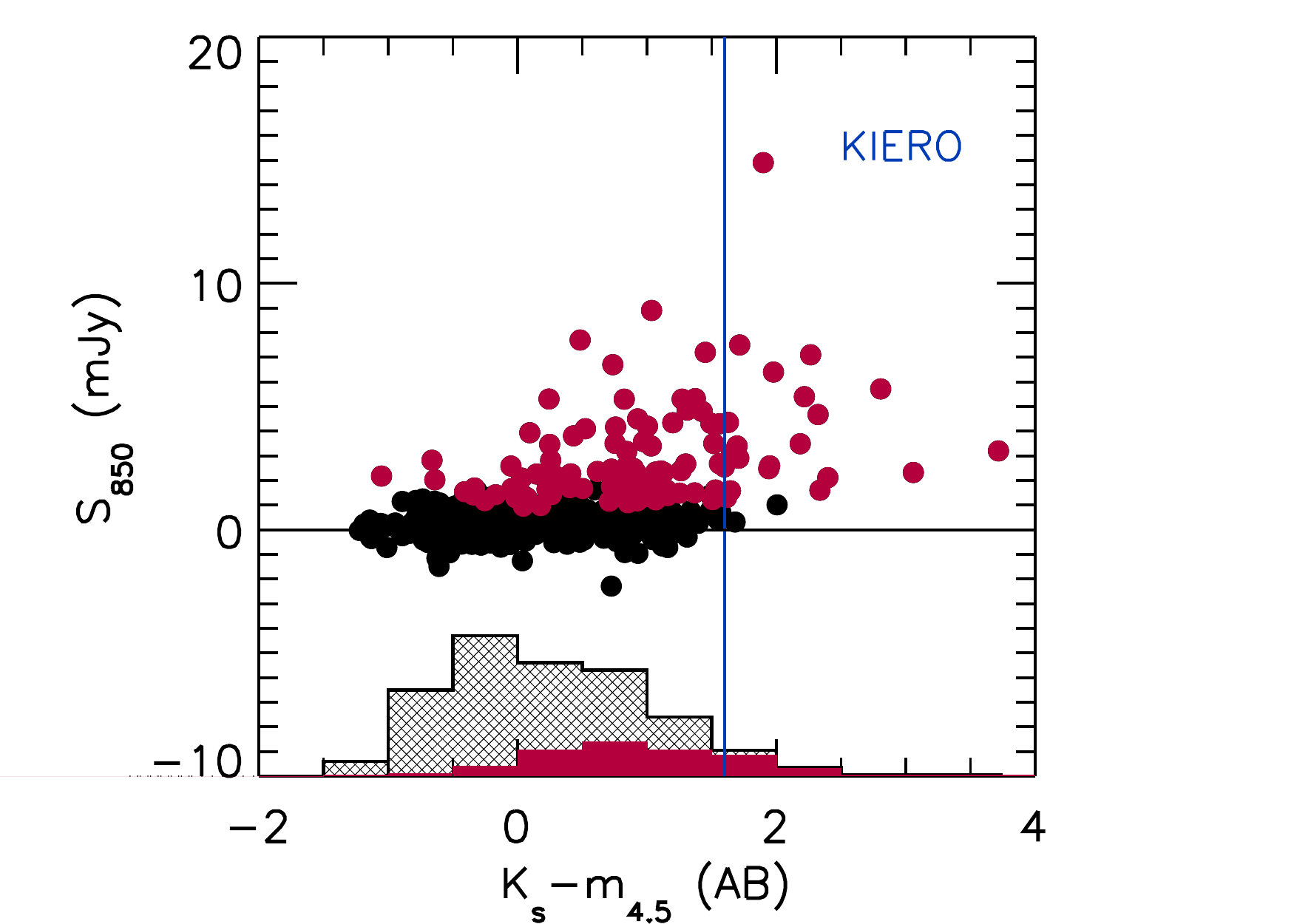}}
\caption{
850\,$\mu$m flux vs. $K_s-4.5\,\mu{\rm m}$ color for the
radio sources that lie in areas of the SCUBA-2 map where the  
850\,$\mu$m rms noise is $<0.75$\,mJy. Red circles show sources detected
above a $3\,\sigma$ threshold at 850\,$\mu$m. 
Above the KIEROs threshold ($K_s-4.5\,\mu$m$>1.6$), 20 of the 22
radio sources are detected in the submillimeter at $>3\,\sigma$.
The histograms at the bottom of the plot show the 
$K_s-4.5\,\mu{\rm m}$ color distributions for the radio sources that
are either submillimeter detected ($>3\,\sigma$; red) or undetected (black shading).
\label{kiero_radio}
}
\end{figure}

Using the radio sample, we investigated how well the KIEROs selection 
works at picking out galaxies that can be detected at submillimeter 
wavelengths. In Figure~\ref{kiero_radio}, we show 850\,$\mu$m flux versus 
$K_s-4.5\,\mu$m$>1.6$ color for the radio sources that lie in areas of 
the SCUBA-2 map where the 850\,$\mu$m rms noise is $<0.75$\,mJy. 
Impressively, 20 of the 22 sources that lie in the KIEROs region have
submillimeter detections at the $>3\,\sigma$ level (red circles).
We show the color distributions for the submillimeter detected ($>3\,\sigma$; red) 
and undetected (black shading) radio sources at the bottom of the plot.

We do not consider the optical/NIR colors further is this paper.
We postpone a more detailed discussion of the SEDs in this wavelength range 
to a later paper in the series, where we will use the combined GOODS-N and 
GOODS-S datasets.

\subsection{MIR and FIR Fluxes}
\label{ffir}
\begin{figure}
\vskip -5cm
\centerline{\includegraphics[width=4in,angle=90]{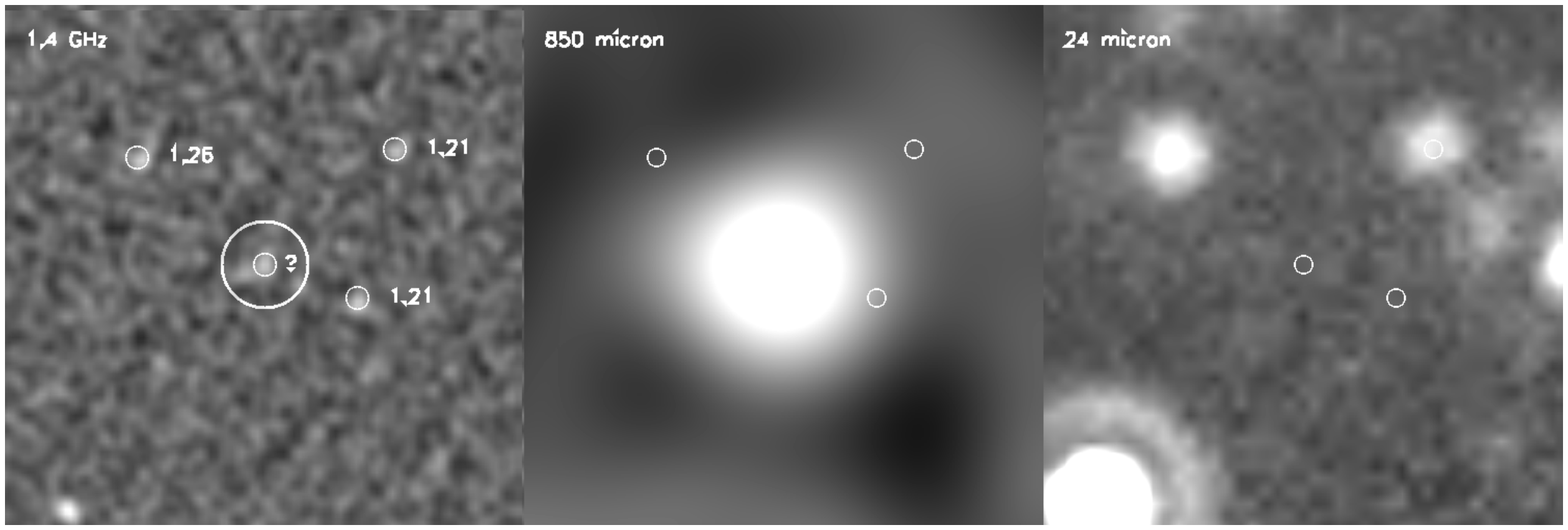}}
\vskip -6.25cm
\centerline{\includegraphics[width=4in,angle=90]{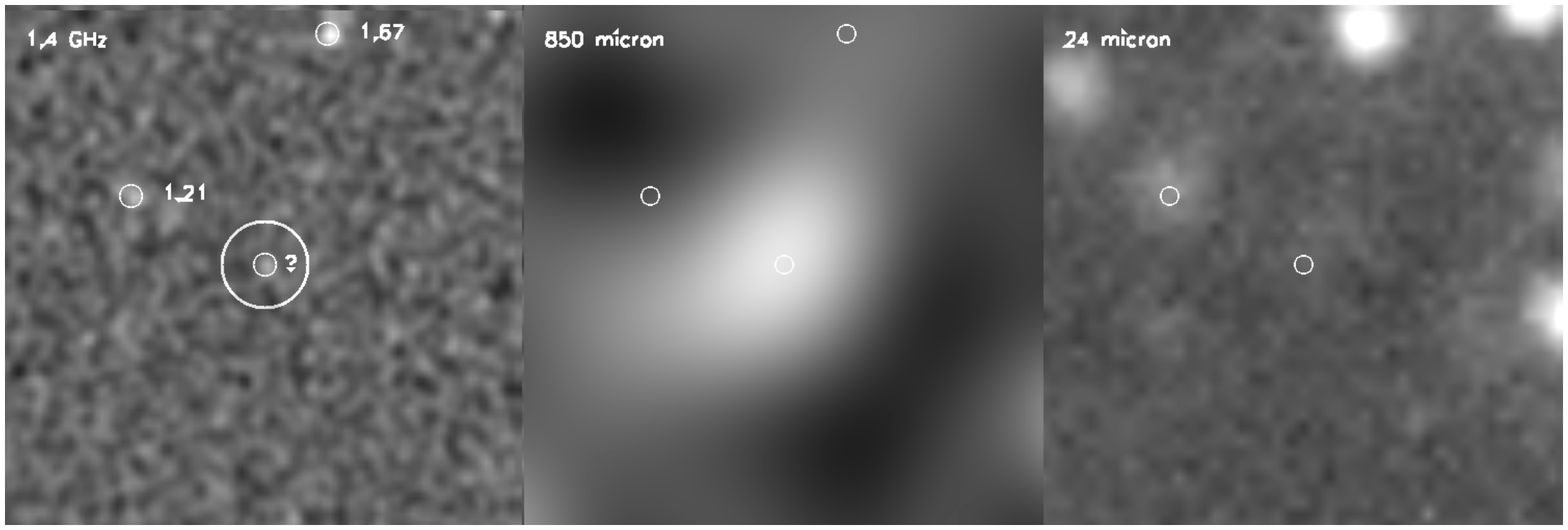}}
\caption{Brightest 2 of the 3 SCUBA-2 sources without detected 24\,$\mu$m 
counterparts (i.e., identification numbers 7 and 29 in Table~5). 
(Left) Radio image with the detected radio sources marked with small circles
and their known spectroscopic redshifts labeled. 
(Center) SCUBA-2 image.
(Right) 24\,$\mu$m image with the detected radio sources marked with small 
circles.
Thumbnails are $50''$ on a side, and the larger circle is $4''$ in radius, which
is roughly the SCUBA-2 positional uncertainty.
\label{no_24}
}
\end{figure}

We used the catalogs of Magnelli et al.\ (2011) to obtain the {\em Spitzer\/}
MIPS 24\,$\mu$m and 70\,$\mu$m fluxes. These catalogs were constructed
using 24\,$\mu$m data from the {\em Spitzer\/} 
GOODS Legacy program (PI: M. Dickinson) and 70\,$\mu$m data 
from a combination of two {\em Spitzer\/} GO programs (PI:  D. Frayer),
the {\em Spitzer\/} Far-Infrared Deep Extragalactic Legacy (FIDEL) 
program (PI: M. Dickinson), and {\em Spitzer\/} guaranteed time observer (GTO) 
programs (PI: G. Rieke). 
We used the catalogs of Magnelli et al.\ (2013) to obtain
the {\em Herschel\/} PACS 100\,$\mu$m and 160\,$\mu$m fluxes. 
These catalogs were constructed using the combined data 
sets of the PACS Evolutionary Probe (PEP; Lutz et al.\ 2011) 
guaranteed time key program and the GOODS-{\em Herschel\/} 
(Elbaz et al.\ 2011) open time key program. 
We also used the wider area {\em Herschel\/} SPIRE 
250\,$\mu$m, 350\,$\mu$m, and 500\,$\mu$m
catalogs from Elbaz et al.\ (2011).
Elbaz et al.\ (2011) and Magnelli et al.\ (2013) both used 24\,$\mu$m priors 
to deblend the {\em Herschel\/} data when constructing their catalogs.
They provide a detailed discussion of the robustness
of the 24 priors in deblending the longer wavelength data.
They also provide flags to assess whether sources are contaminated
by nearby brighter sources.
All but three of the SCUBA-2 sources in the GOODS-{\em Herschel\/} region
have 24\,$\mu$m counterparts, allowing us to obtain directly the fluxes 
from the Elbaz et al.\ and Magnelli et al.\ catalogs for our SCUBA-2 sample.
In Figure~\ref{no_24}, we show the two brightest examples out of the
three SCUBA-2 sources that do not have 24\,$\mu$m counterparts.
Since these sources are isolated, following Barger et al.\ (2015),
we used matched-filter extractions to measure the FIR fluxes for these sources.

\subsection{Redshifts}
\label{redshifts}
In our efforts to obtain redshifts for the SCUBA-2 sources with accurate 
positions, we draw on a new compilation of known spectroscopic
redshifts in the region (A. Barger et al.\ 2017, in preparation; 
see also Cohen et al.\ 2000; Cowie et al.\ 2004; Wirth et al.\ 2004, 2015; 
Swinbank et al.\ 2004; Chapman et al.\ 2005; Treu et al.\ 2005; 
Reddy et al.\ 2006; Barger et al.\ 2008; Pope et al.\ 2008; Trouille et al.\ 2008; 
Cooper et al.\ 2011). This compilation includes
our own reductions of the {\em HST\/} grism data of the GOODS-N (PI: B. Weiner;
Momcheva et al.\ 2016) (see Cowie et al.\ 2016 for details), but 
only four SCUBA-2 sources with accurate positions have grism redshifts,
and all four already had ground-based optical spectroscopic redshifts.
We also use CO spectroscopic redshifts from
Daddi et al.\ (2009a,b), Walter et al.\ (2012), and Bothwell et al.\ (2013).

Additionally,
we made observations with MOSFIRE on Keck~I of 39 of the SCUBA-2 
sources. We focused these observations on those sources without redshift
identifications but with accurate positions. We used a $0\farcs7$ slit width
and observed in the $J$ (24~min exposure), $H$ (16~min),
and $K$ (24~min) bands using an ABBA stepping pattern, where the
A position is $1\farcs25$ below the nominal position, and the B position
is $1\farcs25$ above the nominal position. A description of the reduction
procedures that we used can be found in Cowie et al.\ (2016).
We spectroscopically identified 9 SCUBA-2 sources based on the 
[NII]6584 and H$\alpha$ lines and the [OIII]5007/H$\beta$ complex. 
 
Where there is no spectroscopic redshift, we use photometric 
redshifts from Rafferty et al.\ (2011). 
Half of the SCUBA-2 galaxies with accurate positions (57 of 114)
have spectroscopic or photometric redshifts.
A significant number of the spectroscopic redshifts come from the work of 
Swinbank et al.\ (2004) and Chapman et al.\ (2005), who used radio priors to
identify SCUBA galaxies in the CDF-N. However, it is
worth noting that a substantial fraction of the sources
in these samples turn out not to be detected in our SCUBA-2 map.
In Table~4, we take from the Chapman et al.\ paper the right ascension and 
declination (Columns~1 and 2), 
the radio flux (from Richards 2000) (Column~3), 
the SCUBA flux, error, and S/N (Columns~4--6), and the redshift (Column~7) for 
the galaxies that lie in our SCUBA-2 field, followed by the current radio flux (Column~8), 
the SCUBA-2 flux, error, and S/N (Column~9--11), and the current redshift (Column~12). 
A small number of the previous redshifts were incorrect
and were updated in subsequent work.

One of the Chapman et al.\ (2005) sources is not present in the current radio
observations, and a further eight (out of the total of twenty) are not confirmed 
as SCUBA-2 sources. These eight are high radio power sources 
without genuine submillimeter detections, either because the submillimeter 
emission came from another radio source, or else
because it was simply a noise spike in the SCUBA data.
This effect only slightly increases the median redshift of the
submillimeter detected sources in this sample (from $z=2.0$ to $z=2.1$).

\begin{deluxetable*}{ccccccccccccc}
\tabletypesize{\scriptsize}
\renewcommand\baselinestretch{1.0}
\tablewidth{0pt}
\tablecaption{Chapman et al.\ (2005) Sample}
\tablehead{
R.A. & Decl.& 20\,cm & SCUBA & SCUBA & SCUBA & $z$ & Current 20\,cm & SCUBA-2 & SCUBA-2 & SCUBA-2 & Current $z$ \\ (J2000.0) & (J2000.0) & Flux & Flux & Error & S/N & & Flux & Flux & Error & S/N & \\
& & ($\mu$Jy) &  (mJy) & (mJy) &  &  & ($\mu$Jy) & (mJy) & (mJy) &  & & \\ (1) & (2) & (3) & (4) & (5) & (6) & (7) & (8) & (9) & (10) & (11) & (12)}
\startdata
       189.07637 &        62.264027 &  151. &  7.30 &  1.10 &  6.63 &   1.865 &   169 &   7.1 &  0.69 &   10. &   2.000\cr
       189.09438 &        62.274918 &  70.9 &  7.70 &  1.30 &  5.92 &   2.466 &   81. &   4.0 &  0.41 &   9.7 &   1.790\cr
       189.21568 &        62.205917 &  49.3 &  4.60 & 0.800 &  5.75 &  0.2980 &   62. &  0.33 &  0.29 &   1.1 &  0.3000\cr
       189.06729 &        62.253807 &  53.9 &  5.80 &  1.10 &  5.27 &   2.578 &   38. &   3.4 &  0.60 &   5.6 &   2.578\cr
       189.30020 &        62.203415 &  21.0 &  8.00 &  1.80 &  4.44 &   2.914 &   32. &   5.3 &  0.89 &   5.8 &   2.914\cr
       189.13586 &        62.133358 &  90.6 &  5.50 &  1.30 &  4.23 &   1.993 &   63. &  0.15 &  0.48 &  0.32 &   1.994\cr
       188.97192 &        62.227139 &  58.4 &  8.80 &  2.10 &  4.19 &   2.098 &   40. &   4.3 &  0.80 &   5.3 &   2.098\cr
       189.08862 &        62.285667 &  148. &  7.80 &  1.90 &  4.10 &   1.988 &   164 &   3.4 &  0.60 &   5.6 &   1.988\cr
       189.14830 &        62.240025 &  87.8 &  5.50 &  1.40 &  3.92 &   2.005 &   78. &   4.8 &  0.35 &   13. &   2.005\cr
       189.12137 &        62.179386 &  81.4 &  5.00 &  1.30 &  3.84 &   1.013 &   91. &  0.27 &  0.40 &  0.68 &   1.013\cr
       189.15312 &        62.198914 &  39.0 &  7.00 &  2.10 &  3.33 &  0.5570 & \nodata & \nodata & \nodata & \nodata & \nodata\cr
       188.95601 &        62.260220 &  74.6 &  8.30 &  2.50 &  3.32 &   2.203 &   87. &   3.3 &  0.80 &   4.1 &   2.203\cr
       189.02855 &        62.172611 &  74.4 &  11.6 &  3.50 &  3.31 &   2.505 &   72. & -0.37 &  0.54 & -0.68 &   2.505\cr
       189.00063 &        62.179779 &  131. &  7.90 &  2.40 &  3.29 &   1.994 &   123 & -0.18 &  0.70 & -0.26 &   2.002\cr
       189.02798 &        62.264084 &  24.0 &  4.40 &  1.40 &  3.14 &   2.416 &   13. &   1.7 &  0.50 &   3.4 &   2.413\cr
       189.28004 &        62.235580 &  45.0 &  4.70 &  1.50 &  3.13 &   2.484 &   58. &   7.0 &   1.4 &   5.0 &   2.490\cr
       189.34113 &        62.176472 &  41.0 &  12.0 &  3.90 &  3.07 &  0.9790 &   58. &  0.56 &  0.41 &   1.3 &  0.9781\cr
       189.14380 &        62.211388 &  230. &  4.30 &  1.40 &  3.07 &   1.219 &   188 &   2.8 &  0.34 &   8.2 &   1.224\cr
       189.29991 &        62.223808 &  53.9 &  4.20 &  1.40 &  3.00 &   1.996 &   55. & -0.60 &  0.34 &  -1.7 &   1.996\cr
       188.97975 &        62.150475 &  212. &  5.40 &  1.90 &  2.84 &   1.875 &   179 & -0.24 &  0.81 & -0.30 &   1.875\cr
\enddata
\label{chapman_table}
\end{deluxetable*}

\subsection{Catalog}
\label{catalog}
In Table~5, we give the SCUBA-2 850\,$\mu$m catalog.
While submillimeter and millimeter
samples are often extended to lower significance,
we restrict to sources detected at or above a $4\,\sigma$ threshold, 
relative to the statistical noise,
in order to provide a robust, well-determined 
sample that minimizes the inclusion of spurious sources. 
The catalog is ordered by 850\,$\mu$m flux. 

We give an identification number (Column~1), a name based on the SCUBA-2 
coordinate (Column~2), the right ascension and declination measured from the 
SCUBA-2 image (Columns~3 and 4), and the 850\,$\mu$m flux, error, and S/N
(Columns~5--7).
For each source, we have found all the SMA and radio sources within 
a $4\farcs5$ radius of the SCUBA-2 source. If there is an SMA detection 
or a single radio source in the area, then we give this accurate right ascension
and declination (Columns~8 and 9). SMA identifications are the most
robust and can be identified by the SMA flux given in Column~10. We give 
the 20\,cm flux in Column~11 for all the sources with accurate coordinates. 
For SCUBA-2 sources with no radio source in the area, we give 5$\,\sigma$
upper limits on the 20\,cm flux ($<12\,\mu$Jy). The
remaining sources without radio fluxes have multiple
radio sources in the area, so a unique identification
is only possible with SMA follow-up. Thus, we do not give a radio flux for these.
For sources with accurate positions, we give the $K_s$ magnitude
measured from the image of Wang et al.\ (2010) (Column~12), a spectroscopic
(or photometric) redshift, if available, with the corresponding literature source 
(Column~13), and a
250\,$\mu$m/850\,$\mu$m redshift estimate (Column~14; see Section~\ref{reds}
for details). We use only 2 decimal places for the photometric redshifts given in
Column~13 in order to to distinguish them from the spectroscopic redshifts.

In Table~6, we give the SCUBA-2 450\,$\mu$m catalog.
We list only the sources detected at or above a $4\,\sigma$ threshold in the region 
where the 450\,$\mu$m noise is less than 10\,mJy (see Figure~\ref{noise_radius_450}).
The format is the same as in Table~5, except that in
the final column, we give the 850\,$\mu$m flux measured at the
450\,$\mu$m position, rather than the 250\,$\mu$m/850\,$\mu$m redshift estimate.

\subsection{Comparison with Previous Wide-field Observations}
\label{previous}
Out of the 186 $\ge4\,\sigma$ 850\,$\mu$m sources given in Table~5,
38 had previously been found (with 850\,$\mu$m SCUBA 
observations, Wang et al.\ 2004
and Pope et al.\ 2005; 1.1/1.2\,mm AzTEC and MAMBO 
observations, Penner et al.\ 2011;
and 2\,mm GISMO observations, Staguhn et al.\ 2014). We summarize
these previous results in Table~7. In general, the agreement in the
fluxes is good for the high flux sources, but some of the fainter flux
sources have significant upward biases in the SCUBA fluxes.

\begin{figure}
\vskip -1.25cm
\centerline{\includegraphics[width=3.8in,angle=90]{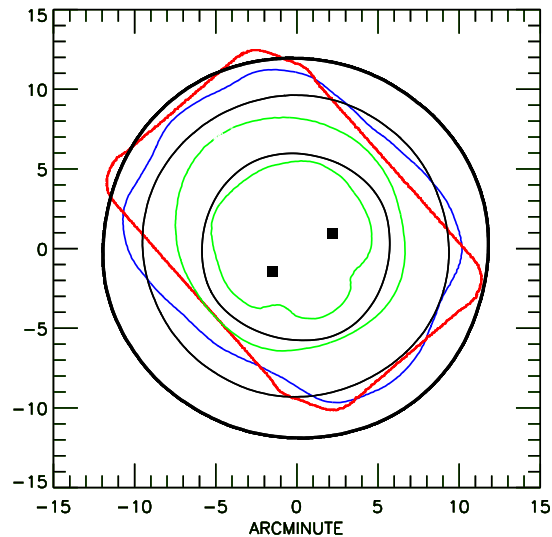}}
\vskip -1.25cm
\caption{Comparison of the areas covered by recent wide-field
surveys. The black contours show the present SCUBA-2 
image (center contour---850\,$\mu$m rms noise level of 0.5\,mJy; 
middle contour---1\,mJy; outer contour---1.5\,mJy). 
The central rms noise is 0.28\,mJy. The green contours show the
S2CLS image (center contour---850\,$\mu$m rms noise level of 1\,mJy; 
outer contour---1.25\,mJy). The central rms noise is 0.83\,mJy.
The outer counter corresponds roughly to the outside edge of the S2CLS image.
The blue contour shows the combined AzTEC+MAMBO image at a 1.16\,mm rms
noise level of 1\,mJy. The central rms noise is 0.53\,mJy. We also show
the area observed with the {\em Herschel\/} PACS survey (red contour), as well as
the two sources detected above the $4\sigma$ threshold in the GISMO 2\,mm 
survey of the central region.
\label{layout}}
\end{figure}

More recently, the SCUBA2 Cosmology Legacy Survey (S2CLS) has 
presented a SCUBA-2 based catalog of the region (Geach et al.\ 2017). 
This survey is considerably shallower than the present data and also
covers a smaller area. In Figure~\ref{layout}, we show the
SC2LS area (green) compared with our survey (black).
For comparison,
we also show the 1.16\,mm area of Penner et al.\ (2011; blue)
and the {\em Herschel\/} PACS area of Elbaz et al.\ (2011; red).
As summarized in Table~7, nearly all of the brighter sources in
the overlap areas are in good agreement.

As we discussed in Section~\ref{redshifts} (see Table~4), there are a number of
SCUBA sources---some even with moderately high significance---that 
are not detected in the present SCUBA-2 data.
For example, of the five sources in the HDF-N SCUBA image of Hughes et al.\ (1998; 
they use HDF850 names), only two 
(HDF850.1 and HDF850.2) appear to be real, with the remaining 
sources being noise spikes in the SCUBA data. 
This carries through to subsequent, more extensive
catalogs in the larger CDF-N region. 

\begin{figure}
\centerline{\includegraphics[width=3.8in,angle=90]{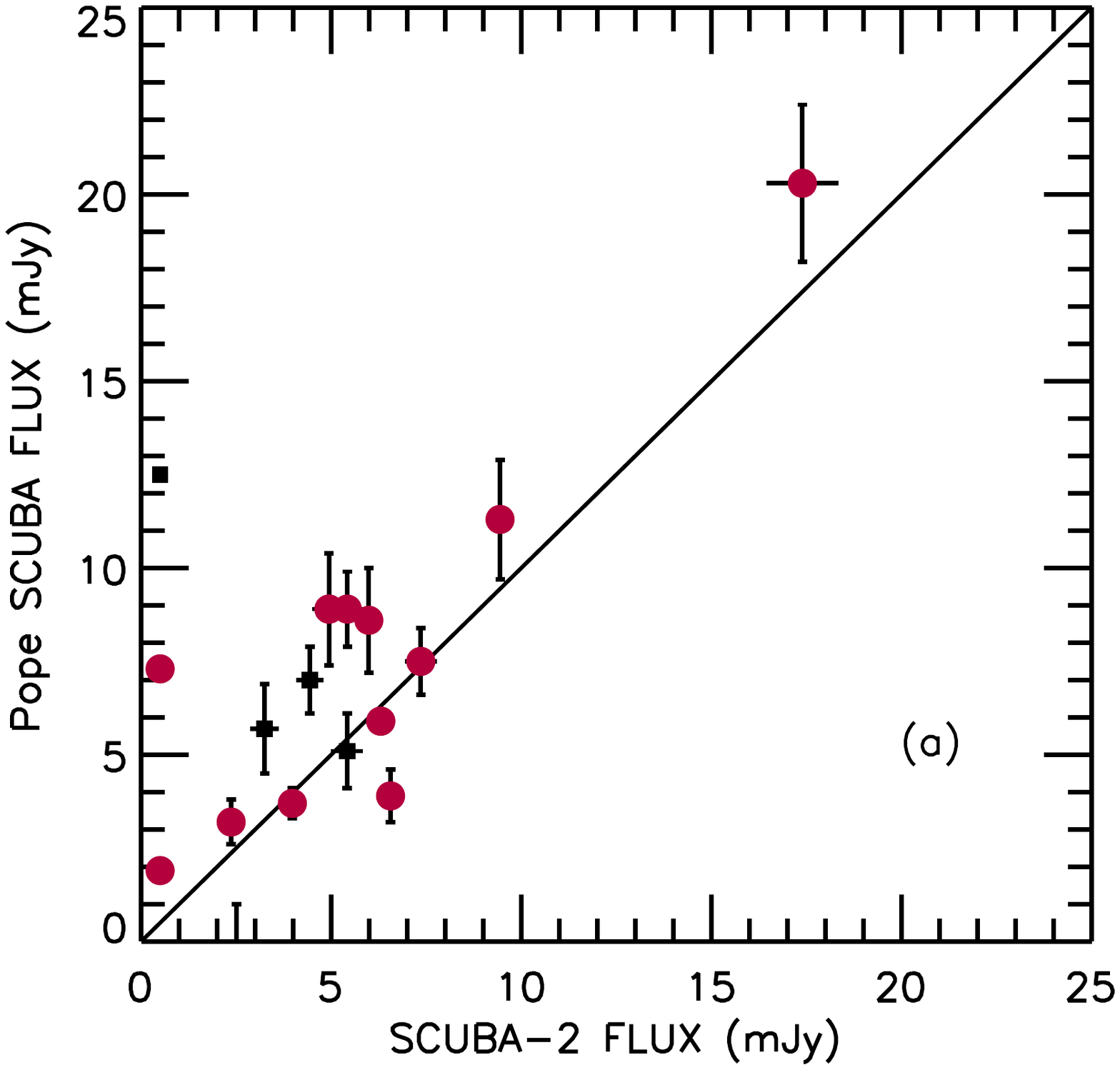}
\hspace{-4cm}
\includegraphics[width=3.8in,angle=90]{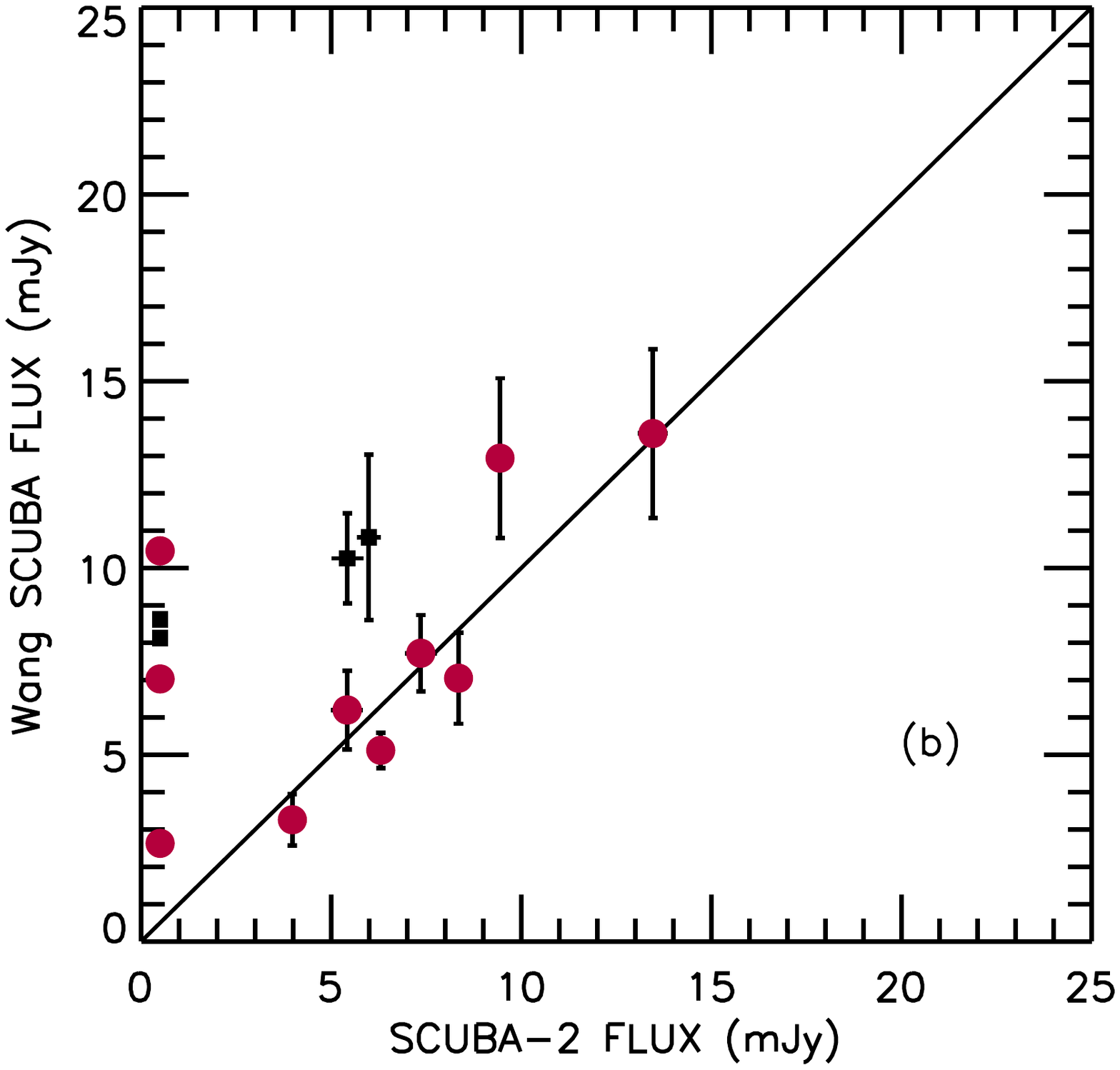}}
\vskip -0.75cm
\centerline{\includegraphics[width=3.2in,angle=90]{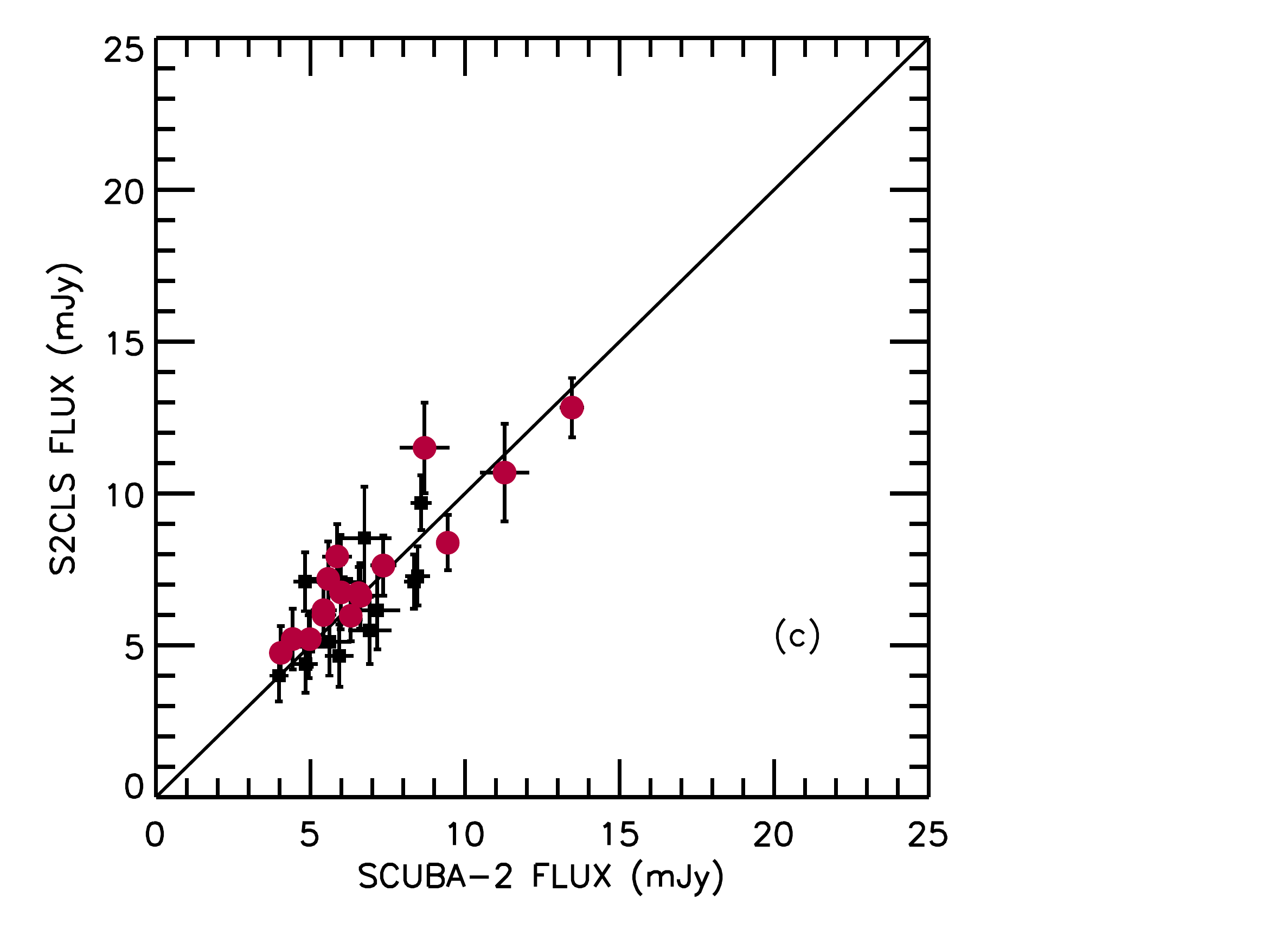}}
\caption{
SCUBA 850\,$\mu$m flux from the S/N$\ge4.5$ SCUBA samples of
(a) Pope et al.\ (2005) and
(b) Wang et al.\ (2004) vs. our SCUBA-2 850\,$\mu$m flux.
There are three undetected sources in (a) and five in (b).
(c) S2CLS 850\,$\mu$m flux from Geach et al.\ (2017) 
for the 29  sources in their sample with S/N$\ge4.5$ vs. our SCUBA-2 
850\,$\mu$m flux.
All of these sources are detected.
In each case, we mark sources with radio counterparts in the
beam with red circles and other sources with black squares. 
Literature sources not detected in our SCUBA-2 data are shown 
at a nominal value of 0.5\,mJy in the $x$-axis.
The error bars are $1\,\sigma$. 
The SCUBA-2 ($x$-axis) errors are smaller than the symbol
sizes for many of the sources.
\label{compare_scuba}}
\end{figure}

In Figure~\ref{compare_scuba},
we compare the fluxes of some literature samples with our SCUBA-2 sample.
We show the $\ge4.5\,\sigma$ SCUBA samples 
of (a) Pope et al.\ (2005; they use GN names) and (b) Wang et al.\ (2004; they use GOODS850
names), as well as the $\ge4.5\,\sigma$ S2CLS sample of 29 sources from (c) Geach et al.\ (2017).
When a literature source is also in our SCUBA-2 catalog, then
we plot the literature flux versus our SCUBA-2 flux.  
When a literature source is not also in our SCUBA-2 catalog, then we use
a nominal SCUBA-2 flux of 0.5\,mJy. We mark sources with radio counterparts 
within the beam with red circles and other sources with black squares. 

In each of the SCUBA samples, there
are SCUBA-2 undetected sources: GN1, GN8, and 
GN13 (GN!3 is also known as GOODS850-10 and HDF850.4),
GOODS850-4, GOODS850-8, GOODS850-13, and GOODS850-14. 
(Note that the two samples were constructed from heavily overlapped data sets.) 
Since none of these sources is seen in the combined AzTEC+MAMBO 1.16\,mm
image of Penner et al. (2011), they are likely all spurious. 

The average SCUBA flux of the sources detected by both SCUBA and SCUBA-2
relative to the average SCUBA-2 flux for these same sources
is 20\% higher for the Pope et al.\ (2005) measurements
and 10\% higher for the Wang et al.\ (2004) measurements.
The slightly higher fluxes measured by 
SCUBA are likely a consequence of the Eddington bias in the
SCUBA selection, since we use raw rather than corrected (de-boosted)
fluxes in the comparison. 

All 29 S2CLS sources with S/N$\ge4.5$
are in our SCUBA-2 catalog. (Below this threshold, we detect only 
about 70\% of the S2CLS sources.)  
Our 850\,$\mu$m fluxes 
are about 2.5\% lower, on average, than those in S2CLS, which is well 
within the calibration uncertainty of the SCUBA-2 data.  

Recently, Shu et al.\ (2016) used a comparison of the 24\,$\mu$m {\em Spitzer\/} 
observations with the 500\,$\mu$m {\em Herschel\/} observations to construct 
a sample of high-redshift 500\,$\mu$m selected galaxies. 
33 of the 36 sources that they selected are in our
850\,$\mu$m sample, while two of the remaining three sources lie close to bright 
SCUBA-2 sources and hence are contaminated. 
The final source is not strongly detected in the SCUBA-2 
image. Their high-redshift candidate sample can therefore be considered a 
subsample of the present data.

\subsection{Measuring Submillimeter Fluxes for Other Samples}
\label{othersamp}
In determining the SCUBA-2 fluxes for other samples,
such as X-ray and optical, we
used priors from the SMA and radio data to
minimize contamination by neighboring bright sources in the field.
The procedure is as follows. For target sample sources
within $2''$ of an SMA detected source (see Table~3),
we used the SMA flux. We constructed a de-blended SCUBA-2 image
with sources at the SMA positions removed using the
measured SCUBA-2 PSFs and fluxes.
We measured the SCUBA-2 fluxes for the radio sample in 
this cleaned image, and, as a function of flux
down to a detection level of $3\,\sigma$,
iteratively removed these fluxes from the SCUBA-2 image.
For sources in the target sample within $2''$ of a SCUBA-2 
detected radio source, we assigned the SCUBA-2 flux
of the radio source. Finally, we measured the fluxes of the remaining 
sources directly at their positions in the SCUBA-2 image 
cleaned of both SMA and radio sources. The procedure can
also be run on blank field (random) samples to measure the
biasing and contamination levels. For the 450\,$\mu$m data,
where the image is shallower and where we are far from
the confusion limit, we simply measured the 450\,$\mu$m flux
in the matched filter SCUBA-2 image at the nominal position of 
the target sample source.
Although other samples are not used in the current
paper, they will be used in subsequent papers in this series.

\section{Discussion}
\label{disc}
\subsection{Spectral Energy Distributions: Converting to Star Formation Rates}
\label{seddisc}
For the higher redshift galaxies (i.e., $z>1.2$), we expect that the observed-frame 
850\,$\mu$m flux will approximate the FIR luminosity independent
of redshift, since the strong negative $K$-correction almost
exactly offsets the effects of distance (e.g., Blain et al.\ 2002; Casey et al.\ 2014).
The exact conversion depends on the SED of the source.

\begin{figure}
\centerline{\includegraphics[width=3.8in]{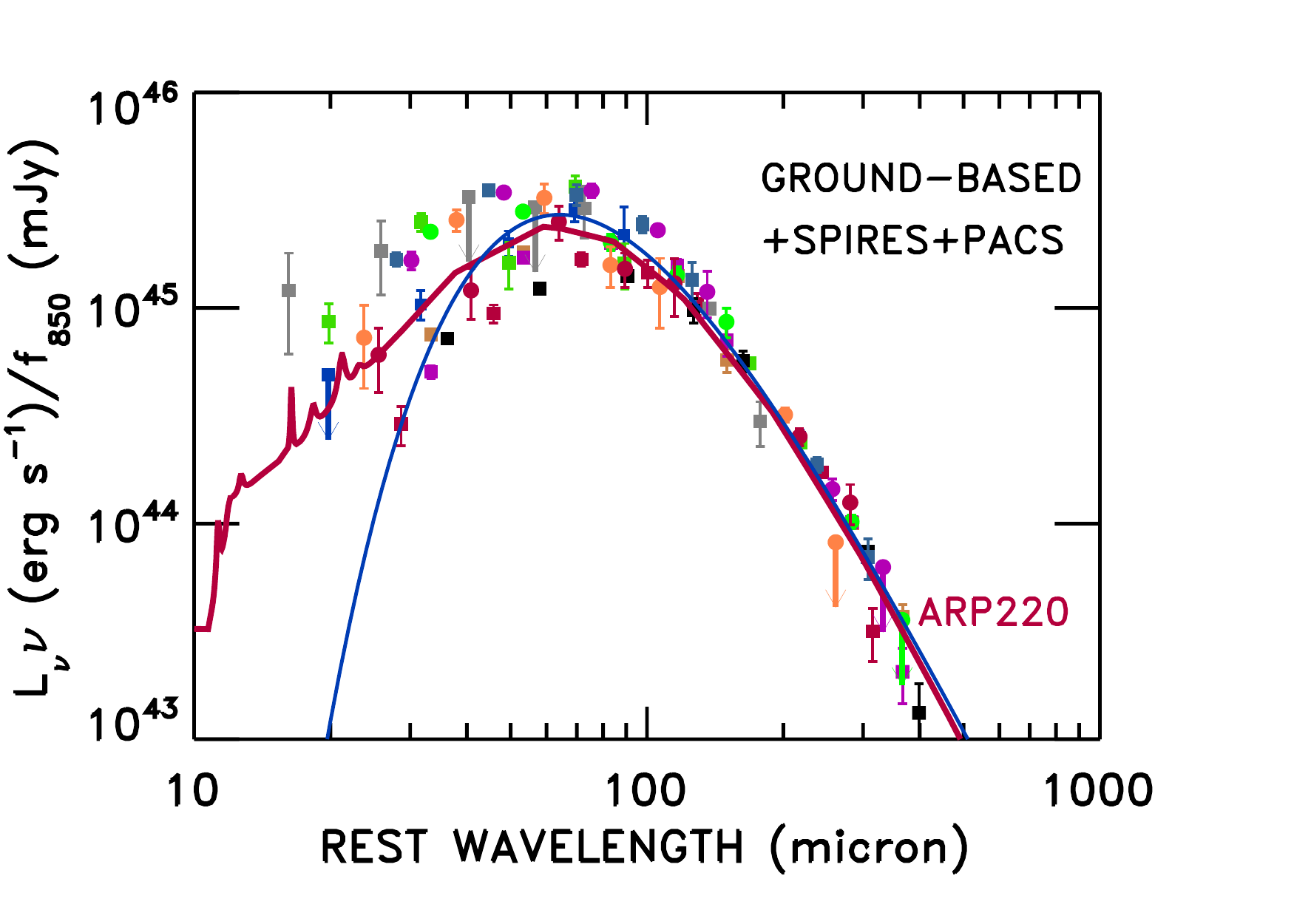}}
\caption{SEDs for the 12 SCUBA-2 galaxies
with spectroscopic redshifts $z>1.2$, 850\,$\mu$m fluxes $>3$\,mJy,
and S/N$>5$. (Identification number 10 in Table~5, also known as GN20.2, is excluded, 
since it lies close to identification number 2, also known as GN20.) All of the SEDs are  
normalized to the observed-frame 850\,$\mu$m flux.
Each galaxy is individually color coded and shows seven $L_{\nu}\nu$ points
at the rest-frame wavelengths corresponding to the 100, 160, 250, 350, 450, 850, 
and 1100\,$\mu$m bands. Non-detections in a band are shown with a $2\,\sigma$
upper limit and a downward pointing arrow.
The red curve shows the corresponding shape of Arp~220
from Silva et al.\ (1998), while the blue curve shows a graybody fit to Arp~220
with a temperature of 43~K and $\beta=1.25$.
\label{mir_lum_sed}}
\end{figure}

\begin{figure}
\vskip -5cm
\centerline{\includegraphics[width=5.5in,angle=90]{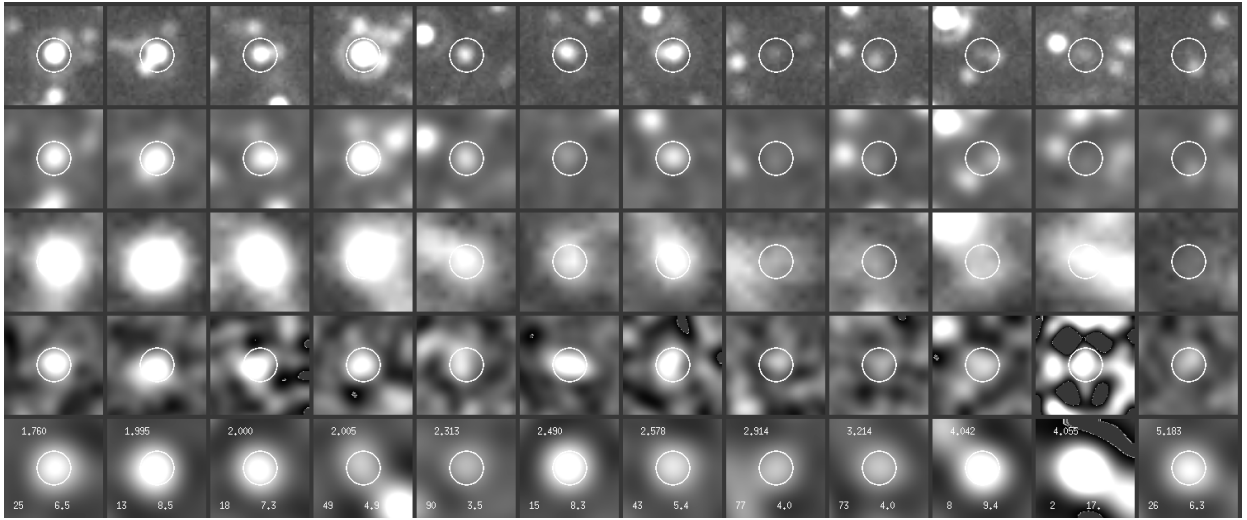}}
\caption{Thumbnails for  the 12 SCUBA-2 galaxies
with spectroscopic redshifts $z>1.2$, 850\,$\mu$m fluxes $>3$\,mJy,
and S/N$>5$ that are shown in Figure~\ref{mir_lum_sed}. The panels 
from bottom to top show the observations at 850, 450, 250, 100,
and 24\,$\mu$m. The identification number (from Table~5), SCUBA-2 
850\,$\mu$m flux, and redshift are given for each source in the bottom 
panel. The panels are shown in ascending order of redshift from left
to right. The fluxes at shorter wavelengths drop
rapidly with increasing redshift. 
\label{bright_thumbs}}
\end{figure}

Barger et al.\ (2014) showed that
the SEDs of nearly all high-redshift SCUBA-2 galaxies are very similar 
to those of Arp~220  at least at the longer wavelengths. 
We illustrate this again in Figure~\ref{mir_lum_sed}, where 
we show the quantity $L_\nu \nu$ normalized to the observed-frame 850\,$\mu$m
flux for the 12 brightest SCUBA-2 galaxies with spectroscopic redshifts $z>1.2$, 
850\,$\mu$m fluxes $>3$~mJy, and S/N$>5$ and that
lie in the more sensitive regions of the GOODS-{\em Herschel\/} observations.
Nearly all of these are isolated sources, as we illustrate in
Figure~\ref{bright_thumbs} where we show the thumbnail images
at 850, 450, 250, 100, and 24\,$\mu$m. 
The deblending using the 24\,$\mu$m priors is only important in a small
number of sources (identification numbers 8 and 26).
Each galaxy in Figure~\ref{mir_lum_sed}
is color coded. For cases where there is no detection in a band,
we show a $2\,\sigma$ upper limit and a downward pointing arrow. For comparison,
we show the Arp~220 SED given by Silva et al.\ (1998) (red curve).
We also show a graybody fit to Arp~220 (a single temperature of 43~K and an emissivity 
index of $\beta=1.25$; blue curve), which is a very good approximation
to the observed SEDs (all of the points lie within a multiplicative factor of 1.8)
down to $\sim50\,\mu$m.
This is similar to da Cunha et al.\ (2015)'s result that the typical ALESS 
(i.e., ALMA follow-up survey of the single-dish LABOCA 870\,$\mu$m survey of the Extended
Chandra Deep Field South) source has a luminosity-averaged dust temperature of $43\pm2$\,K.
In order to fit the shorter wavelengths that are dominated by hot dust emission, the
fit would need to be extended with a MIR power law (e.g., Casey 2012).

\begin{figure}
\centerline{\includegraphics[width=4in,angle=90]{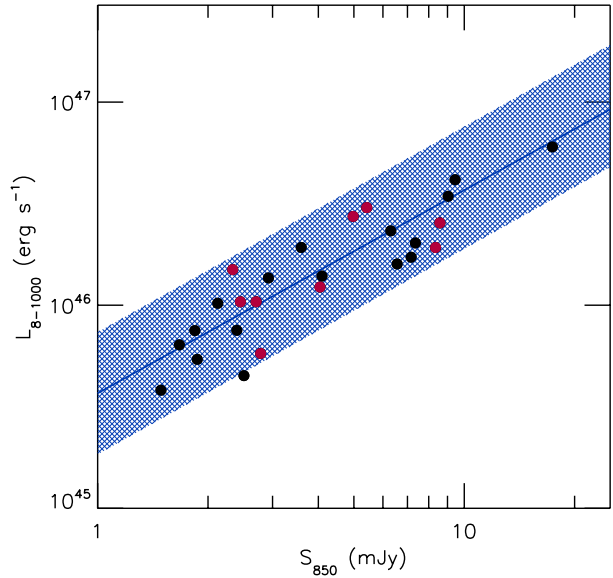}}
\vskip -1.25cm
\caption{
FIR ($8-1000\,\mu$m) luminosity vs. 850\,$\mu$m
flux (circles) for the 26 SCUBA-2 galaxies in the
more sensitive regions of the GOODS-{\em Herschel\/} 
coverage (less than 2 times the minimum noise) 
with accurate positions and spectroscopic redshifts. The blue
line shows a linear relation based on the median ratio, while
the shaded region shows a spread by a multiplicative
factor of two about this line. Red circles show galaxies that
would be classified as containing an AGN based on having 
rest-frame $2-8$~keV luminosities $>10^{43}$~erg~s$^{-1}$.
\label{sed_lum}}
\end{figure}

To determine the conversion from observed-frame 850\,$\mu$m flux
($S_{850\,\mu{\rm m}}$) to $L_{8-1000\,\mu{\rm m}}$ luminosity, we interpolated 
between the individual $\log L_\nu \nu$ points and then integrated to
obtain $L_{8-1000\,\mu{\rm m}}$. The number of data points
is sufficiently large to make this procedure robust, and it
avoids fitting model or template SEDs. In Figure~\ref{sed_lum},
we show $L_{8-1000\,\mu{\rm m}}$ versus $S_{850\,\mu{\rm m}}$ 
for the 26 SCUBA-2 galaxies
with accurate positions and spectroscopic redshifts (circles). 
We determine a mean conversion of
\begin{equation}
\log L_{8-1000\,\mu {\rm m}}\ ({\rm erg\ s^{-1}}) = \log {S_{850\,\mu{\rm m}}\,({\rm mJy})} + 45.60\pm0.05 \,,
\label{lfir850}
\end{equation}
where we have computed the $1\,\sigma$ error using the jackknife resampling statistic.
The median would give a conversion of 45.57 (blue line in Figure~\ref{sed_lum}).
All of the sources lie within a multiplicative factor of 2 of this relation (shaded region).

For the SCUBA-2 galaxies, $L_{8-1000\,\mu{\rm m}}$ approximates the bolometric luminosity. 
Thus, to get a SFR from $S_{850\,\mu{\rm m}}$, we adopt the theoretical conversion of 
Murphy et al.\ (2011) between $L_{8-1000\,\mu{\rm m}}$ and SFR (their Equation~4), namely,
\begin{equation}
\log {\rm SFR} (M_\odot~{\rm yr}^{-1}) = \log {L_{8-1000\,\mu{\rm m}}}~({\rm erg~s^{-1}}) - 43.41 \,.
\label{sfr_murphy}
\end{equation}
They computed this conversion from Starburst99 (Leitherer et al.\ 1999)
for a constant SFR and a Kroupa (2001) IMF. Madau \& Dickinson (2014)
adopted a slightly lower conversion of $-43.55$ (after converting 
from their adopted Salpeter 1955 IMF to a Kroupa IMF), 
which should be taken into account when
comparing with their SFRs (see Section~\ref{sfr_dist}).
Now, combining Equations~\ref{lfir850} and \ref{sfr_murphy}, we get
\begin{equation}
{\rm SFR} (M_\odot~{\rm yr}^{-1}) = 143 \pm 20 \times  S_{850\,\mu{\rm m}}~({\rm mJy}) \,.
\label{sfr_850}
\end{equation}
This conversion should be adequate for a Chabrier (2003) IMF, as well,
but for a simple power law Salpeter (1955) IMF with power law
index $-2.35$ and mass limits 0.1 and 100 solar masses, the normalization
would rise to $230\pm 32$. This Salpeter conversion is consistent
with that given in Barger et al.\ (2014).

Some portion of the shorter wavelength SCUBA-2 SEDs could be contributed
by AGN activity. As can be seen from Figure~\ref{sed_lum}, there is no clear 
difference in the bolometric luminosities of the SCUBA-2
galaxies with known X-ray detections (red circles have rest-frame $2-8$~keV
luminosities $>10^{43}$~erg~s$^{-1}$)---suggesting they contain AGNs---versus 
those without,
so this effect is not large for AGNs that are not very Compton thick.
Emission from an AGN torus is also not likely to contribute significantly above 
$\sim100~\mu$m (Fritz et al.\ 2006; Netzer et al.\ 2007; Schartmann et al.\ 2008; 
H\"onig \& Kishimoto 2010; Mullaney et al.\ 2011; Siebenmorgen et al.\ 2015),
so the longer wavelength contributions to the bolometric luminosity should not
be affected by AGN activity. Thus, the maximum correction
for any AGN contributions would likely be smaller than the
multiplicative factor of two already ascribed. We will
return to the issue of the AGN contributions in Paper~3
of the series using the much deeper X-ray observations in
the CDF-S.

\subsection{Redshift Estimators}
\label{reds}
Historically, the most popular method for estimating redshifts for
submillimeter detected galaxies
without spectroscopic or photometric redshifts was to use the
ratio of the 20\,cm flux to the 850\,$\mu$m flux
(Carilli \& Yun 1999; Barger et al.\ 2000, who called these millimetric redshifts). 
It was thought that using a FIR flux to submillimeter flux ratio 
to estimate redshifts would not work, given the expectation at the
time of a wide range of temperatures in submillimeter 
detected galaxies. However, since we have found that
these bright SCUBA-2 selected galaxies have relatively uniform SEDs
at the longer rest-frame wavelengths 
(see Section~\ref{seddisc}), this opens up the possibility of using the shape 
of the SED to estimate redshifts (e.g., Chakrabarti et al.\ 2013).

We only have full SED information in the GOODS-{\em Herschel\/} region.
However, the 250\,$\mu$m, 350\,$\mu$m, and 
500\,$\mu$m data from Elbaz et al.\ (2011) cover the full SCUBA-2 field.
Since the sources outside the GOODS-{\em Herschel\/} region are not 
contained in the Elbaz et al.\ catalogs, following Barger et al.\ (2015), 
we used matched-filter extractions to measure the FIR fluxes for these sources.
Here we investigate how well a two-color redshift estimator works for 
determining redshifts for the SCUBA-2 sample; specifically, we use the 
ratio of the 250\,$\mu$m flux to the 850\,$\mu$m flux.

\begin{figure}
\centerline{\includegraphics[width=4.0in,angle=90]{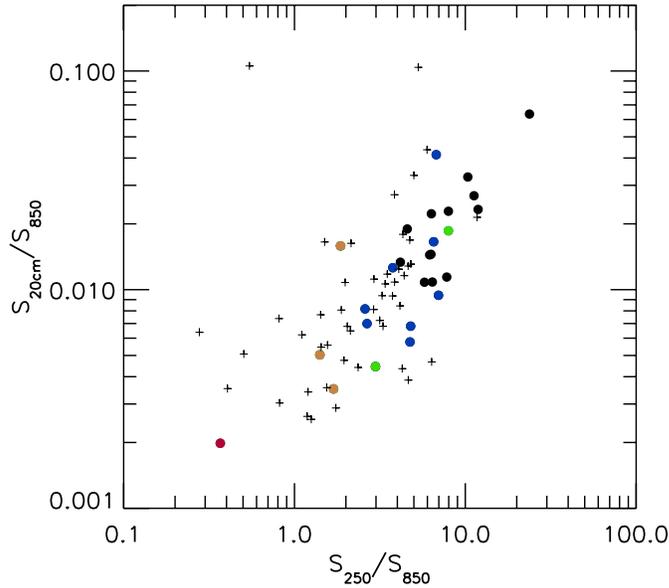}}
\vskip -0.5cm
\caption{
20\,cm/850\,$\mu$m flux vs. 250\,$\mu$m/850\,$\mu$m flux.
There is a good relation between the two flux ratios, and
the sources with spectroscopic redshifts are well segregated by redshift
(circles: black---$z<2$, blue---$z=2-3$, green---$z=3-4$, 
gold---$z=4-5$, red---$z=5-6$).
SCUBA-2 sources without spectroscopic redshifts but with accurate
positions are denoted by crosses.
The two sources at the top of the plot that lie off the correlation are high radio
power sources where the radio power is not related to star formation.
\label{color_reds}
}
\end{figure}

In Figure~\ref{color_reds}, we compare the 20\,cm/850\,$\mu$m
flux ratio with the 250\,$\mu$m/850\,$\mu$m flux ratio. Both ratios are well 
correlated, which suggests that both provide a measure of the redshifts.
Sources without spectroscopic redshifts but with accurate positions are denoted by 
crosses, while sources with spectroscopic redshifts are color-coded by redshift 
interval (circles; see figure caption for intervals). 
The sources with spectroscopic redshifts are indeed well segregated by
redshift interval in the plot.
There are two high radio power sources at the top of the plot that lie off the correlation,
which illustrates how there can be submillimeter detected
sources where the radio power is not produced by star formation
(see Barger et al.\ 2017).

\begin{figure}
\centerline{
\includegraphics[width=3.4in,angle=0]{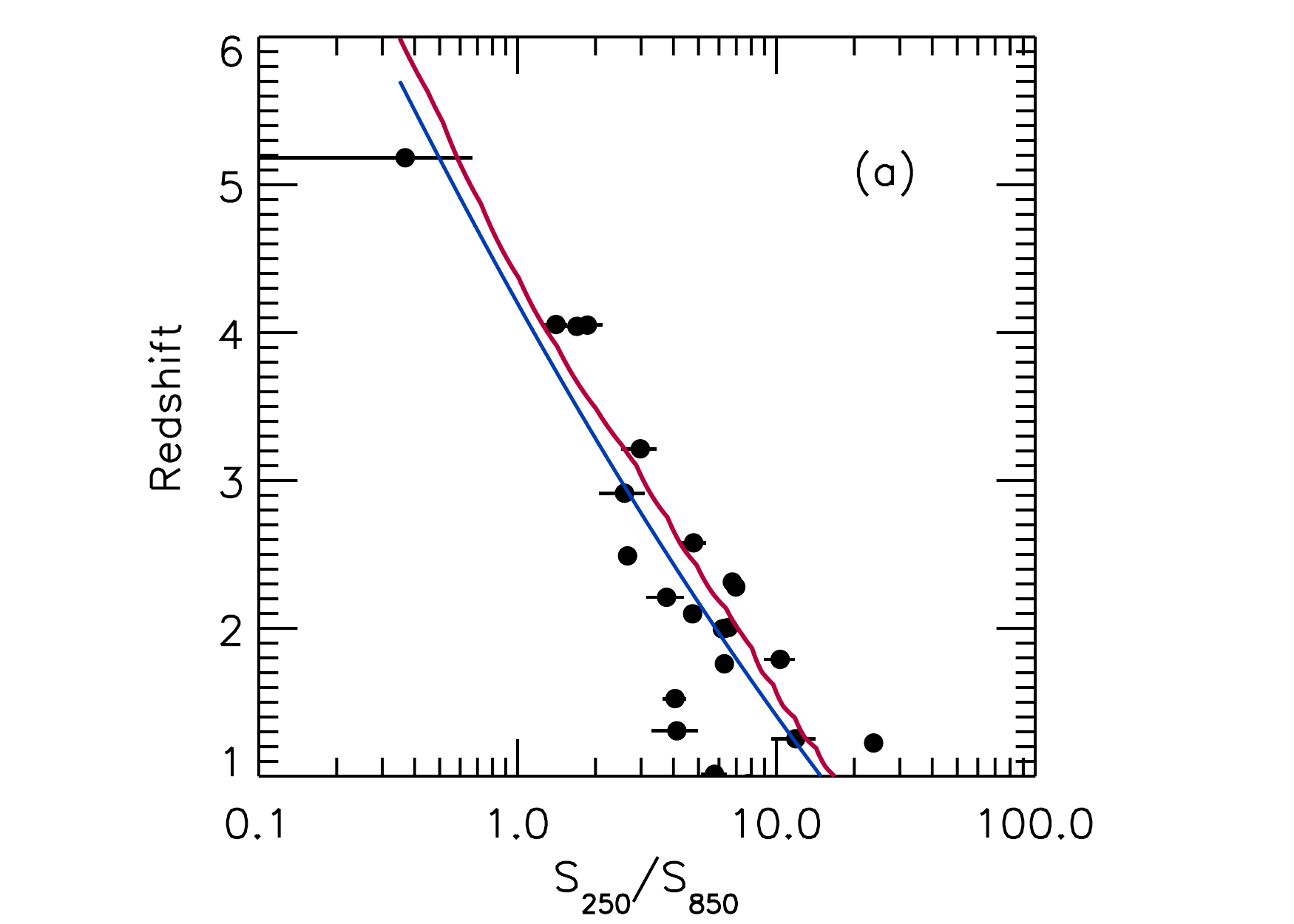}
\includegraphics[width=3.4in,angle=0]{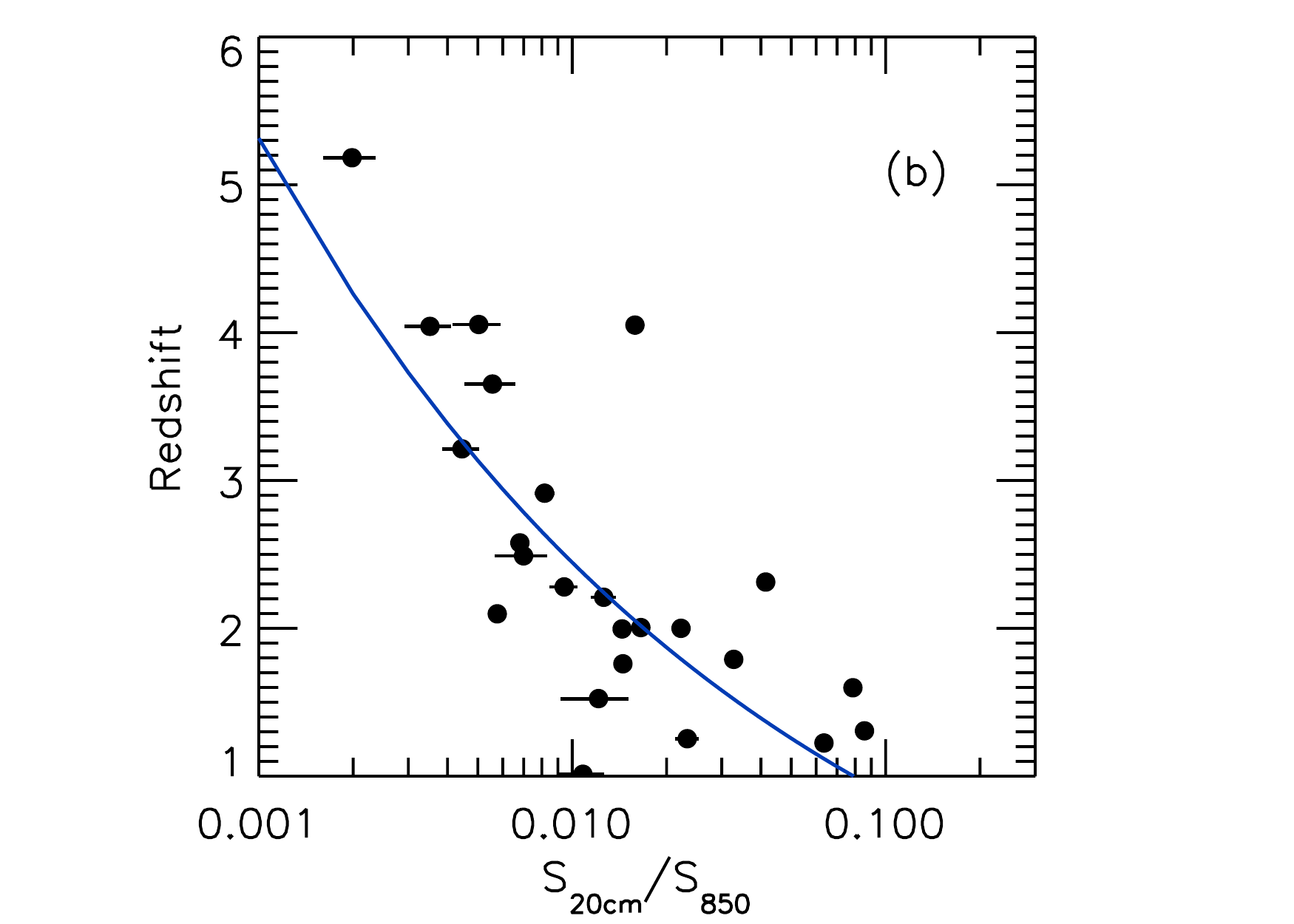}
}
\caption{
(a) Spectroscopic redshift vs. 250\,$\mu$m/850\,$\mu$m flux ratio
(black circles). The red curve shows the redshift vs. the same flux
ratio for an Arp~220 SED, while the blue curve shows
a fit of the form given in Equation~\ref{z_250}. 
(b) Spectroscopic redshift vs. 20\,cm/850\,$\mu$m flux ratio.
The blue curve shows the redshift relation from Barger et al.\ (2000)
for an Arp~220 SED.
\label{col_redshifts}
}
\end{figure}

To see more clearly how well each flux ratio does at estimating reliable redshifts,
in Figure~\ref{col_redshifts}, we plot spectroscopic redshift versus
(a) 250\,$\mu$m/850\,$\mu$m flux ratio and (b) 20\,cm/850\,$\mu$m
flux ratio. In (a), the red curve shows the 250\,$\mu$m/850\,$\mu$m 
flux ratio versus redshift for an Arp~220 SED, while the blue curve 
shows the relation
\begin{equation}
z_{250} = 4.20-3.14\,\rm{log}(S_{250\,\mu{\rm m}}/S_{850\,\mu{\rm m}}) 
+ 0.352\,\rm{log}(S_{250\,\mu{\rm m}}/S_{850\,\mu{\rm m}})^2 \,,
\label{z_250}
\end{equation}
which is based on a quadratic fit to the logarithmic flux ratio. 
In (b), the blue curve shows the redshift relation from Barger et al.\ (2000;
the ratio of their Equations~2 to 4) for an Arp~220 SED.
The scatter in Figure~\ref{col_redshifts}(b) is considerably larger than 
the scatter in Figure~\ref{col_redshifts}(a).
It is clear that the radio power is not always an accurate measure
of the star formation.
The scatter in Figure~\ref{col_redshifts}(a) is primarily
caused by the variations in the SEDs of the sources (see 
Figure~\ref{mir_lum_sed}).

\begin{figure}
\centerline{
\hspace{0.5cm}\includegraphics[width=2.9in,angle=0]{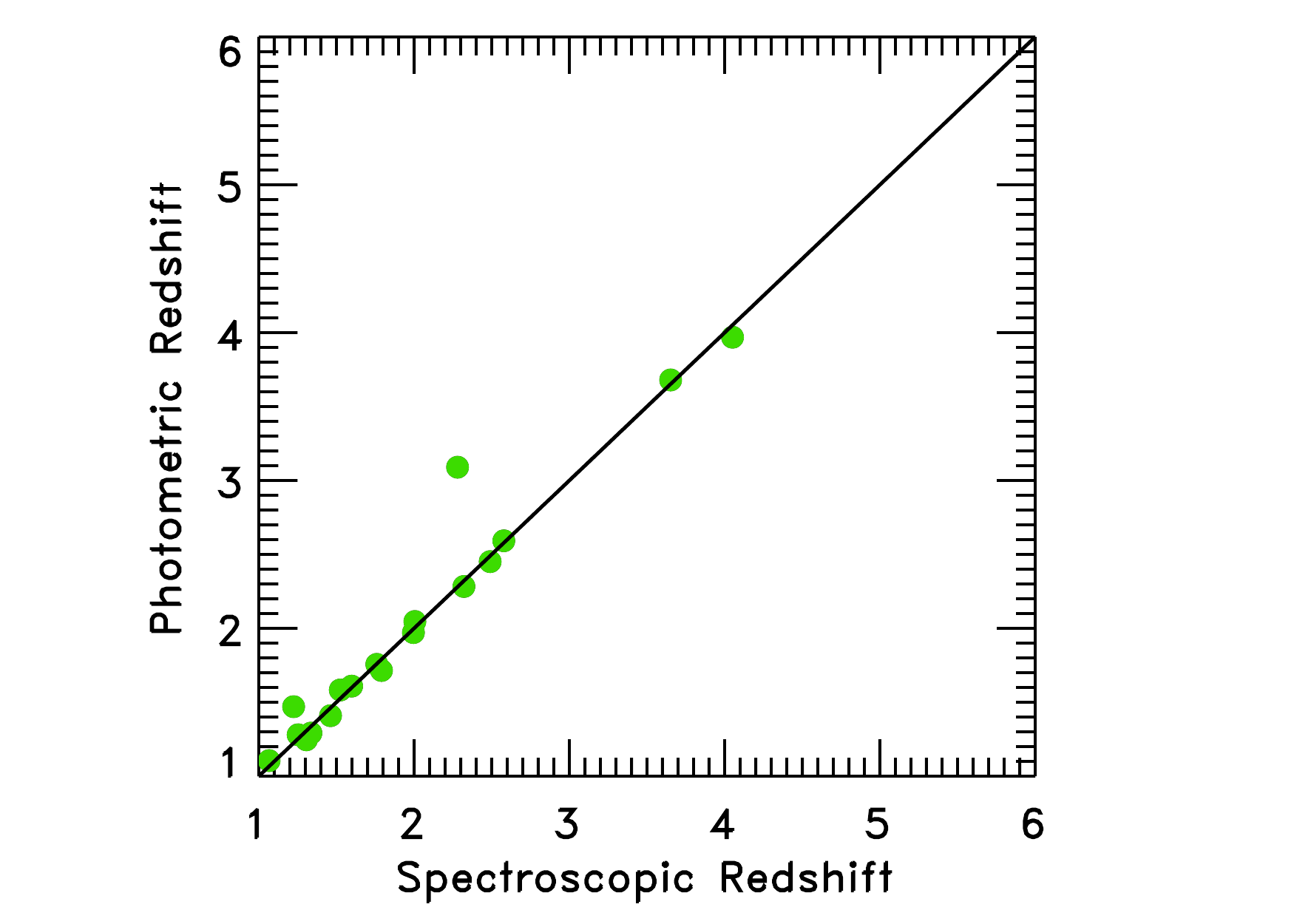}
\hspace{-1.4cm}\includegraphics[width=2.9in,angle=0]{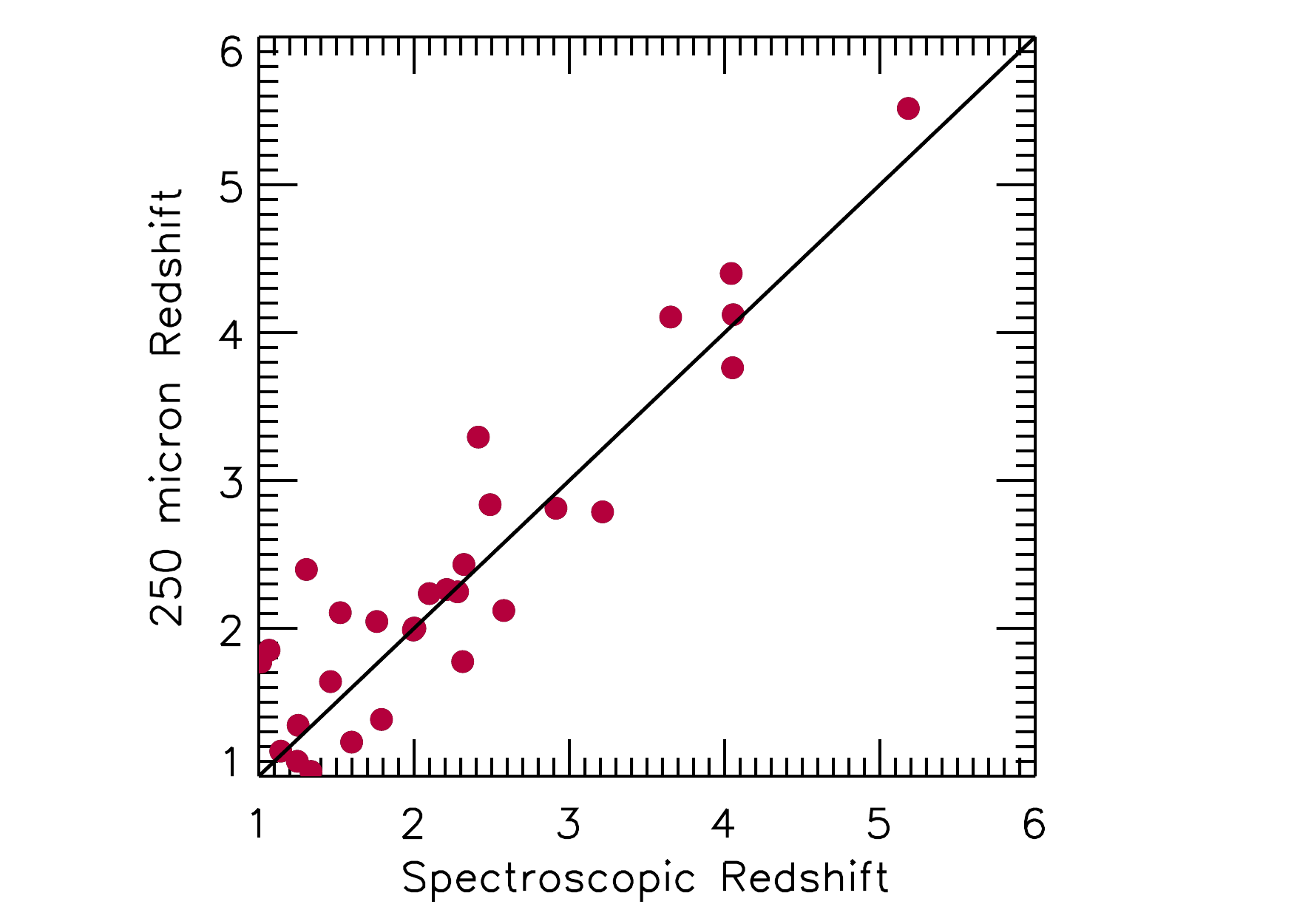}
\hspace{-1.4cm}\includegraphics[width=2.9in,angle=0]{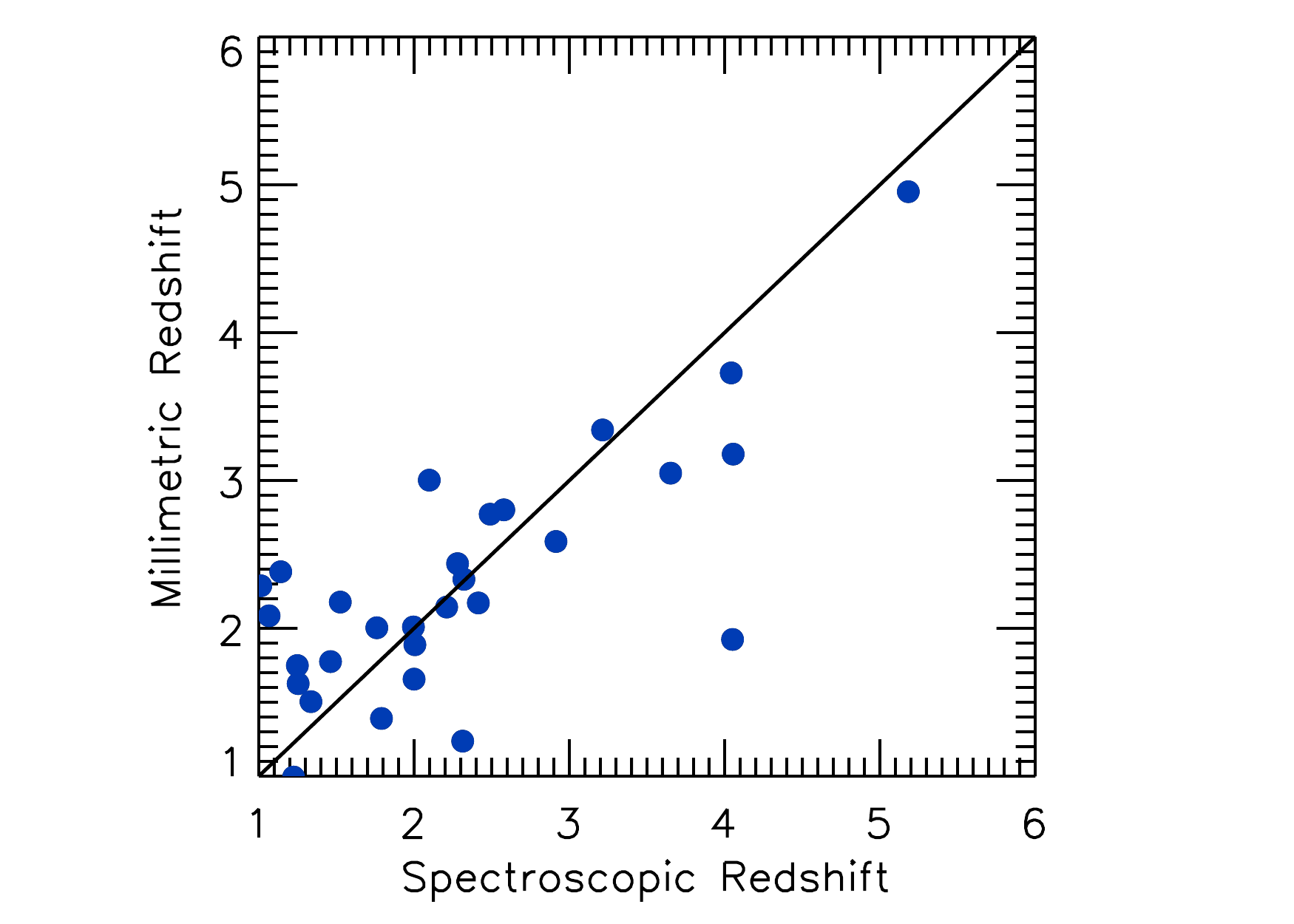}
}
\caption{
Comparison of (a) photometric (green circles), 
(b) $250\,\mu$m/$850\,\mu$m (red circles), and 20\,cm/850\,$\mu$m 
(blue circles) redshift estimates vs. spectroscopic
redshifts for the SCUBA-2 galaxies with $z>1.2$ and
accurate positions.
\label{compare_redshifts}}
\end{figure}

In Figure~\ref{compare_redshifts}, we show (a) photometric,
(b) $z_{250}$ (using Equation~\ref{z_250}, so there is no SED
assumption), and (c) millimetric redshift estimates
for the SCUBA-2 galaxies with $z>1.2$ and accurate positions. The 
photometric redshift estimates provide the best approximation to the 
spectroscopic redshifts, but they cannot be computed for the highest 
redshift galaxies detected in the submillimeter, which are too faint in 
the optical/NIR. The $z_{250}$ redshift estimates can be extended to 
all of the SCUBA-2 galaxies (though some are not detected at 
$250\,\mu$m, providing only lower limits on the redshifts). Note
that these will be most reliable for galaxies at $z\gg1$.
Although the millimetric redshift estimates are not as good as $z_{250}$, 
they can still be used to test $z_{250}$.

\begin{figure}
\centerline{\includegraphics[width=3.8in,angle=0]{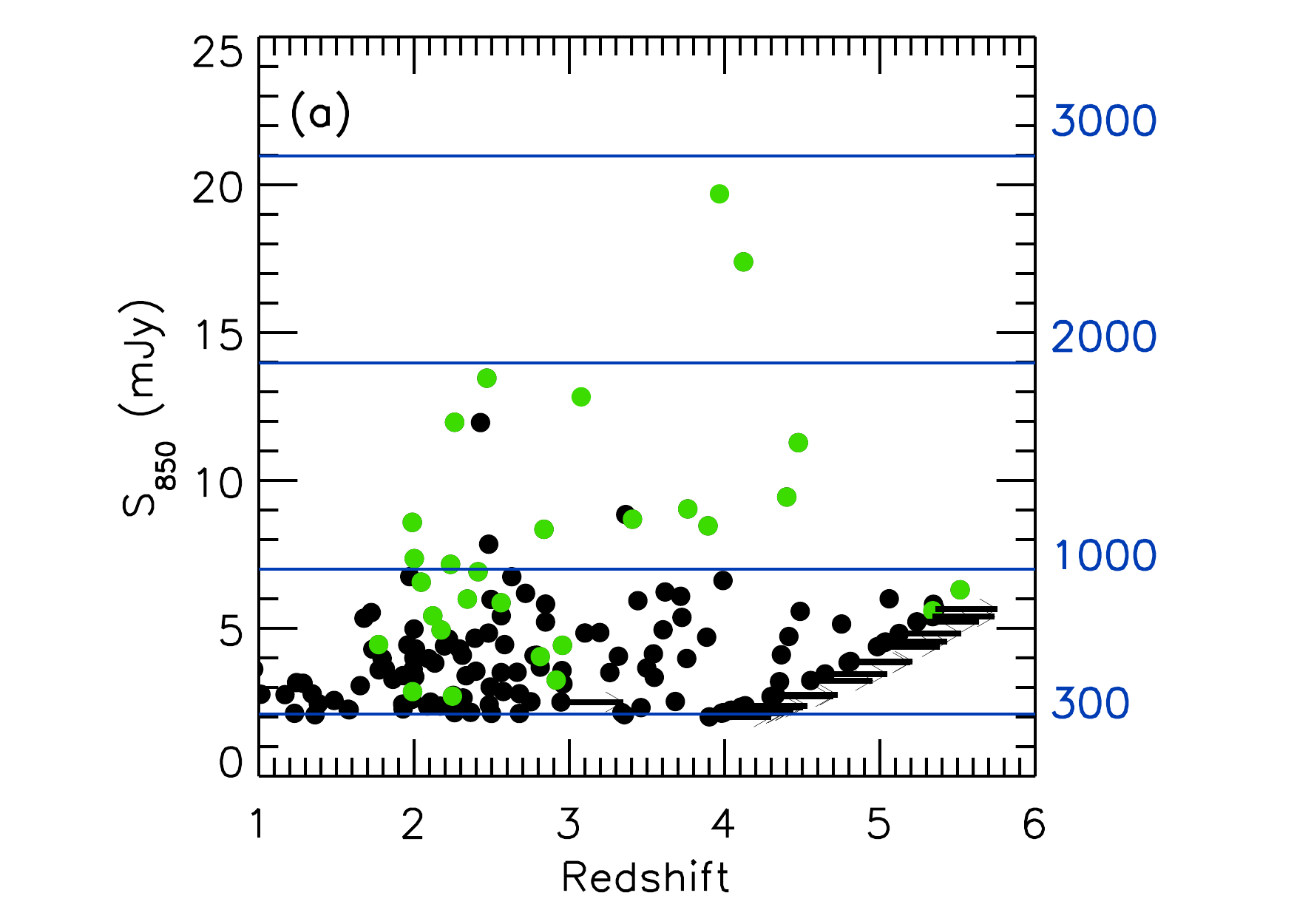}
\hspace{-1cm}
\includegraphics[width=3.8in,angle=0]{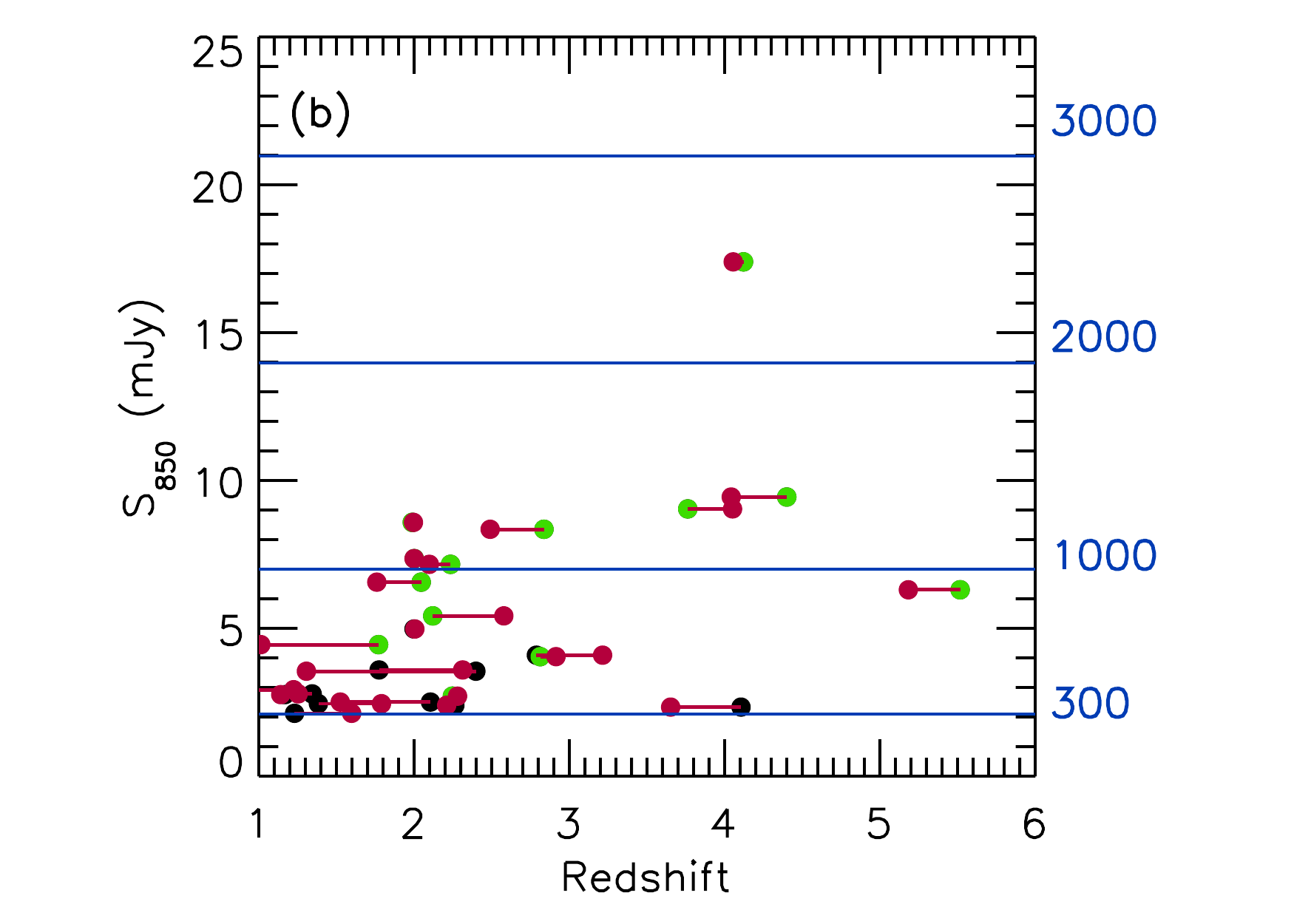}}
\caption{
(a) 850\,$\mu$m flux vs. $z_{250}$
for the full sample of SCUBA-2 sources with 850\,$\mu$m fluxes 
$>2$\,mJy. Where there is (is not) an SMA measurement,
the source is shown with a green (black) circle. 
The green circles include nearly all of the brighter SCUBA-2 galaxies.
Sources where there is no 250\,$\mu$m detection are shown with 
right-pointing arrows. 
(b) The sample with spectroscopic redshifts. The red circles are plotted at
the spectroscopic redshifts and then connected with red lines to the 
corresponding black or green circles plotted at $z_{250}$. 
The blue horizontal lines and right-hand axis labels in both panels show 
the SFRs for a Kroupa (2001) IMF.
\label{z_250_flux}
}
\end{figure}

\subsection{Redshift Distributions}
\label{z_dist}
By adopting $z_{250}$, we can use all the sources from the full 
SCUBA-2 field in measuring the star formation history, as well as avoid 
any biases that might be introduced
from radio matching. In Figure~\ref{z_250_flux}(a), we plot
850\,$\mu$m flux versus $z_{250}$ (green circles for sources with SMA
measurements; black circles otherwise). Sources without significant
250\,$\mu$m detections have only lower limits on z$_{250}$ and are
shown with right-pointing arrows in the figure based on the
$1\sigma$ limit on the 250\,$\mu$m flux.
We show with blue horizontal lines and right-hand axis labels the SFRs 
that correspond to the 850\,$\mu$m fluxes for a Kroupa (2001) IMF
based on Equation~\ref{sfr_850}.
For sources with a spectroscopic redshift, we plot them again
in Figure~\ref{z_250_flux}(b) as red circles at the values
of the spectroscopic redshifts.
We also plot red connectors from the spectroscopic redshifts to
$z_{250}$ (as in (a), green circles for sources with SMA
measurements; black circles otherwise).
Fortunately, the differences between $z_{250}$ and the spectroscopic
redshifts are generally too small to affect the determinations
of the SFR distribution functions. 

We note several features from Figure~\ref{z_250_flux}(a). First,
there is a strong deficiency of luminous SCUBA-2 sources at $z\ll2$. 
Nearly all of the luminous sources lie in the $z\sim 2-5$ range. Second,
there is a tendency for the SCUBA-2 sources to lie in redshift sheets,
such as the well-known spectroscopic
features at $z=1.99$ (Chapman et al.\ 2009) and 
$z=4.05$ (Daddi et al.\ 2009b). Unfortunately, $z_{250}$ 
is not accurate enough to decide whether additional sources lie in
these sheets; only with spectroscopic redshifts is there enough velocity
resolution for that.
The presence of redshift sheets emphasizes the problem of
cosmic variance in determining the star formation history,
and we postpone a more detailed discussion of this issue to a 
subsequent paper in the series. 

\begin{figure}
\centerline{\includegraphics[width=3.4in,angle=90]{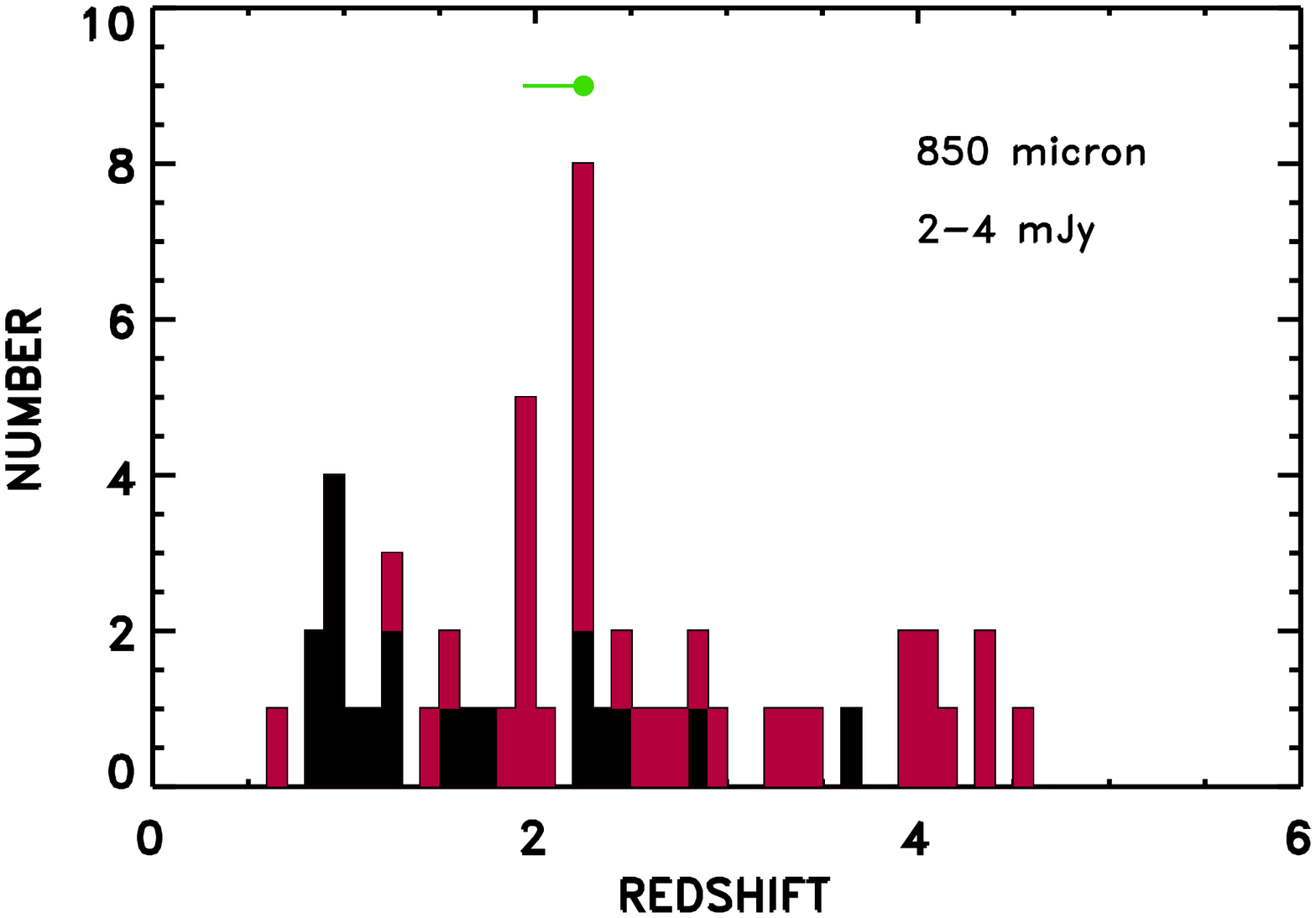}
\hskip -2cm
\includegraphics[width=3.4in,angle=90]{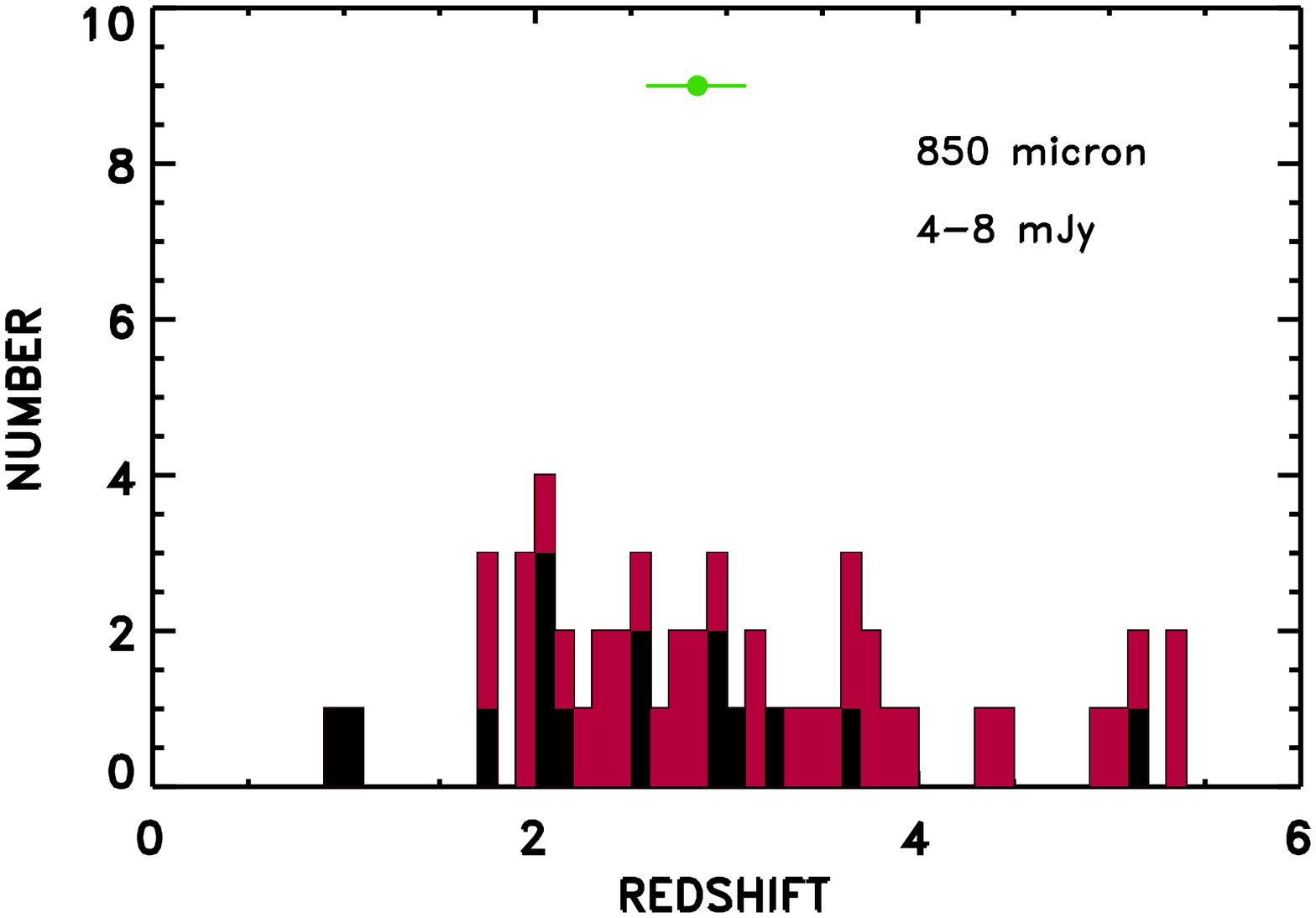}}
\vskip -2cm
\centerline{\includegraphics[width=3.4in,angle=90]{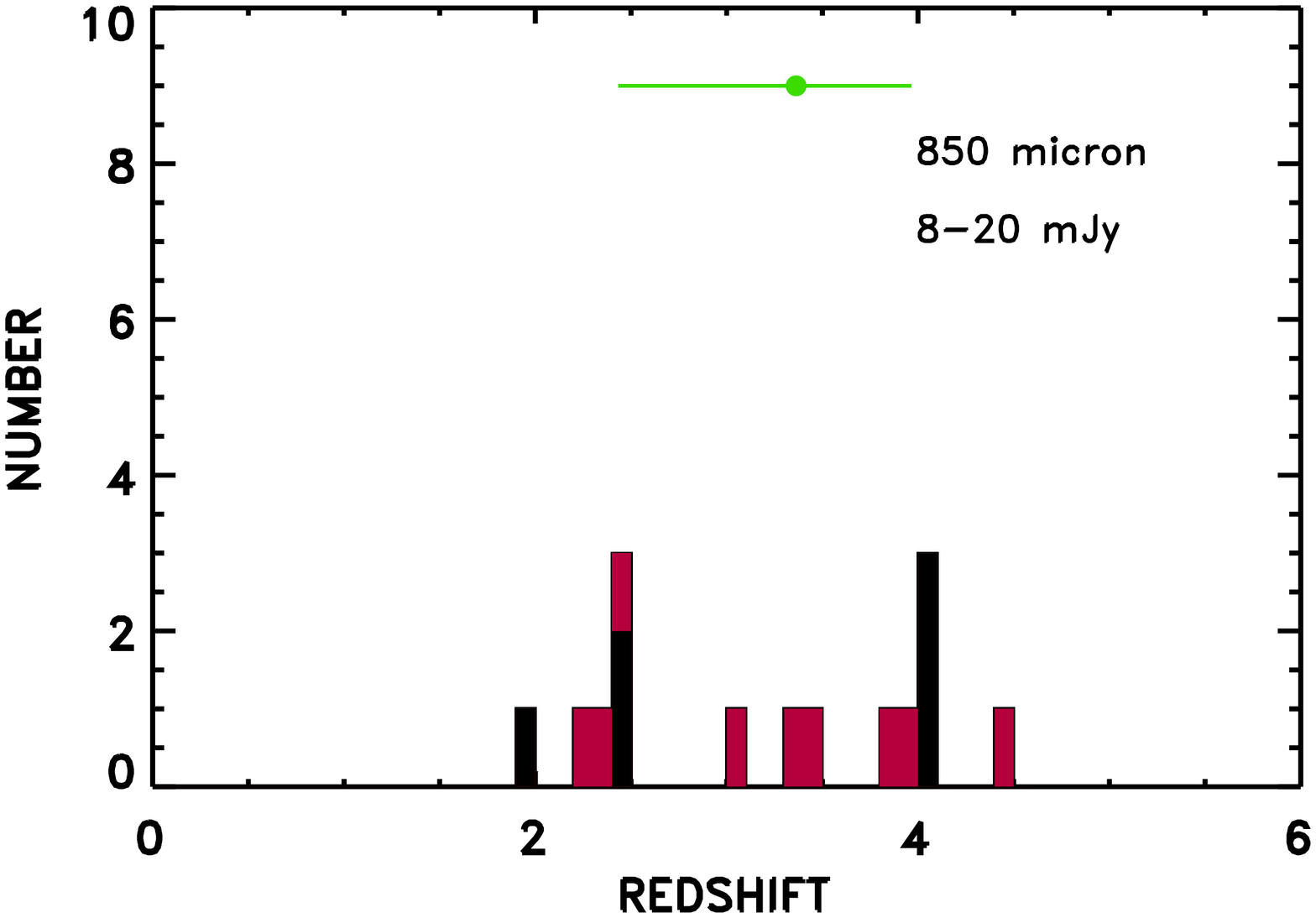}
\hskip -2cm
\includegraphics[width=3.4in,angle=90]{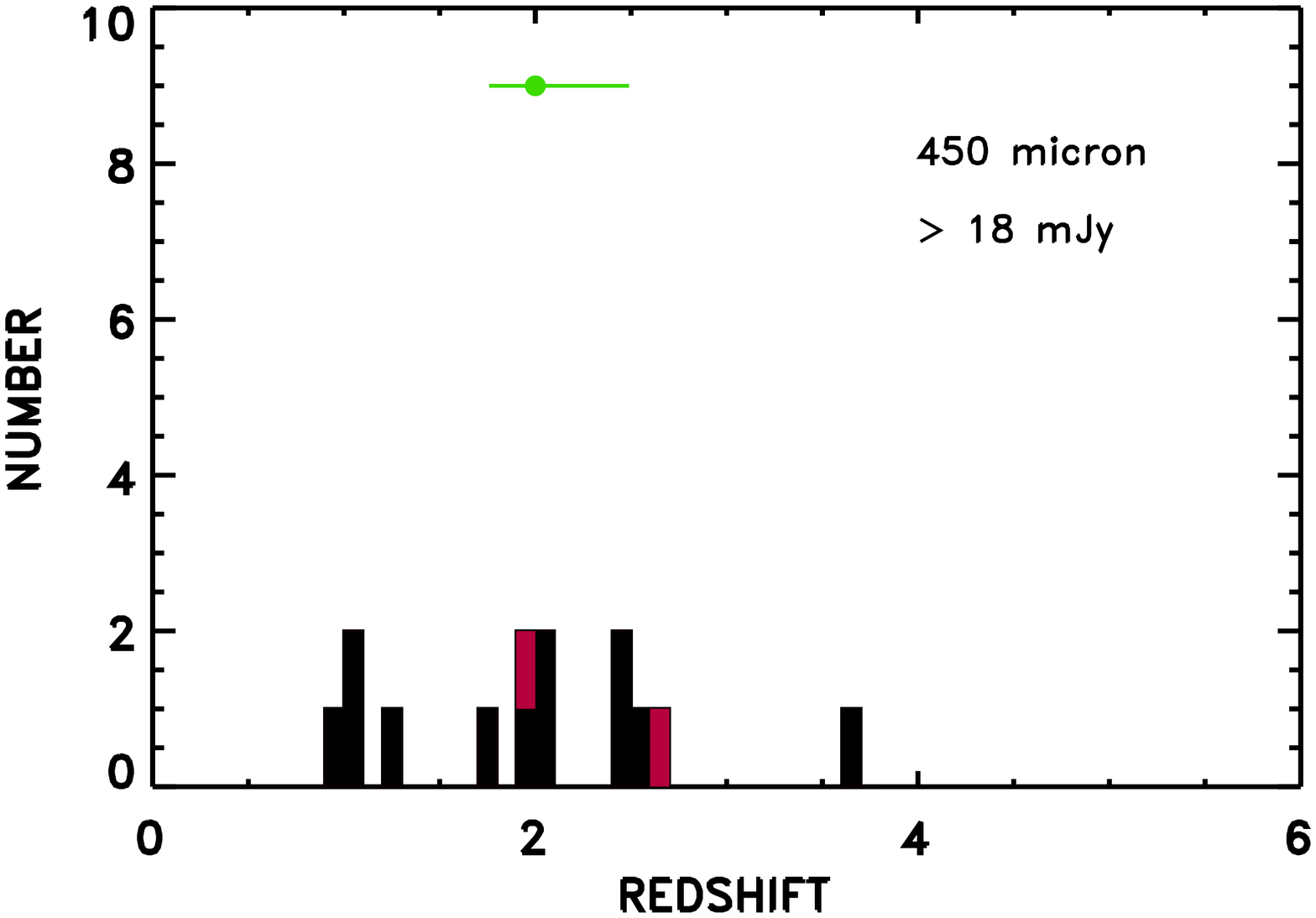}}
\vskip -0.5cm
\caption{Redshift distributions for three intervals of 850\,$\mu$m flux:
(a) 2--4\,mJy, (b) 4--8\,mJy, and (c) 8--20\,mJy, as well as for
(d) the full 450\,$\mu$m sample $>18$\,mJy. For each flux interval,
sources are only shown if they lie in areas
where the errors are low enough that all $4\,\sigma$ sources
down to the flux limit are included (i.e., a complete sample).
Black shows spectroscopic or photometric redshifts,
and red shows $z_{250}$ redshifts. The green circle shows the median
redshift, and the green bar the 68\% confidence interval on the median.
\label{redshift_hist}
}
\end{figure}

In Figure~\ref{redshift_hist}, we show the redshift distribution
for the 850\,$\mu$m sample divided into three flux intervals: 
(a) 2--4\,mJy, (b) 4--8\,mJy, and (c) 8--20\,mJy.
In Figure~\ref{redshift_hist}(d), we also show the redshift
distribution for the full 450\,$\mu$m sample $>18$\,mJy.
As expected, because of the smaller $K$-correction,
the 450\,$\mu$m sample has a relatively low redshift distribution
with a median value of $z=2.0^{+0.4}_{-0.2}$, where the superscript
and subscript give the 68\% confidence interval on the median.
(See B{\'e}thermin et al.\ 2015 for an extensive discussion of
the wavelength dependence of the redshift distribution of dusty,
star-forming galaxies.)
The 450\,$\mu$m sample is too small for an analysis of the dependence 
of the redshift distribution on the flux. 
However, the median value can be compared with the median redshift of 
$z=1.95\pm0.19$ found by Casey et al.\ (2013) and the mean
redshift of $z=1.3$ found by Geach et al.\ (2013)
for samples limited at fainter fluxes of 13\,mJy and 5\,mJy, respectively.
These results suggest that the redshift distribution at 450\,$\mu$m is 
increasing with increasing flux.

\begin{figure}
\centerline{\includegraphics[width=3.8in,angle=0]{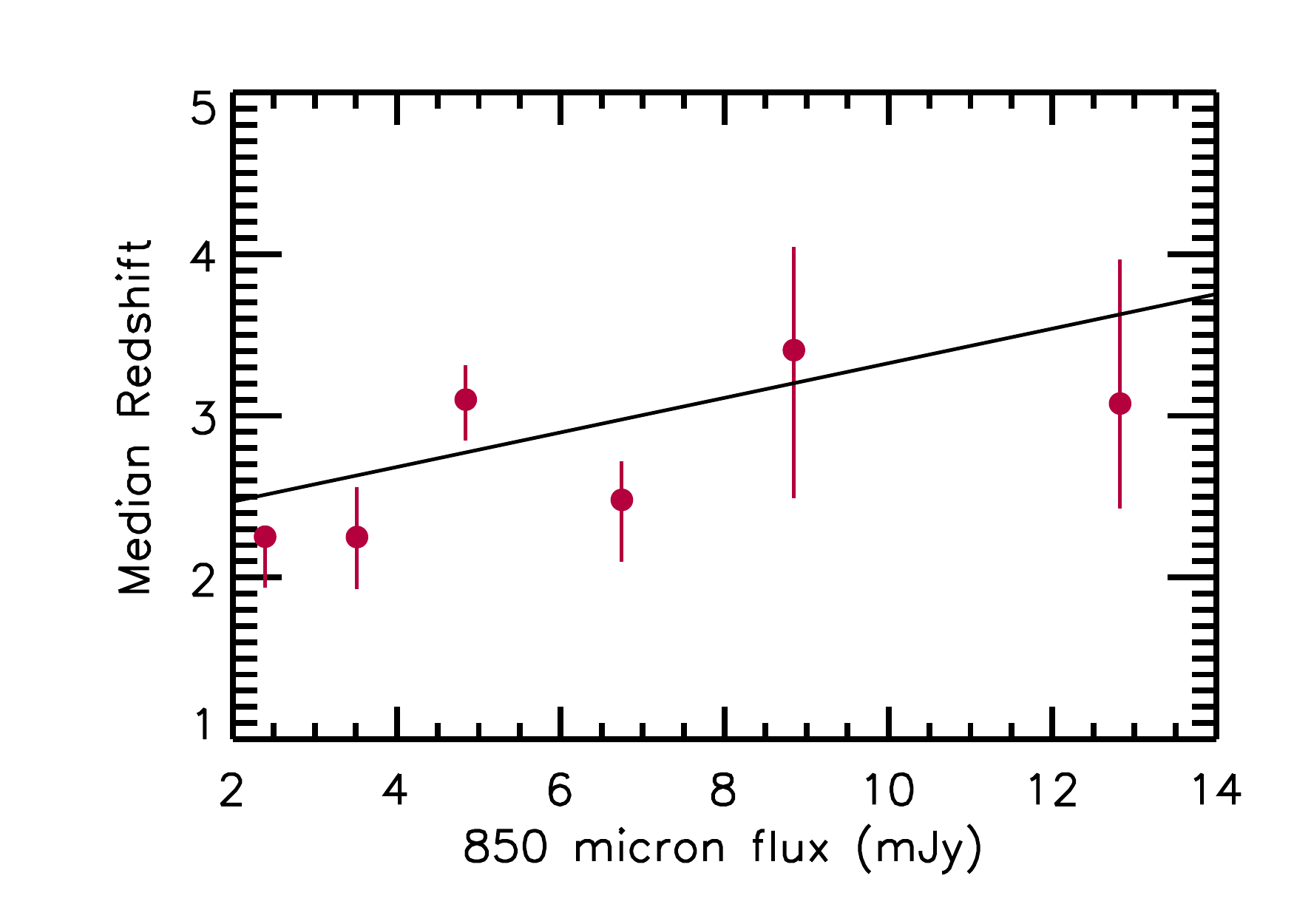}}
\caption{Median redshift vs. 850\,$\mu$m flux (red circles).
The error bars show the 68\% confidence intervals on the medians.
The black line shows a linear least squares fit (Equation~\ref{red_fit}).
\label{redshift_fit}
}
\end{figure}

The situation is much clearer for the larger 850\,$\mu$m
sample, where the redshift distribution is clearly increasing with increasing
flux, as shown in Figure~\ref{redshift_fit}.
A Mann-Whitney test gives only a 0.15\% probability
that the 8--16\,mJy sample (Figure~\ref{redshift_hist})
is drawn from the same redshift distribution as the 2--4\,mJy
sample. A linear least squares fit gives the relation 
\begin{equation}
z_{median}=2.26 + 0.11 (\pm0.03) \times S_{850} \, {\rm (mJy)} \,,
\label{red_fit}
\end{equation}
which we show as the black line in Figure~\ref{redshift_fit}.
These results are broadly consistent with recent modeling by 
B{\'e}thermin et al.\ (2015), who find lower redshift distributions
for lower fluxes and shorter wavelengths.

\subsection{Evolution of the SFR Density Distribution Functions with Redshift}
\label{sfr_dist}

\begin{figure}
\centerline{\includegraphics[width=4.0in,angle=0]{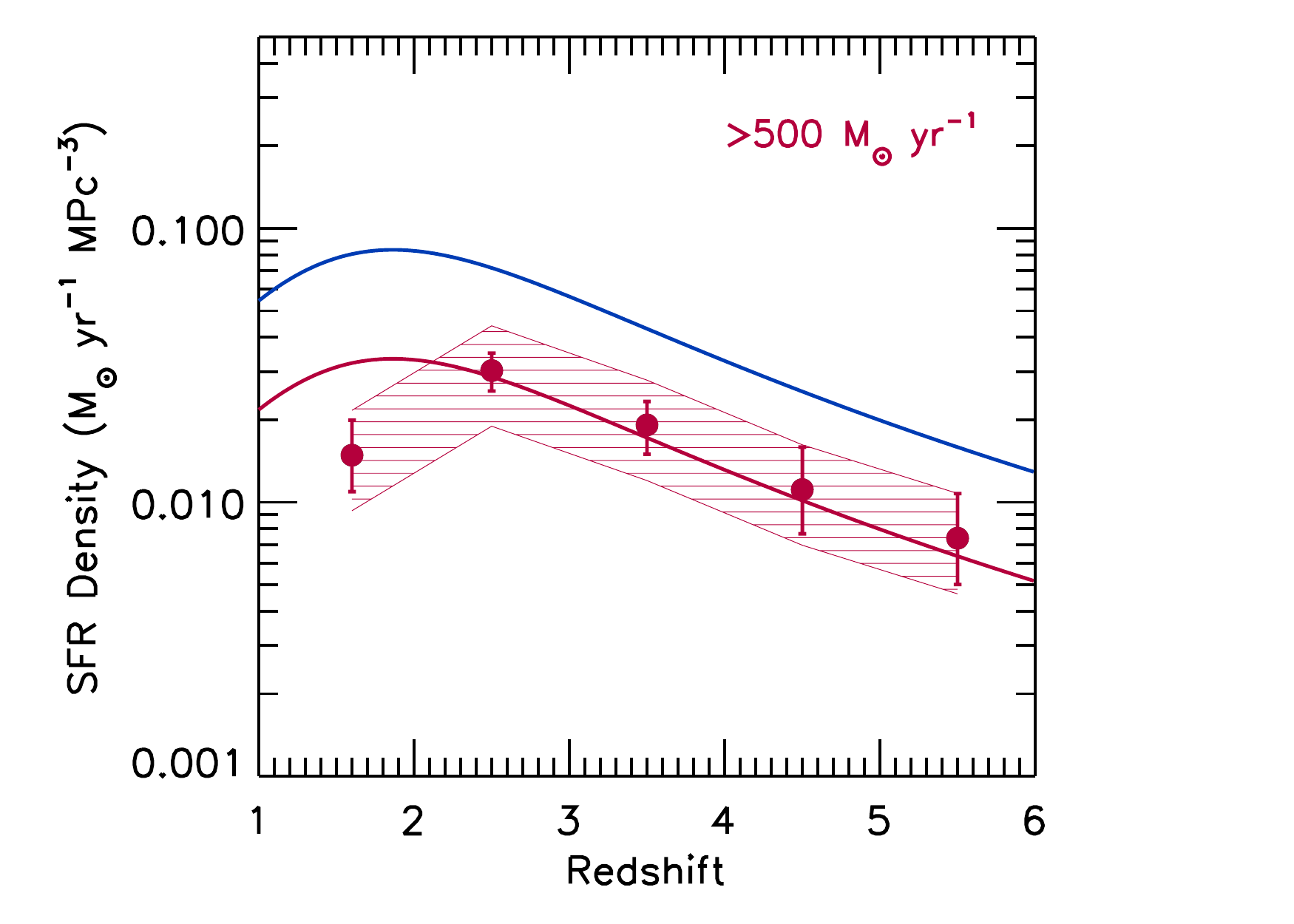}}
\centerline{\includegraphics[width=4.0in,angle=0]{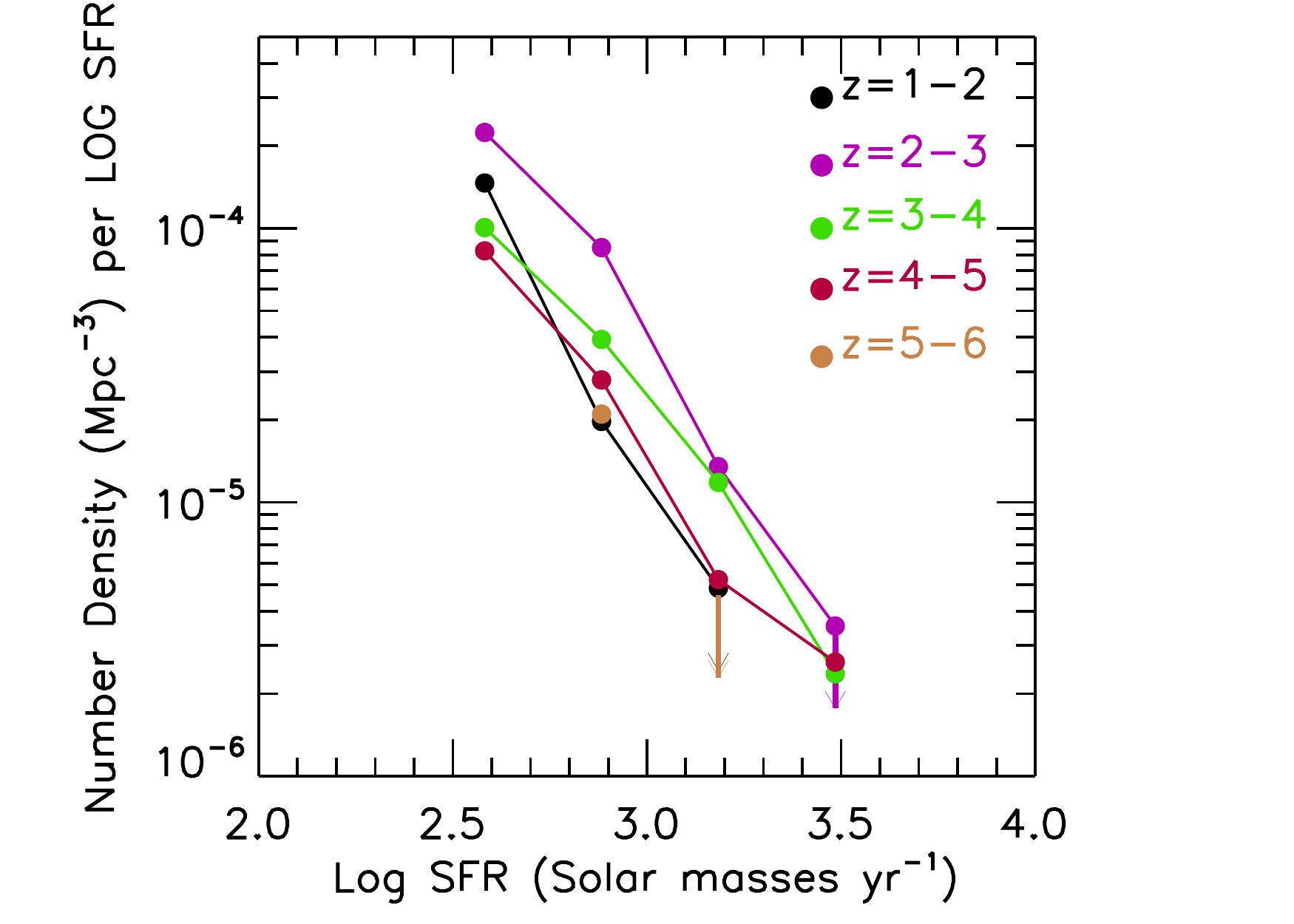}}
\caption{
(a) SFR density per unit comoving volume vs. redshift
for the SCUBA-2 sources 
with SFRs $\gtrsim500\,M_\odot$~yr$^{-1}$ for a Kroupa (2001) IMF 
(red circles; the $1\,\sigma$ error bars are Poissonian 
based on the number of sources in each bin).
The computations were made at $z=1-2$, $2-3$, $3-4$, $4-5$,
and $5-6$ and are plotted at the mean redshift of each bin.
The blue curve shows the SFR density history computed by
Madau \& Dickinson (2014), after we converted it to a Kroupa IMF,
and the red curve shows it multiplied by 0.4 to match roughly 
the SFR density history of the SCUBA-2 sources. However,
once we also take into account the different FIR calibrations that were used in
this work versus in Madau \& Dickinson, we find that the SCUBA-2 data are 
only 0.29 times the Madau \& Dickinson curve. The red
shaded region shows the 68$\%$ confidence range when allowance
is made for the systematic uncertainties in the SFR calibrations. 
(b) Number density per unit comoving volume per unit $\log$ SFR 
vs. $\log$ SFR in five redshift intervals (see legend).
\label{sfr_history}
}
\end{figure}

\begin{figure}
\centerline{\includegraphics[width=4.0in,angle=0]{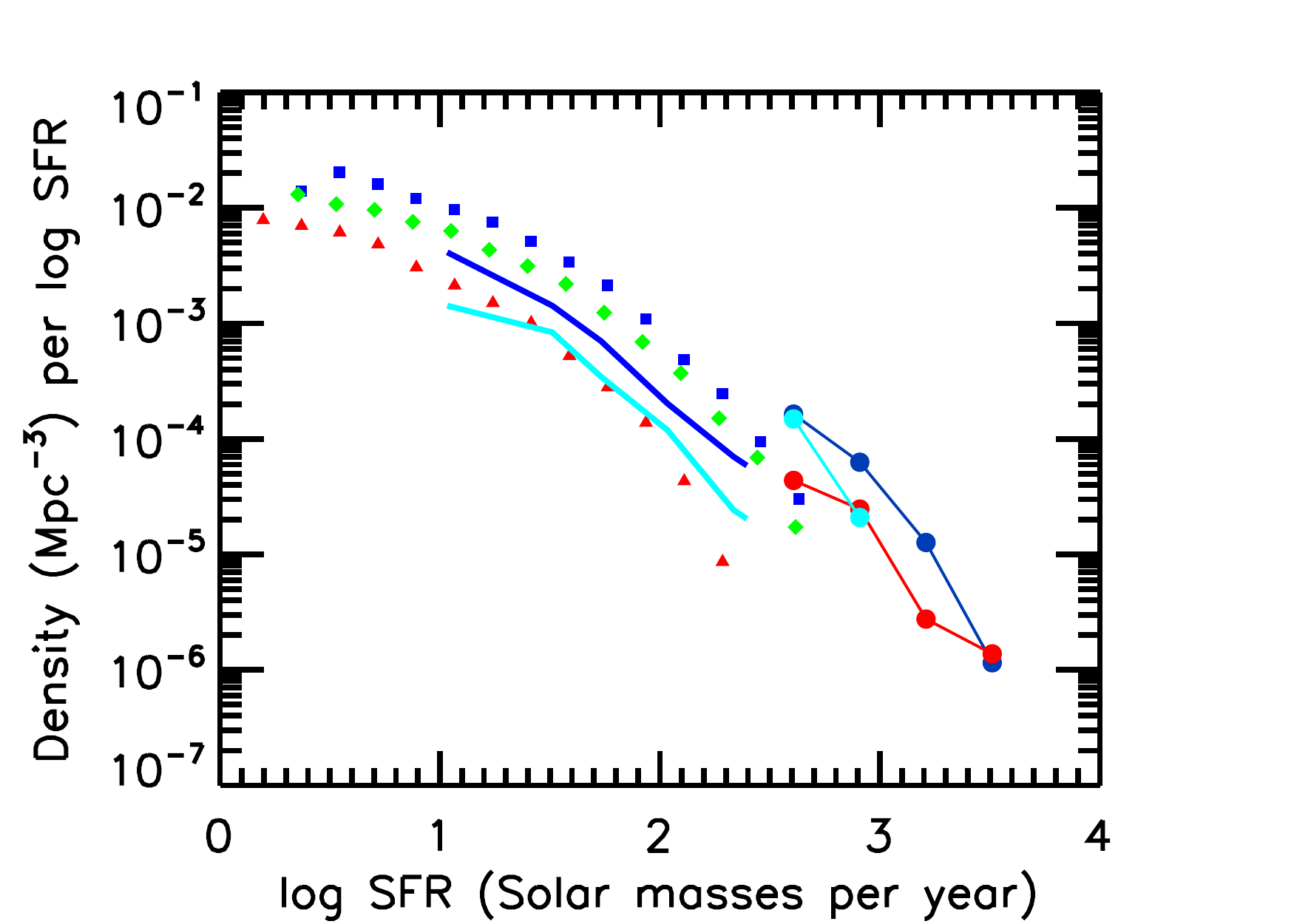}}
\caption{
Number density per unit comoving volume per unit $\log$ SFR vs.
$\log$ SFR from Figure~\ref{sfr_history}, but now in three coarser
redshift intervals
(circles:  red---$z=1.25-2$; blue---$z=2-4$; cyan---$z=4-6$).
For comparison, the small symbols and curves show
extinction-corrected UV results from
Reddy \& Steidel (2009; cyan curve---$z\sim3$, blue curve---$z\sim2$)
and van der Burg (2010; red triangles---$z=4.8$,
green diamonds---$z=3.8$, blue squares---$z=3.1$),
assuming a Kennicutt (1998) conversion of UV luminosity to SFR
for a Kroupa (2001) IMF.
The higher van der Burg et al.\ than Reddy \& Steidel values
reflect their larger adopted extinction corrections.
\label{sfr_comparison}}
\end{figure}

We computed the SFR density history and the SFR density distribution functions 
for the SCUBA-2 sources with SFRs $\gtrsim500\,M_\odot$~yr$^{-1}$
following Barger et al.\ (2014). In Figure~\ref{sfr_history}(a),
we show the SFR density history computed for the redshift intervals
$z=1-2$, $2-3$, $3-4$, $4-5$, and $5-6$ (red circles). 
Below a redshift of $z\sim2$, the SCUBA-2 contribution drops rapidly.
Above $z\sim2$, this SFR density history corresponds to $\sim40$\% (red curve) 
of the total SFR density history compiled by Madau \& Dickinson (2014), after
converting theirs to a Kroupa IMF (blue curve). Note, however, that we 
show our results for our adopted conversion from $L_{8-1000\,\mu{\rm m}}$ 
to SFR. Since Madau \& Dickinson adopted a slightly lower conversion
(see Section~\ref{seddisc}), we also need to apply that correction in order 
to do a proper comparison. That correction reduces the percentage
from $\sim40$\% to $\sim29$\%.  
(Although the present SFR density history is consistent with that 
found by Barger et al.\ 2014 using a subset of 
the current SCUBA-2 sample, in that paper we compared with the 
Hopkins \& Beacom 2006 compilation, which is higher than
the Madau \& Dickinson compilation. Thus, the contribution to the
total that we quoted there was a lower 16\%.)
Clearly, the contribution of dusty, powerfully star-forming galaxies to 
the overall star formation history is impressively large.

The error bars in Figure~\ref{sfr_history}(a) are purely statistical.
However, systematic errors in the SFR conversion
and in the redshift estimates may also be important. In order
to estimate these potential errors, we recomputed the
SFR density history randomly, reassigning the SFRs and redshifts within
the potential error ranges. 
Based on these Monte Carlo calculations,
we computed the 68\% error range produced by the systematics,
which we show as the red shaded region. For
the highest and lowest redshifts, where there are fewer objects, the systematic
and statistical errors are similar, but for redshifts between
2 and 4, the systematic error dominates. 

In Figure~\ref{sfr_history}(b), we show the SCUBA-2 SFR density distribution 
functions computed for the same redshift intervals vs. $\log$ SFR. 
Although the normalizations of the distribution functions are changing---they 
rise to reach a peak at $z=2-3$ before dropping at higher redshifts---the shapes 
over $z=2-5$ are remarkably similar; that is, the number density of high SFR galaxies 
relative to that of lower SFR galaxies is about the same in each redshift 
interval. However, at some point this similarity in shape must 
break down, as there cannot be large numbers of powerfully star-forming 
galaxies at very high redshifts. The present data show that this
must occur at redshifts higher than $z \sim 5$.

In Figure~\ref{sfr_comparison}, we show the SCUBA-2 distribution functions 
of Figure~\ref{sfr_history}(b) recomputed for the coarser redshift intervals 
$z=1.25-2, 2-4$, and $4-6$ (colored circles; note that in the $z=4-6$ interval,
there are not enough sources to plot the two highest SFR bins).
We compare these with those computed at lower SFRs
from the extinction-corrected UV samples of Reddy \& Steidel (2009)
and van der Burg et al.\ (2010) (small colored symbols and curves),
assuming a Kennicutt (1998) conversion of UV luminosity to SFR
for a Kroupa (2001) IMF. This emphasizes the disjoint nature of the SFRs
of galaxies selected by UV and submillimeter observations (e.g.,
Barger et al.\ 2014): only submillimeter observations can find the most 
powerfully star-forming galaxies at redshifts $z\gtrsim1$.
Both the UV and SCUBA-2 distribution functions have similar shapes
in the $z=2-5$ interval. Thus, the number density of high SFR galaxies measured
from the SCUBA-2 sample relative to that of lower SFR galaxies from
UV-selected samples is about the same in this redshift interval. 
However, the normalizations of the SCUBA-2  
functions appear slightly high relative to the UV functions at the 
overlap point. This may suggest that even at these lower SFRs, the 
UV samples are missing some star-forming galaxies.

\section{Summary}
\label{secconcl}
In this paper, we presented $\ge4\,\sigma$ 850\,$\mu$m and 450\,$\mu$m
catalogs from our uniform and deep SCUBA-2 survey of the 
GOODS-N/CANDELS/CDF-N field. 
We used submillimeter interferometry and 20\,cm data
to identify counterparts to the SCUBA-2 sources, which resulted in the
localization of 114 of the 186 850\,$\mu$m sources,
including 26 of the 29 sources with 850\,$\mu$m fluxes above 6\,mJy.

We obtained new spectroscopic redshifts with Keck and utilized 
spectroscopic and photometric redshifts
from the literature, where possible. We also estimated redshifts
from the 20\,cm to 850\,$\mu$m flux ratio (millimetric redshifts) and
from the 250\,$\mu$m to 850\,$\mu$m flux ratio ($z_{250}$ redshifts).
The redshift distribution of the submillimeter sample increases with both flux and
wavelength, consistent with recent model predictions (B{\'e}thermin et al. 2015).
We parameterized this dependence on flux.
We showed how $z_{250}$ is good enough 
(and better than millimetric) when compared with the spectroscopic 
redshifts for determining the SFR density history and the SFR density distribution 
functions for the SCUBA-2 sources with SFRs $\gtrsim500\,M_\odot$~yr$^{-1}$ 
for a Kroupa (2001) IMF.

We found that above $z\sim2$, the contribution from the SCUBA-2
sources is an impressively large 29\% of the total SFR density history 
compiled by Madau \& Dickinson (2014).
However, below $z\sim2$, the SCUBA-2 contribution drops rapidly.
We computed the SCUBA-2 SFR density distribution functions for five
redshift intervals and found that, although the normalizations rise to a peak
around $z=2-3$ before dropping at higher redshifts, the shapes over
$z=2-5$ remain strikingly invariant. In other words, the number of high
SFR galaxies relative to the number of lower SFR galaxies remains about
the same in each redshift interval above $z\sim2$. This shape invariance
cannot be maintained to the highest redshifts, as eventually powerfully
star-forming galaxies must disappear. We will investigate this
further in a later paper in the series, where we will use the combined
GOODS-N and GOODS-S datasets along with wider field samples. 

We also compared the SCUBA-2
SFR density distribution functions in coarser redshift bins with results
from extinction-corrected UV samples. As previously shown by Barger et al.\ (2014),
the UV and submillimeter selected samples are nearly disjoint, with
only the submillimeter observations able to find the most powerfully 
star-forming galaxies in the universe. We found that at both wavelengths,
the shapes of the distribution functions are similar from $z=2-5$; however,
the normalizations of the SCUBA-2 functions are slightly high relative to the UV 
functions in the overlap region. This may suggest that the UV samples are still 
missing some star-forming galaxies, even at these lower SFRs. However, it
may also simply be due to uncertain extinction corrections in the UV and different 
calibrations of the SFRs in the two populations.

\acknowledgements
We thank the referee for a very useful report that improved the paper. 
We gratefully acknowledge support from
NSF grants AST-0709356 (L.~L.~C., L.-Y.~H.) and 
AST-1313150 (A.~J.~B.), the John Simon Guggenheim Memorial
Foundation and Trustees of the William F. Vilas Estate (A.~J.~B.), and
the ERC Advanced Investigator programme DUSTYGAL 321334
(C.-C.~C.). 
The James Clerk Maxwell Telescope has historically been operated by
the Joint Astronomy Centre on behalf of the Science and Technology
Facilities Council of the United Kingdom, the National Research Council
of Canada and the Netherlands Organisation for Scientific Research.
Additional funds for the construction of SCUBA-2 were provided by the
Canada Foundation for Innovation.
We acknowledge the cultural significance that the summit of 
Maunakea has to the indigenous Hawaiian community.


\newpage
\begin{deluxetable*}{ccccrrrcccrclc}
\tabletypesize{\scriptsize}
\renewcommand\baselinestretch{1.0}
\tablewidth{0pt}
\tablecaption{SCUBA-2 850 Micron Sample ($4\,\sigma$)}
\tablehead{No. & Name & R.A. & Decl.&  Flux  & Error & S/N  & R.A. & Decl. & SMA & 20\,cm & $K_{s}$ & $z$ & $z_{250}$\\  & & \multicolumn{2}{c}{(SCUBA-2)} & & & & \multicolumn{2}{c}{(accurate)} & & & & \\ & & (J2000.0) & (J2000.0) & (mJy) & (mJy) &  & (J2000.0) & (J2000.0) & (mJy) & ($\mu$Jy) & (AB) & & \\ (1) & (2) & (3) & (4) & (5) & (6) & (7) & (8) & (9) & (10) & (11) & (12) & (13) & (14)}
\startdata
 1& SMM123551622145 &       12       35 51.71 &       62       21 45.2 & 19.6 &  1.49 &  13.1 &       12       35 51.37 &       62       21 47.2 & 13.7 &  51 &     22.1 & \nodata &   3.96\cr
 2& SMM123712622210 &       12       37 12.05 &       62       22 10.9 & 17.3 & 0.94 &  18.4 &       12       37 11.92 &       62       22 12.0 & 23.9 &  87 &     22.4 &   4.055 f1,g &   4.12\cr
 3& SMM123730621258 &       12       37 30.69 &       62       12 58.6 & 13.4 & 0.39 &  33.7 &       12       37 30.80 &       62       12 58.7 & 14.9 &  126 &     22.9 &   2.43 &   2.46\cr
 4& SMM123555622238 &       12       35 55.85 &       62       22 38.2 & 12.8 &  1.72 &  7.4 &       12       35 55.88 &       62       22 39.0 & 17.0 &  56 &     22.1 & \nodata &   3.07\cr
 5& SMM123546622012 &       12       35 46.87 &       62       20 12.0 & 11.9 &  1.21 &  9.8 &       12       35 46.64 &       62       20 13.3 & 14.0 &  46 &     22.7 & \nodata &   2.26\cr
 6& SMM123623620334 &       12       36 23.76 &       62 03 34.7 & 11.9 &  1.18 &  10.1 & \nodata & \nodata &\nodata &\nodata & \nodata & \nodata &   2.42\cr
 7& SMM123627620605 &       12       36 27.28 &       62 06 5.89 & 11.2 & 0.80 &  14.0 &       12       36 27.21 &       62 06 5.50 & 11.5 &  34 &     24.8 & \nodata &   4.47\cr
 8& SMM123633621407 &       12       36 33.46 &       62       14 7.91 & 9.4 & 0.36 &  26.0 &       12       36 33.42 &       62       14 8.50 & 12.0 &  33 &     25.5 &   4.042 f2 &   4.40\cr
 9& SMM123817620900 &       12       38 17.69 &       62 09 0.29 & 9.0 &  1.32 &  6.8 &       12       38 18.21 &       62 08 58.0 &\nodata &  105 &     21.5 & \nodata &   2.32\cr
  10  &  SMM123709622200 &       12       37 9.320 &       62       22 0.89 & 9.0 & 0.90 &  9.9 &       12       37 8.822 &       62       22 1.90 & 9.2 &  143 &     23.7 &   4.051 f1 &   3.76\cr
11& SMM123549621905 &       12       35 49.07 &       62       19 5.09 & 8.8 &  1.01 &  8.7 &       12       35 48.84 &       62       19 4.91 &\nodata &  71 &     22.8 & \nodata &   3.36\cr
12& SMM123550621041 &       12       35 50.51 &       62       10 41.2 & 8.6 & 0.80 &  10.8 &       12       35 50.35 &       62       10 41.9 & 10.1 &  25 &     23.4 & \nodata &   3.40\cr
13& SMM123711621329 &       12       37 11.53 &       62       13 29.9 & 8.5 & 0.34 &  24.9 &       12       37 11.34 &       62       13 30.9 & 6.7 &  123 &     20.4 &   1.995 d,g &   1.98\cr
14& SMM123631621712 &       12       36 31.99 &       62       17 12.9 & 8.4 & 0.40 &  20.9 &       12       36 31.94 &       62       17 14.7 & 7.1 &  21 &     23.0 & \nodata &   3.89\cr
  15  &  SMM123707621408 &       12       37 7.390 &       62       14 8.89 & 8.3 & 0.33 &  25.2 &       12       37 7.177 &       62       14 8.19 & 7.1 &  58 &     21.4 &   2.490 d,g &   2.83\cr
16& SMM123629620255 &       12       36 29.30 &       62 02 55.9 & 7.8 &  1.29 &  6.0 &       12       36 29.21 &       62 02 53.9 &\nodata &  213 &     21.1 & \nodata &   2.48\cr
17& SMM123515621510 &       12       35 15.27 &       62       15 10.9 & 7.4 &  1.71 &  4.3 &       12       35 14.91 &       62       15 9.99 &\nodata &  237 &     22.1 & \nodata &   2.33\cr
18& SMM123618621549 &       12       36 18.54 &       62       15 49.7 & 7.3 & 0.42 &  17.4 &       12       36 18.35 &       62       15 50.4 & 7.2 &  163 &     22.0 &   2.000 e,g &   2.00\cr
19& SMM123553621338 &       12       35 53.84 &       62       13 38.3 & 7.1 & 0.73 &  9.7 &       12       35 53.26 &       62       13 37.5 & 4.3 &  41 &     22.8 &   2.098 c &   2.23\cr
20& SMM123634621921 &       12       36 34.84 &       62       19 21.9 & 6.9 & 0.70 &  9.8 &       12       36 34.92 &       62       19 23.5 & 8.9 &  85 &     21.7 & \nodata &   2.41\cr
21& SMM123721620709 &       12       37 21.29 &       62 07 9.79 & 6.7 & 0.74 &  9.0 &       12       37 21.40 &       62 07 8.30 &\nodata &  294 &     20.9 & \nodata &   1.97\cr
22& SMM123722620539 &       12       37 22.97 &       62 05 39.7 & 6.7 & 0.87 &  7.7 &       12       37 23.00 &       62 05 39.5 &\nodata &  71 &     22.1 & \nodata &   2.62\cr
23& SMM123638620226 &       12       36 38.55 &       62 02 26.9 & 6.6 &  1.37 &  4.8 & \nodata & \nodata &\nodata &$<12$ & \nodata & \nodata &   1.64\cr
24& SMM123644621938 &       12       36 44.75 &       62       19 38.9 & 6.6 & 0.69 &  9.5 & \nodata & \nodata &\nodata &$<12$ & \nodata & \nodata &   3.98\cr
  25  &  SMM123701621145 &       12       37 1.501 &       62       11 45.9 & 6.5 & 0.32 &  20.2 &       12       37 1.578 &       62       11 46.4 & 4.8 &  95 &     20.5 &   1.760 e &   2.04\cr
26& SMM123652621224 &       12       36 52.07 &       62       12 24.9 & 6.3 & 0.29 &  21.4 &       12       36 52.03 &       62       12 25.9 & 7.8 &  12 &     99.0 &   5.183 h &   5.51\cr
27& SMM123539621241 &       12       35 39.71 &       62       12 41.7 & 6.2 & 0.88 &  7.0 &       12       35 39.58 &       62       12 44.0 &\nodata &  22 &     22.6 & \nodata &   3.61\cr
28& SMM123812621453 &       12       38 12.39 &       62       14 53.5 & 6.1 & 0.97 &  6.3 &       12       38 12.45 &       62       14 55.3 &\nodata &  44 &     19.4 & \nodata &   2.71\cr
29& SMM123632620621 &       12       36 32.98 &       62 06 21.9 & 6.0 & 0.76 &  7.9 &       12       36 32.69 &       62 06 21.1 &\nodata &  33 &     23.8 & \nodata &   3.71\cr
  30  &  SMM123803620631 &       12       38 3.468 &       62 06 31.7 & 5.9 &  1.28 &  4.6 & \nodata & \nodata &\nodata &$<12$ & \nodata & \nodata &   5.05\cr
31& SMM123646621447 &       12       36 46.05 &       62       14 47.9 & 5.9 & 0.32 &  18.5 &       12       36 46.08 &       62       14 48.6 & 4.2 &  101 &     22.4 &   3.63 &   2.34\cr
32& SMM123616620701 &       12       36 16.71 &       62 07 1.69 & 5.9 & 0.77 &  7.7 &       12       36 16.54 &       62 07 2.79 &\nodata &  26 &     21.6 & \nodata &   2.49\cr
33& SMM123617622315 &       12       36 17.54 &       62       23 15.7 & 5.9 &  1.35 &  4.3 & \nodata & \nodata &\nodata &$<12$ & \nodata & \nodata &   3.72\cr
34& SMM123714621824 &       12       37 14.02 &       62       18 24.9 & 5.9 & 0.46 &  12.7 &       12       37 13.89 &       62       18 26.2 &\nodata &  625 &     24.4 & \nodata &   3.44\cr
35& SMM123741621221 &       12       37 41.10 &       62       12 21.4 & 5.8 & 0.47 &  12.2 &       12       37 41.16 &       62       12 21.0 & 7.1 &  27 &     23.3 &   3.02 &   2.55\cr
36& SMM123610620646 &       12       36 10.01 &       62 06 46.5 & 5.8 & 0.84 &  6.9 & \nodata & \nodata &\nodata &\nodata & \nodata & \nodata &   2.84\cr
37& SMM123633620257 &       12       36 33.85 &       62 02 57.9 & 5.6 &  1.24 &  4.5 & \nodata & \nodata &\nodata &$<12$ & \nodata & \nodata &   5.35\cr
38& SMM123738621731 &       12       37 38.10 &       62       17 31.5 & 5.6 & 0.71 &  7.8 &       12       37 38.09 &       62       17 32.2 &\nodata &  19 &     24.5 & \nodata &   5.34\cr
39& SMM123539621308 &       12       35 39.70 &       62       13 8.79 & 5.5 & 0.88 &  6.3 &       12       35 39.49 &       62       13 11.0 & 8.5 &  35 &     24.0 & \nodata &   5.34\cr
40& SMM123648622104 &       12       36 48.33 &       62       21 4.00 & 5.5 & 0.79 &  6.9 &       12       36 48.29 &       62       21 6.99 &\nodata &  41 &     22.6 & \nodata &   4.48\cr
41& SMM123610622043 &       12       36 10.12 &       62       20 43.5 & 5.5 & 0.93 &  5.9 & \nodata & \nodata &\nodata &\nodata & \nodata & \nodata &   1.72\cr
42& SMM123622621620 &       12       36 22.26 &       62       16 20.7 & 5.4 & 0.41 &  12.9 & \nodata & \nodata &\nodata &$<12$ & \nodata & \nodata &   2.56\cr
43& SMM123616621514 &       12       36 16.12 &       62       15 14.7 & 5.4 & 0.42 &  12.8 &       12       36 16.10 &       62       15 13.7 & 3.4 &  36 &     22.2 &   2.578 c &   2.12\cr
44& SMM123553620930 &       12       35 53.26 &       62 09 30.2 & 5.3 & 0.81 &  6.5 &       12       35 53.03 &       62 09 29.5 &\nodata &  41 &     22.9 & \nodata &   3.72\cr
45& SMM123636620708 &       12       36 36.25 &       62 07 8.90 & 5.3 & 0.56 &  9.3 &       12       36 35.89 &       62 07 7.50 &\nodata &  61 &     19.3 &  0.9500 a1 &   1.67\cr
46& SMM123528621428 &       12       35 28.61 &       62       14 28.4 & 5.2 &  1.10 &  4.7 & \nodata & \nodata &\nodata &$<12$ & \nodata & \nodata &   5.23\cr
47& SMM123620620637 &       12       36 20.71 &       62 06 37.8 & 5.2 & 0.78 &  6.6 &       12       36 20.47 &       62 06 39.6 &\nodata &  42 &     21.5 & \nodata &   2.84\cr
  48  &  SMM123803620840 &       12       38 3.980 &       62 08 40.7 & 5.1 &  1.02 &  5.0 & \nodata & \nodata &\nodata &$<12$ & \nodata & \nodata &   4.75\cr
49& SMM123635621424 &       12       36 35.76 &       62       14 24.9 & 4.9 & 0.35 &  13.9 &       12       36 35.59 &       62       14 24.0 &\nodata &  82 &     20.1 &   2.005 a2,c &   1.99\cr
50& SMM123645622019 &       12       36 45.32 &       62       20 19.0 & 4.9 & 0.75 &  6.5 &       12       36 45.30 &       62       20 19.7 &\nodata &  27 &     23.6 & \nodata &   3.60\cr
51& SMM123621621706 &       12       36 21.25 &       62       17 6.79 & 4.9 & 0.44 &  11.0 &       12       36 21.28 &       62       17 8.40 & 3.4 &  164 &     21.9 & \nodata &   2.17\cr
52& SMM123651620500 &       12       36 51.64 &       62 05 0.99 & 4.8 & 0.86 &  5.6 &       12       36 51.71 &       62 05 3.00 &\nodata &  79 &     24.0 & \nodata &   3.19\cr
53& SMM123712621035 &       12       37 12.64 &       62       10 35.9 & 4.8 & 0.39 &  12.2 &       12       37 12.48 &       62       10 35.6 &\nodata &  23 &     23.5 & \nodata &   3.10\cr
54& SMM123726620822 &       12       37 26.73 &       62 08 22.7 & 4.8 & 0.55 &  8.7 &       12       37 26.66 &       62 08 23.2 &\nodata &  52 &     21.2 &   2.59 &   2.47\cr
55& SMM123627621218 &       12       36 27.61 &       62       12 18.9 & 4.8 & 0.37 &  12.9 &       12       36 27.55 &       62       12 18.0 &\nodata &  17 &     24.7 & \nodata &   5.12\cr
56& SMM123542620826 &       12       35 42.73 &       62 08 26.9 & 4.7 &  1.01 &  4.6 & \nodata & \nodata &\nodata &$<12$ & \nodata & \nodata &   4.41\cr
57& SMM123719621218 &       12       37 19.37 &       62       12 18.8 & 4.6 & 0.38 &  12.1 &       12       37 18.96 &       62       12 17.5 &\nodata &  16 &     23.1 & \nodata &   3.88\cr
58& SMM123743620752 &       12       37 43.97 &       62 07 52.4 & 4.6 & 0.83 &  5.5 &       12       37 44.11 &       62 07 54.0 &\nodata &  39 &     23.1 & \nodata &   2.39\cr
59& SMM123617621929 &       12       36 17.47 &       62       19 29.7 & 4.6 & 0.78 &  5.9 &       12       36 17.05 &       62       19 31.9 &\nodata &  60 &     21.6 & \nodata &   2.22\cr
  60  &  SMM123802621853 &       12       38 2.808 &       62       18 53.9 & 4.5 &  1.04 &  4.3 & \nodata & \nodata &\nodata &$<12$ & \nodata & \nodata &   5.03\cr
61& SMM123634620529 &       12       36 34.69 &       62 05 29.9 & 4.5 & 0.83 &  5.4 & \nodata & \nodata &\nodata &$<12$ & \nodata & \nodata &   5.02\cr
62& SMM123637620853 &       12       36 37.22 &       62 08 53.0 & 4.4 & 0.42 &  10.4 &       12       36 37.03 &       62 08 52.4 &\nodata &  79 &     22.0 &   2.13 &   2.58\cr
63& SMM123629621045 &       12       36 29.21 &       62       10 45.8 & 4.4 & 0.40 &  11.0 &       12       36 29.03 &       62       10 45.5 & 7.7 &  48 &     19.7 &   1.013 a1,c &   1.77\cr
64& SMM123636621156 &       12       36 36.92 &       62       11 56.9 & 4.4 & 0.35 &  12.4 & \nodata & \nodata &\nodata &$<12$ & \nodata & \nodata &   1.95\cr
65& SMM123658620932 &       12       36 58.64 &       62 09 32.0& 4.4 & 0.38 &  11.4 &       12       36 58.55 &       62 09 31.4 & 4.6&  27 &     23.5 & \nodata &   2.95\cr
  66  &  SMM123800621616 &       12       38 0.128 &       62       16 16.9& 4.4 & 0.84 &  5.2 & \nodata & \nodata &\nodata&\nodata & \nodata & \nodata &   2.19\cr
 67& SMM123620620551 &       12       36 20.15 &       62 05 51.7& 4.3 & 0.86 &  5.0 & \nodata & \nodata &\nodata&$<12$ & \nodata & \nodata &   4.98\cr
 68& SMM123814621422 &       12       38 14.37 &       62       14 22.4& 4.3 & 0.99 &  4.3 & \nodata & \nodata &\nodata&$<12$ & \nodata & \nodata &   2.00\cr
 69& SMM123655620418 &       12       36 55.34 &       62 04 18.0& 4.3 & 0.94 &  4.5 &       12       36 55.42 &       62 04 18.0 &\nodata&  56 &     22.9 & \nodata &   2.29\cr
 70& SMM123540621439 &       12       35 40.07 &       62       14 39.9& 4.2 & 0.89 &  4.7 &       12       35 39.96 &       62       14 42.0 &\nodata&  79 &     22.8 &   2.93 &   1.73\cr
  71  &  SMM123807621046 &       12       38 7.210 &       62       10 46.6& 4.1 & 0.94 &  4.3 & \nodata & \nodata &\nodata&$<12$ & \nodata & \nodata &   3.54\cr
 72& SMM123734620527 &       12       37 34.37 &       62 05 27.5& 4.1 & 0.97 &  4.2 & \nodata & \nodata &\nodata&$<12$ & \nodata & \nodata &   4.36\cr
  73  &  SMM123702621426 &       12       37 2.658 &       62       14 26.9& 4.0 & 0.31 &  12.8 &       12       37 2.600 &       62       14 26.9 &\nodata&  18 &     22.2 &   3.214 a1,a3 &   2.78\cr
  74  &  SMM123808621425 &       12       38 8.942 &       62       14 25.7& 4.0 & 0.90 &  4.4 & \nodata & \nodata &\nodata&$<12$ & \nodata & \nodata &   2.31\cr
 75& SMM123728621920 &       12       37 28.39 &       62       19 20.7& 4.0 & 0.75 &  5.4 &       12       37 28.12 &       62       19 20.1 &\nodata&  26 &     24.0 & \nodata &   2.77\cr
 76& SMM123643622058 &       12       36 43.73 &       62       20 58.0& 4.0 & 0.79 &  5.0 & \nodata & \nodata &\nodata&\nodata & \nodata & \nodata &   3.31\cr
 77& SMM123712621211 &       12       37 12.09 &       62       12 11.9& 4.0 & 0.36 &  11.0 &       12       37 12.05 &       62       12 11.9 & 5.3&  33 &     22.0 &   2.914 c &   2.81\cr
 78& SMM123658621451 &       12       36 58.65 &       62       14 51.0& 3.9 & 0.32 &  12.4 & \nodata & \nodata &\nodata&$<12$ & \nodata & \nodata &   1.99\cr
 79& SMM123542620956 &       12       35 42.96 &       62 09 56.9& 3.9 & 0.90 &  4.3 &       12       35 43.12 &       62 09 55.4 &\nodata&  66 &     22.1 & \nodata &   1.79\cr
 80& SMM123656621205 &       12       36 56.35 &       62       12 5.00& 3.9 & 0.30 &  13.0 &       12       36 56.59 &       62       12 7.40 &\nodata&  42 &     25.0 & \nodata &   3.75\cr
  81  &  SMM123801621302 &       12       38 1.442 &       62       13 2.89& 3.9 & 0.81 &  4.8 &       12       38 1.856 &       62       13 0.90 &\nodata&  44 &     21.0 &   1.11 &   2.09\cr
  82  &  SMM123605620836 &       12       36 5.841 &       62 08 36.4& 3.8 & 0.75 &  5.1 &       12       36 5.621 &       62 08 37.7 &\nodata&  19 &     22.7 & \nodata &   4.81\cr
 83& SMM123623620520 &       12       36 23.15 &       62 05 20.8& 3.8 & 0.90 &  4.2 & \nodata & \nodata &\nodata&$<12$ & \nodata & \nodata &   4.79\cr
 84& SMM123659622211 &       12       36 59.26 &       62       22 11.9& 3.8 & 0.91 &  4.1 &       12       36 58.96 &       62       22 15.4 &\nodata&  27 &     22.8 & \nodata &   2.13\cr
  85  &  SMM123608621438 &       12       36 8.408 &       62       14 38.5& 3.6 & 0.45 &  8.0 &       12       36 8.598 &       62       14 35.4 &\nodata&  41 &     22.6 & \nodata &   2.81\cr
 86& SMM123634620940 &       12       36 34.80 &       62 09 40.9& 3.6 & 0.41 &  8.7 & \nodata & \nodata &\nodata&$<12$ & \nodata & \nodata &   3.49\cr
 87& SMM123540621217 &       12       35 40.86 &       62       12 17.9& 3.6 & 0.87 &  4.1 &       12       35 40.71 &       62       12 18.9 &\nodata&  71 &     22.1 & \nodata &   1.81\cr
  88  &  SMM123709620839 &       12       37 9.909 &       62 08 39.9& 3.6 & 0.43 &  8.3 & \nodata & \nodata &\nodata&\nodata & \nodata & \nodata &  0.967\cr
 89& SMM123550621538 &       12       35 50.05 &       62       15 38.2& 3.6 & 0.80 &  4.5 & \nodata & \nodata &\nodata&$<12$ & \nodata & \nodata &   1.99\cr
 90& SMM123631620958 &       12       36 31.21 &       62 09 58.9& 3.5 & 0.41 &  8.6 &       12       36 31.26 &       62 09 57.6 &\nodata&  148 &     21.7 &   2.313 a3 &   1.77\cr
  91  &  SMM123609620800 &       12       36 9.001 &       62 08 0.59& 3.5 & 0.76 &  4.6 &       12       36 8.869 &       62 08 3.90 &\nodata&  26 &     22.8 & \nodata &   2.95\cr
 92& SMM123754621102 &       12       37 54.21 &       62       11 2.09& 3.5 & 0.77 &  4.5 &       12       37 54.37 &       62       10 59.3 &\nodata&  303 &     20.0 &   1.306 a1 &   2.39\cr
  93  &  SMM123701622024 &       12       37 1.842 &       62       20 24.0& 3.5 & 0.75 &  4.6 &       12       37 1.560 &       62       20 24.7 &\nodata&  60 &     22.1 &   3.49 &   2.66\cr
 94& SMM123652620643 &       12       36 52.21 &       62 06 43.9& 3.5 & 0.70 &  4.9 & \nodata & \nodata &\nodata&$<12$ & \nodata & \nodata &   2.56\cr
 95& SMM123719621021 &       12       37 19.35 &       62       10 21.7& 3.5 & 0.41 &  8.5 &       12       37 19.55 &       62       10 21.2 &\nodata&  23 &     24.1 & \nodata &   3.26\cr
 96& SMM123557621726 &       12       35 57.59 &       62       17 26.2& 3.4 & 0.79 &  4.3 &       12       35 57.18 &       62       17 25.0 &\nodata&  32 &     22.4 &  0.8263 a1 &   4.64\cr
 97& SMM123728621422 &       12       37 28.42 &       62       14 22.7& 3.4 & 0.39 &  8.6 & \nodata & \nodata &\nodata&$<12$ & \nodata & \nodata &   1.92\cr
  98  &  SMM123608620852 &       12       36 8.122 &       62 08 52.5& 3.3 & 0.72 &  4.7 &       12       36 8.356 &       62 08 52.6 &\nodata&  29 &     21.9 & \nodata &   2.33\cr
 99& SMM123741621903 &       12       37 41.01 &       62       19 3.40& 3.3 & 0.80 &  4.1 &       12       37 41.48 &       62       19 3.49 &\nodata&  68 &     22.5 &  0.8805 a1 &   2.00\cr
100& SMM123718620654 &       12       37 18.87 &       62 06 54.8& 3.3 & 0.75 &  4.4 &       12       37 18.90 &       62 06 56.9 &\nodata&  21 &     24.3 & \nodata &   3.54\cr
 101  &  SMM123601621804 &       12       36 1.450 &       62       18 4.40& 3.2 & 0.80 &  4.0 &       12       36 0.963 &       62       18 7.00 &\nodata&  72 &     21.5 &   1.77 &   1.86\cr
102& SMM123713621156 &       12       37 13.94 &       62       11 56.9& 3.2 & 0.38 &  8.5 &       12       37 14.05 &       62       11 56.5 & 3.2&  22 &     24.2 & \nodata &   2.91\cr
103& SMM123732620728 &       12       37 32.98 &       62 07 28.5& 3.2 & 0.78 &  4.1 & \nodata & \nodata &\nodata&$<12$ & \nodata & \nodata &   4.55\cr
104& SMM123711621118 &       12       37 11.78 &       62       11 18.9& 3.1 & 0.38 &  8.3 & \nodata & \nodata &\nodata&$<12$ & \nodata & \nodata &   4.35\cr
105& SMM123717620801 &       12       37 17.32 &       62 08 1.90& 3.1 & 0.51 &  6.1 &       12       37 17.46 &       62 08 4.40 &\nodata&  67 &     20.8 & \nodata &   1.24\cr
106& SMM123725620857 &       12       37 25.03 &       62 08 57.6& 3.1 & 0.48 &  6.4 &       12       37 25.00 &       62 08 56.5 &\nodata&  84 &     19.6 &  0.9367 a1 &   1.28\cr
107& SMM123612620834 &       12       36 12.69 &       62 08 34.7& 3.1 & 0.70 &  4.4 &       12       36 12.63 &       62 08 35.2 &\nodata&  21 &     21.8 &   2.25 &   2.95\cr
108& SMM123620621908 &       12       36 20.07 &       62       19 8.79& 3.0 & 0.74 &  4.0 & \nodata & \nodata &\nodata&\nodata & \nodata & \nodata &   1.65\cr
 109  &  SMM123600621051 &       12       36 0.640 &       62       10 51.4& 3.0 & 0.70 &  4.2 &       12       36 0.450 &       62       10 53.3 &\nodata&  51 &     22.9 & \nodata &   2.49\cr
110& SMM123634621239 &       12       36 34.47 &       62       12 39.9& 2.9 & 0.34 &  8.3 &       12       36 34.51 &       62       12 40.9 &\nodata&  185 &     19.9 &   1.224 a1,g &  0.686\cr
111& SMM123650620822 &       12       36 50.21 &       62 08 22.0& 2.8 & 0.42 &  6.6 & \nodata & \nodata &\nodata&\nodata & \nodata & \nodata &   2.57\cr
 112  &  SMM123700620910 &       12       37 0.058 &       62 09 10.0& 2.8 & 0.39 &  7.2 &       12       37 0.270 &       62 09 9.70 & 3.1&  297 &     21.8 & \nodata &   1.99\cr
113& SMM123723621713 &       12       37 23.76 &       62       17 13.7& 2.7 & 0.44 &  6.3 &       12       37 23.56 &       62       17 13.7 &\nodata&  26 &     23.8 & \nodata &   2.67\cr
 114  &  SMM123703620755 &       12       37 3.907 &       62 07 55.9& 2.7 & 0.46 &  5.9 &       12       37 4.108 &       62 07 55.0 &\nodata&  64 &     19.4 &   1.253 a2 &   1.34\cr
115& SMM123646620833 &       12       36 46.51 &       62 08 33.0& 2.7 & 0.42 &  6.5 & \nodata & \nodata &\nodata&\nodata & \nodata & \nodata &   1.01\cr
116& SMM123735621056 &       12       37 35.64 &       62       10 56.5& 2.7 & 0.45 &  6.0 & \nodata & \nodata &\nodata&\nodata & \nodata & \nodata &   1.16\cr
117& SMM123624621014 &       12       36 24.07 &       62       10 14.7& 2.7 & 0.41 &  6.5 &       12       36 24.30 &       62       10 17.1 &\nodata&  38 &     24.3 & \nodata &   4.32\cr
118& SMM123713621545 &       12       37 13.27 &       62       15 45.9& 2.7 & 0.38 &  7.1 & \nodata & \nodata &\nodata&\nodata & \nodata & \nodata &   2.25\cr
119& SMM123714621210 &       12       37 14.79 &       62       12 10.9& 2.7 & 0.37 &  7.1 &       12       37 14.28 &       62       12 8.50 & 4.1&  25 &     21.0 &   2.280 a1 &   2.24\cr
120& SMM123648621116 &       12       36 48.78 &       62       11 16.0& 2.6 & 0.34 &  7.8 & \nodata & \nodata &\nodata&$<12$ & \nodata & \nodata &   4.29\cr
121& SMM123739621601 &       12       37 39.06 &       62       16 1.49& 2.6 & 0.52 &  5.0 &       12       37 39.52 &       62       15 58.5 &\nodata&  43 &     23.0 & \nodata &   2.31\cr
122& SMM123653621112 &       12       36 53.35 &       62       11 12.9& 2.5 & 0.33 &  7.7 &       12       36 53.26 &       62       11 16.7 &\nodata&  28 &     19.5 &  0.9380 a1,k &   1.98\cr
123& SMM123637621747 &       12       36 37.44 &       62       17 47.9& 2.5 & 0.41 &  6.1 & \nodata & \nodata &\nodata&\nodata & \nodata & \nodata &   1.48\cr
124& SMM123645621845 &       12       36 45.76 &       62       18 45.0& 2.5 & 0.46 &  5.4 & \nodata & \nodata &\nodata&$<12$ & \nodata & \nodata &   3.68\cr
125& SMM123613620900 &       12       36 13.39 &       62 09 0.70& 2.5 & 0.55 &  4.5 & \nodata & \nodata &\nodata&$<12$ & \nodata & \nodata &   2.75\cr
 126  &  SMM123608621251 &       12       36 8.737 &       62       12 51.5& 2.5 & 0.44 &  5.5 &       12       36 8.671 &       62       12 51.0 &\nodata&  39 &     23.0 & \nodata &   2.94\cr
127& SMM123722621835 &       12       37 22.92 &       62       18 35.7& 2.5 & 0.55 &  4.5 &       12       37 22.53 &       62       18 38.2 &\nodata&  30 &     20.2 &   1.524 a3 &   2.10\cr
128& SMM123623621629 &       12       36 23.12 &       62       16 29.7& 2.4 & 0.41 &  5.8 &       12       36 22.67 &       62       16 29.7 &\nodata&  80 &     21.7 &   1.790 e &   1.38\cr
129& SMM123731621022 &       12       37 31.05 &       62       10 22.5& 2.4 & 0.44 &  5.4 &       12       37 31.02 &       62       10 18.7 &\nodata&  19 &     20.1 &  0.8601 a1 &   1.92\cr
 130  &  SMM123709620753 &       12       37 9.180 &       62 07 53.9& 2.4 & 0.49 &  4.9 &       12       37 9.569 &       62 07 53.7 &\nodata&  15 &     23.7 & \nodata &   2.48\cr
131& SMM123619621003 &       12       36 19.51 &       62       10 3.79& 2.3 & 0.43 &  5.5 &       12       36 19.11 &       62       10 4.30 &\nodata&  30 &     21.9 &   2.210 e &   2.26\cr
132& SMM123731620847 &       12       37 31.87 &       62 08 47.5& 2.3 & 0.56 &  4.1 & \nodata & \nodata &\nodata&$<12$ & \nodata & \nodata &   4.13\cr
133& SMM123642621719 &       12       36 42.18 &       62       17 19.0& 2.3 & 0.38 &  6.1 &       12       36 42.18 &       62       17 22.5 &\nodata&  30 &     20.6 &   1.09 &   2.08\cr
 134  &  SMM123702621302 &       12       37 2.512 &       62       13 2.99& 2.3 & 0.30 &  7.8 &       12       37 2.570 &       62       13 2.40 &\nodata&  27 &     25.6 & \nodata &   2.16\cr
135& SMM123629621511 &       12       36 29.45 &       62       15 11.9& 2.3 & 0.39 &  5.9 &       12       36 29.45 &       62       15 13.1 &\nodata&  13 &     21.9 &   3.652 a1 &   4.10\cr
136& SMM123711621243 &       12       37 11.09 &       62       12 43.9& 2.3 & 0.34 &  6.6 & \nodata & \nodata &\nodata&$<12$ & \nodata & \nodata &   3.46\cr
137& SMM123655620814 &       12       36 55.48 &       62 08 14.0& 2.2 & 0.43 &  5.2 & \nodata & \nodata &\nodata&$<12$ & \nodata & \nodata &  0.659\cr
138& SMM123617621407 &       12       36 17.72 &       62       14 7.70& 2.2 & 0.40 &  5.6 &       12       36 17.83 &       62       14 7.91 &\nodata&  51 &     20.4 &  0.8460 k &   1.92\cr
139& SMM123716621644 &       12       37 16.44 &       62       16 44.9& 2.2 & 0.39 &  5.6 & \nodata & \nodata &\nodata&\nodata & \nodata & \nodata &   1.57\cr
140& SMM123731621614 &       12       37 31.90 &       62       16 14.5& 2.2 & 0.45 &  4.9 & \nodata & \nodata &\nodata&$<12$ & \nodata & \nodata &   1.58\cr
141& SMM123652621856 &       12       36 52.92 &       62       18 56.0& 2.2 & 0.47 &  4.7 & \nodata & \nodata &\nodata&$<12$ & \nodata & \nodata &   4.04\cr
142& SMM123616621232 &       12       36 16.90 &       62       12 32.7& 2.2 & 0.39 &  5.5 &       12       36 17.03 &       62       12 31.3 &\nodata&  21 &     22.6 & \nodata &   4.03\cr
143& SMM123611621033 &       12       36 11.35 &       62       10 33.5& 2.1 & 0.47 &  4.5 &       12       36 11.52 &       62       10 33.5 &\nodata&  26 &     21.4 &   2.40 &   2.25\cr
144& SMM123618620901 &       12       36 18.81 &       62 09 1.79& 2.1 & 0.48 &  4.4 &       12       36 18.80 &       62 09 0.80 &\nodata&  18 &     22.1 &   1.66 &   2.36\cr
145& SMM123651621457 &       12       36 51.92 &       62       14 57.9& 2.1 & 0.32 &  6.7 & \nodata & \nodata &\nodata&$<12$ & \nodata & \nodata &   3.99\cr
 146  &  SMM123607621449 &       12       36 7.690 &       62       14 49.5& 2.1 & 0.46 &  4.5 &       12       36 7.800 &       62       14 49.7 &\nodata&  24 &     22.3 &   2.85 &   3.33\cr
147& SMM123642621545 &       12       36 42.90 &       62       15 45.0& 2.1 & 0.35 &  6.0 & \nodata & \nodata &\nodata&$<12$ & \nodata & \nodata &   2.26\cr
 148  &  SMM123706621251 &       12       37 6.522 &       62       12 51.8& 2.1 & 0.32 &  6.6 & \nodata & \nodata &\nodata&$<12$ & \nodata & \nodata &   3.98\cr
149& SMM123741621253 &       12       37 41.40 &       62       12 53.4& 2.1 & 0.47 &  4.4 &       12       37 41.40 &       62       12 51.0 &\nodata&  167 &     19.8 &   1.598 a1 &   1.23\cr
150& SMM123737621357 &       12       37 37.71 &       62       13 57.5& 2.1 & 0.44 &  4.7 &       12       37 37.91 &       62       13 58.0 &\nodata&  16 &     23.8 & \nodata &   2.49\cr
151& SMM123639621542 &       12       36 39.46 &       62       15 42.9& 2.1 & 0.36 &  5.8 & \nodata & \nodata &\nodata&$<12$ & \nodata & \nodata &   2.67\cr
 152  &  SMM123606621237 &       12       36 6.170 &       62       12 37.4& 2.0 & 0.47 &  4.3 & \nodata & \nodata &\nodata&$<12$ & \nodata & \nodata &   3.35\cr
153& SMM123633620834 &       12       36 33.23 &       62 08 34.9& 2.0 & 0.44 &  4.6 &       12       36 33.23 &       62 08 34.7 &\nodata&  50 &     19.2 &  0.9340 b,a2 &   1.36\cr
154& SMM123644621620 &       12       36 44.47 &       62       16 20.9& 2.0 & 0.36 &  5.5 & \nodata & \nodata &\nodata&$<12$ & \nodata & \nodata &   3.90\cr
155& SMM123654620850 &       12       36 54.49 &       62 08 50.9& 1.9 & 0.40 &  4.9 & \nodata & \nodata &\nodata&$<12$ & \nodata & \nodata &   2.27\cr
156& SMM123612621144 &       12       36 12.04 &       62       11 44.5& 1.9 & 0.43 &  4.5 & \nodata & \nodata &\nodata&$<12$ & \nodata & \nodata &  0.916\cr
 157  &  SMM123702620836 &       12       37 2.907 &       62 08 36.0& 1.9 & 0.42 &  4.6 &       12       37 2.980 &       62 08 33.0 &\nodata&  19 &     21.8 &   1.97 &   2.68\cr
158& SMM123612621221 &       12       36 12.60 &       62       12 21.7& 1.9 & 0.42 &  4.5 & \nodata & \nodata &\nodata&$<12$ & \nodata & \nodata &   3.85\cr
159& SMM123731621256 &       12       37 31.96 &       62       12 56.5& 1.9 & 0.40 &  4.7 & \nodata & \nodata &\nodata&$<12$ & \nodata & \nodata &   1.25\cr
160& SMM123727621706 &       12       37 27.19 &       62       17 6.69& 1.9 & 0.45 &  4.1 &       12       37 27.70 &       62       17 5.90 &\nodata&  14 &     22.5 & \nodata &   2.82\cr
 161  &  SMM123703621635 &       12       37 3.537 &       62       16 35.0& 1.8 & 0.38 &  4.8 & \nodata & \nodata &\nodata&$<12$ & \nodata & \nodata &   2.56\cr
 162  &  SMM123609621141 &       12       36 9.049 &       62       11 41.5& 1.8 & 0.46 &  4.0 &       12       36 8.822 &       62       11 43.7 &\nodata&  51 &     19.9 &   1.336 a1 &   1.03\cr
163& SMM123736621240 &       12       37 36.53 &       62       12 40.5& 1.8 & 0.43 &  4.2 &       12       37 36.80 &       62       12 42.7 &\nodata&  30 &     22.1 &   1.67 &   2.09\cr
164& SMM123649621814 &       12       36 49.35 &       62       18 14.0& 1.8 & 0.42 &  4.3 &       12       36 49.13 &       62       18 13.9 &\nodata&  19 &     21.3 &   2.321 a3 &   2.43\cr
165& SMM123613621436 &       12       36 13.42 &       62       14 36.7& 1.8 & 0.42 &  4.3 & \nodata & \nodata &\nodata&$<12$ & \nodata & \nodata &   2.49\cr
166& SMM123720621102 &       12       37 20.35 &       62       11 2.79& 1.8 & 0.40 &  4.5 &       12       37 20.32 &       62       11 3.20 &\nodata&  23 &     22.8 & \nodata &   2.53\cr
 167  &  SMM123702621402 &       12       37 2.658 &       62       14 2.00& 1.7 & 0.30 &  5.8 & \nodata & \nodata &\nodata&\nodata & \nodata & \nodata &   1.10\cr
168& SMM123624620929 &       12       36 24.94 &       62 09 29.7& 1.7 & 0.43 &  4.1 & \nodata & \nodata &\nodata&$<12$ & \nodata & \nodata &   3.74\cr
169& SMM123627621314 &       12       36 27.46 &       62       13 14.9& 1.7 & 0.37 &  4.7 & \nodata & \nodata &\nodata&$<12$ & \nodata & \nodata &   3.74\cr
170& SMM123639621005 &       12       36 39.51 &       62       10 5.00& 1.7 & 0.40 &  4.3 & \nodata & \nodata &\nodata&$<12$ & \nodata & \nodata &   1.54\cr
171& SMM123724621629 &       12       37 24.03 &       62       16 29.7& 1.7 & 0.41 &  4.2 & \nodata & \nodata &\nodata&$<12$ & \nodata & \nodata &   3.71\cr
172& SMM123710621429 &       12       37 10.96 &       62       14 29.9& 1.7 & 0.35 &  4.8 & \nodata & \nodata &\nodata&$<12$ & \nodata & \nodata &   3.70\cr
 173  &  SMM123706621607 &       12       37 6.690 &       62       16 7.90& 1.7 & 0.37 &  4.5 & \nodata & \nodata &\nodata&$<12$ & \nodata & \nodata &   1.81\cr
174& SMM123655621023 &       12       36 55.21 &       62       10 23.9& 1.6 & 0.36 &  4.6 & \nodata & \nodata &\nodata&$<12$ & \nodata & \nodata &   3.67\cr
175& SMM123713621143 &       12       37 13.37 &       62       11 43.9& 1.6 & 0.38 &  4.3 &       12       37 13.17 &       62       11 45.5 &\nodata&  20 &     23.0 &   2.413 a3 &   3.29\cr
176& SMM123648621214 &       12       36 48.78 &       62       12 14.9& 1.6 & 0.30 &  5.5 &       12       36 48.65 &       62       12 15.7 &\nodata&  22 &     20.9 &   1.066 a1 &   1.85\cr
177& SMM123620621110 &       12       36 20.64 &       62       11 10.7& 1.6 & 0.40 &  4.0 &       12       36 20.98 &       62       11 13.7 &\nodata&  19 &     20.1 &  0.9447 a1 &   1.33\cr
178& SMM123713621251 &       12       37 13.67 &       62       12 51.8& 1.6 & 0.35 &  4.5 &       12       37 13.69 &       62       12 49.4 &\nodata&  12 &     22.5 &  0.8992 a1 &   3.63\cr
179& SMM123657621654 &       12       36 57.52 &       62       16 54.9& 1.5 & 0.38 &  4.1 & \nodata & \nodata &\nodata&$<12$ & \nodata & \nodata &   3.59\cr
 180  &  SMM123707621127 &       12       37 7.928 &       62       11 27.9& 1.5 & 0.36 &  4.3 & \nodata & \nodata &\nodata&$<12$ & \nodata & \nodata &   1.80\cr
 181  &  SMM123701621513 &       12       37 1.237 &       62       15 13.9& 1.5 & 0.34 &  4.5 & \nodata & \nodata &\nodata&$<12$ & \nodata & \nodata &   2.17\cr
182& SMM123645621148 &       12       36 45.35 &       62       11 48.0& 1.5 & 0.32 &  4.6 & \nodata & \nodata &\nodata&$<12$ & \nodata & \nodata &   2.73\cr
183& SMM123657621407 &       12       36 57.51 &       62       14 7.00& 1.4 & 0.29 &  5.0 &       12       36 57.37 &       62       14 7.99 &\nodata&  28 &     21.0 &   1.462 a1,a3 &   1.63\cr
 184  &  SMM123706621156 &       12       37 6.372 &       62       11 56.0& 1.4 & 0.34 &  4.2 &       12       37 5.859 &       62       11 53.6 &\nodata&  44 &     19.1 &  0.9015 a1,a3 &   1.06\cr
185 & SMM123634621215 &       12       36 34.92 &       62       12 15.9& 1.4 & 0.35 &  4.0 & \nodata & \nodata &\nodata&\nodata & \nodata & \nodata & .00891\cr
186 & SMM123656621254 &       12       36 56.07 &       62       12 54.0& 1.4 & 0.28 &  4.9 & \nodata & \nodata &\nodata&$<12$ & \nodata & \nodata &   3.44\cr
\enddata
\tablenotetext{a1}{\,\,Our DEIMOS redshift}
\tablenotetext{a2}{\,\,Our LRIS redshift}
\tablenotetext{a3}{\,\,Our MOSFIRE redshift}
\tablenotetext{b}{Wirth et al.\ 2004}
\tablenotetext{c}{Chapman et al.\ 2005}
\tablenotetext{d}{Swinbank et al.\ 2004}
\tablenotetext{e}{Pope et al.\ 2008}
\tablenotetext{f1}{\,\,Daddi et al.\ 2009b}
\tablenotetext{f2}{\,\,Daddi et al.\ 2009a}
\tablenotetext{g}{Bothwell et al.\ 2013}
\tablenotetext{h}{Walter et al.\ 2012}
\tablenotetext{k}{Cohen et al.\ 2000}
\label{tab5}
\end{deluxetable*}

\begin{deluxetable*}{lcccccccccclcc}
\tabletypesize{\scriptsize}
\renewcommand\baselinestretch{1.0}
\tablewidth{0pt}
\tablecaption{SCUBA-2 450 Micron Sample ($4\,\sigma$)}
\tablehead{No. & Name & R.A. & Decl.& \multicolumn{2}{c}{450\,$\mu$m} & S/N  & R.A. (accurate) & Decl. & SMA & 20\,cm & $K_{s}$ & $z$ & 850\,$\mu$m\\ 
& & & & Flux & Error & & & & & & & & Flux\\
& & (J2000.0) & (J2000.0) & (mJy) & (mJy) &  & (J2000.0) & (J2000.0) & (mJy) & ($\mu$Jy) & (AB) & & (mJy) \\ (1) & (2) & (3) & (4) & (5) & (6) & (7) & (8) & (9) & (10) & (11) & (12) & (13) & (14)}
\startdata
   1   & SMM123730621259 &       12       37 30.69 &       62       12 59.6 &   42  &   4.1 &   10. &       12       37 30.80 &       62       12 58.7 &  14.9 &   123 &     22.9 &   2.43 &   13.\cr
   2   & SMM123553621335 &       12       35 53.70 &       62       13 35.1 &   39 &   9.5 &   4.1 &       12       35 53.86 &       62       13 37.2 & \nodata &   32 &     19.6 &  0.8810 &   5.9\cr
   3   & SMM123721620709 &       12       37 21.01 &       62 07 9.79 &   35 &   6.9 &   5.0 &       12       37 21.40 &       62 07 8.30 & \nodata &   293 &     20.9 & \nodata &   6.4\cr
   4   & SMM123741621222 &       12       37 41.40 &       62       12 22.4 &   33 &   5.3 &   6.3 &       12       37 41.16 &       62       12 21.0 &  7.10 &   28 &     23.3 &   3.02 &   5.4\cr
   5  &  SMM123618621549 &       12       36 18.84 &       62       15 49.7 &   30 &   4.4 &   6.9 &       12       36 18.35 &       62       15 50.4 &  7.20 &   169 &     22.0 &   2.000 &   6.4\cr
   6   & SMM123634621918 &       12       36 34.71 &       62       19 18.9 &   30 &   6.8 &   4.4 & \nodata & \nodata & \nodata & $<12$ & \nodata & \nodata &   5.3\cr
   7  &  SMM123629621045 &       12       36 29.21 &       62       10 45.8 &   28 &   3.9 &   7.2 &       12       36 29.03 &       62       10 45.5 &  7.70 &   91 &     19.7 &   1.013 &   4.3\cr
   8   & SMM123711621328 &       12       37 11.39 &       62       13 28.9 &   28 &   3.3 &   8.4 &       12       37 11.34 &       62       13 30.9 &  6.70 &   126 &     20.4 &   1.995 &   8.1\cr
   9   & SMM123701621146 &       12       37 1.651 &       62       11 46.9 &   26 &   3.0 &   8.7 &       12       37 1.578 &       62       11 46.4 &  4.80 &   95 &     20.5 &   1.760 &   6.3\cr
  10   & SMM123707621407 &       12       37 7.236 &       62       14 7.91 &   26 &   3.2 &   8.0 &       12       37 7.177 &       62       14 8.19 &  7.10 &   28 &     21.4 &   2.45 &   8.0\cr
  11  &  SMM123616621513 &       12       36 16.42 &       62       15 13.7 &   24 &   4.3 &   5.5 &       12       36 16.10 &       62       15 13.7 &  3.40 &   38 &     22.2 &   2.578 &   4.6\cr
  12   & SMM123726620823 &       12       37 26.16 &       62 08 23.7 &   23 &   5.4 &   4.3 &       12       37 26.66 &       62 08 23.2 & \nodata &   51 &     21.2 &   2.59 &   4.3\cr
  13   & SMM123635621423 &       12       36 35.76 &       62       14 23.9 &   22 &   3.5 &   6.2 &       12       36 35.59 &       62       14 24.0 & \nodata &   78 &     20.1 &   2.005 &   3.8\cr
  14   & SMM123622621629 &       12       36 22.70 &       62       16 29.7 &   22 &   4.5 &   4.9 &       12       36 22.67 &       62       16 29.7 & \nodata &   81 &     21.7 &   1.790 &   3.9\cr
  15   & SMM123622621616 &       12       36 22.26 &       62       16 16.7 &   21 &   4.4 &   4.8 &       12       36 22.10 &       62       16 15.9 &  5.40 &   20 &     23.9 & \nodata &   4.6\cr
  16   & SMM123634621239 &       12       36 34.90 &       62       12 39.9 &   20 &   3.3 &   6.0 &       12       36 34.51 &       62       12 40.9 & \nodata &   188 &     19.9 &   1.224 &   2.5\cr
  17   & SMM123716621640 &       12       37 16.14 &       62       16 40.9 &   20 &   4.2 &   4.7 & \nodata & \nodata & \nodata & $<12$ & \nodata & \nodata &   1.9\cr
  18   & SMM123646621448 &       12       36 46.05 &       62       14 48.9 &   19 &   3.2 &   5.9 &       12       36 46.08 &       62       14 48.6 &  4.20 &   103 &     22.4 &   3.63 &   5.9\cr
  19   & SMM123642621721 &       12       36 42.18 &       62       17 21.9 &   18 &   4.0 &   4.5 &       12       36 42.18 &       62       17 22.5 & \nodata &   24 &     20.6 & \nodata &   2.2\cr
  20   & SMM123646620833 &       12       36 46.65 &       62 08 33.9 &   18 &   3.9 &   4.6 &       12       36 46.68 &       62 08 33.2 & \nodata &   95 &     19.0 &  0.9710 &   2.5\cr
  21   & SMM123631621714 &       12       36 31.99 &       62       17 14.9 &   17 &   4.3 &   4.0 &       12       36 31.94 &       62       17 14.7 &  7.10 &   22 &     23.0 & \nodata &   8.4\cr
  22   & SMM123726621328 &       12       37 26.12 &       62       13 28.7 &   17 &   3.9 &   4.5 & \nodata & \nodata & \nodata & $<12$ & \nodata & \nodata &  0.28\cr
  23   & SMM123637620854 &       12       36 37.08 &       62 08 54.0 &   17 &   4.0 &   4.3 &       12       36 37.03 &       62 08 52.4 & \nodata &   90 &     22.0 &   2.13 &   4.3\cr
  24   & SMM123631620957 &       12       36 31.51 &       62 09 57.9 &   16 &   3.9 &   4.1 &       12       36 31.26 &       62 09 57.6 & \nodata &   140 &     21.7 &   2.313 &   2.9\cr
  25   & SMM123713621153 &       12       37 13.79 &       62       11 53.9 &   16  &   3.6 &   4.4 &       12       37 14.05 &       62       11 56.5 &  3.20 &   21 &     24.2 & \nodata &   2.5\cr
  26  &  SMM123628621313 &       12       36 28.17 &       62       13 13.8 &   16 &   3.6 &   4.4 & \nodata & \nodata & \nodata & $<12$ & \nodata & \nodata &   1.4\cr
  27  &  SMM123633621408 &       12       36 33.46 &       62       14 8.89 &   15 &   3.6 &   4.2 &       12       36 33.42 &       62       14 8.50 &  12.0 &   33 &     25.5 &   4.042 &   9.2\cr
  28  &  SMM123700620911 &       12       37 0.487 &       62 09 11.0 &   15 &   3.8 &   4.0 &       12       37 0.270 &       62 09 9.70 &  3.10 &   297 &     21.8 & \nodata &   2.0\cr
  29  &  SMM123712621325 &       12       37 12.53 &       62       13 25.9 &   14 &   3.4 &   4.1 & \nodata & \nodata & \nodata & $<12$ & \nodata & \nodata &   3.2\cr
  30 &  SMM123651621228 &       12       36 51.92 &       62       12 28.0 &   13 &   2.7 &   4.7 &       12       36 52.03 &       62       12 25.9 &  7.80 &   12 &     99.0 &   5.183 &   5.6\cr
  31  & SMM123652621355 &       12       36 52.35 &       62       13 55.0 &   11 &   2.8 &   4.1 &       12       36 52.78 &       62       13 54.3 & \nodata &   21 &     21.1 &   1.355 &  0.58
\enddata
\label{450_table}
\end{deluxetable*}

\begin{deluxetable*}{lllllll}
\renewcommand\baselinestretch{1.0}
\tablewidth{0pt}
\tablecaption{Comparison with Previous Wide-field mm/submm Surveys}
\scriptsize
\tablehead{Name &  Present  & S2CLS &  Wang:SCUBA & Pope:SCUBA & AzTEC+MAMBO & GISMO \\ & (850\,$\mu$m) & 850\,$\mu$m) & (850\,$\mu$m) & (850\,$\mu$m)  & (1.16\,mm) & (2\,mm) \\ & (mJy) & (mJy) &  & (mJy) & (mJy) \\ (1) & (2) & (3) & (4) & (5) & (6) & (7) }
\startdata
   2   SMM123712622210 &  17.3 (0.94) & \nodata & \nodata &  20.3 ( 2.1) &  10.2 (0.68) & \nodata\\
   3   SMM123730621258 &  13.4 (0.39) &  12.8 (0.97) &  13.6 ( 2.2) & \nodata &  4.51 (0.55) & \nodata\\
   7   SMM123627620605 &  11.2 (0.80) &  10.6 ( 1.6) & \nodata & \nodata &  4.81 (0.67) & \nodata\\
   8   SMM123633621407 &  9.44 (0.36) &  8.37 (0.90) &  12.9 ( 2.1) &  11.3 ( 1.6) &  5.24 (0.57) & 0.790 (0.14)\\
  12   SMM123550621041 &  8.69 (0.80) &  11.5 ( 1.4) & \nodata & \nodata &  5.00 (0.71) & \nodata\\
  13   SMM123711621329 &  8.58 (0.34) &  9.69 (0.90) &  4.40 ( 1.3) & \nodata &  4.09 (0.54) & \nodata\\
  14   SMM123631621712 &  8.46 (0.40) &  7.28 (0.97) & \nodata & \nodata &  4.22 (0.56) & \nodata\\
  15   SMM123707621408 &  8.34 (0.33) &  7.09 (0.89) & \nodata &  10.7 ( 2.7) & \nodata & \nodata\\
  18   SMM123618621549 &  7.35 (0.42) &  7.62 (0.99) &  7.72 ( 1.0) &  7.50 (0.90) & \nodata & \nodata\\
  20   SMM123634621921 &  6.91 (0.70) &  5.48 ( 1.1) & \nodata & \nodata &  2.44 (0.61) & \nodata\\
  21   SMM123721620709 &  6.74 (0.74) &  5.70 ( 1.3) & \nodata & \nodata & \nodata & \nodata\\
  22   SMM123722620539 &  6.74 (0.87) &  8.53 ( 1.6) & \nodata & \nodata & \nodata & \nodata\\
  24   SMM123644621938 &  6.61 (0.69) &  6.62 ( 1.0) & \nodata & \nodata &  3.03 (0.59) & \nodata\\
  25   SMM123701621145 &  6.56 (0.32) &  6.71 (0.86) & \nodata &  3.90 (0.70) & \nodata & \nodata\\
  26   SMM123652621224 &  6.30 (0.29) &  5.97 (0.83) &  5.12 (0.47) &  5.90 (0.30) & \nodata & 0.420 (0.13)\\
  27   SMM123539621241 &  6.22 (0.88) & \nodata & \nodata & \nodata &  2.97 (0.75) & \nodata\\
  28   SMM123812621453 &  6.18 (0.97) & \nodata & \nodata & \nodata &  2.78 (0.73) & \nodata\\
  29   SMM123632620621 &  6.08 (0.76) & \nodata & \nodata & \nodata &  3.31 (0.66) & \nodata\\
  31   SMM123646621447 &  5.99 (0.32) &  6.74 (0.92) &  10.8 ( 2.2) &  8.60 ( 1.4) & \nodata & \nodata\\
  32   SMM123616620701 &  5.97 (0.77) &  7.08 ( 1.5) & \nodata & \nodata & \nodata & \nodata\\
  34   SMM123714621824 &  5.93 (0.46) &  4.65 ( 1.0) & \nodata & \nodata &  3.46 (0.57) & \nodata\\
  35   SMM123741621221 &  5.86 (0.47) &  7.92 ( 1.0) & \nodata & \nodata &  2.41 (0.57) & \nodata\\
  36   SMM123610620646 &  5.81 (0.84) &  7.52 ( 1.7) & \nodata & \nodata & \nodata & \nodata\\
  38   SMM123738621731 &  5.61 (0.71) & \nodata & \nodata & \nodata &  2.62 (0.58) & \nodata\\
  40   SMM123648622104 &  5.57 (0.79) &  7.18 ( 1.2) & \nodata & \nodata &  3.05 (0.63) & \nodata\\
  42   SMM123622621620 &  5.42 (0.41) &  6.15 (0.98) &  10.2 ( 1.2) & \nodata & \nodata & \nodata\\
  43   SMM123616621514 &  5.42 (0.42) &  5.99 (0.99) &  6.20 ( 1.0) & \nodata &  2.61 (0.55) & \nodata\\
  45   SMM123636620708 &  5.34 (0.56) & \nodata & \nodata & \nodata &  3.35 (0.61) & \nodata\\
  49   SMM123635621424 &  4.97 (0.35) &  5.20 (0.91) & \nodata & \nodata & \nodata & \nodata\\
  51   SMM123621621706 &  4.94 (0.44) &  4.95 ( 1.0) &  8.69 ( 1.9) &  8.90 ( 1.5) & \nodata & \nodata\\
  52   SMM123651620500 &  4.85 (0.86) &  6.76 ( 1.6) & \nodata & \nodata &  3.74 (0.88) & \nodata\\
  53   SMM123712621035 &  4.84 (0.39) &  4.38 (0.94) & \nodata & \nodata & \nodata & \nodata\\
  55   SMM123627621218 &  4.82 (0.37) &  7.09 (0.97) & \nodata & \nodata &  2.61 (0.57) & \nodata\\
  57   SMM123719621218 &  4.69 (0.38) &  3.98 (0.92) & \nodata & \nodata & \nodata & \nodata\\
  58   SMM123743620752 &  4.66 (0.83) &  5.79 ( 1.5) & \nodata & \nodata & \nodata & \nodata\\
  62   SMM123637620853 &  4.44 (0.42) &  3.54 ( 1.0) & \nodata & \nodata & \nodata & \nodata\\
  63   SMM123629621045 &  4.44 (0.40) & \nodata &  5.72 ( 1.3) & \nodata & \nodata & \nodata\\
  64   SMM123636621156 &  4.43 (0.35) &  3.67 (0.90) & \nodata &  7.00 (0.90) & \nodata & \nodata\\
  65   SMM123658620932 &  4.42 (0.38) &  5.20 (0.99) & \nodata & \nodata & \nodata & \nodata\\
  73   SMM123702621426 &  4.09 (0.31) &  3.51 (0.88) & \nodata & \nodata & \nodata & \nodata\\
  75   SMM123728621920 &  4.07 (0.75) &  4.35 ( 1.1) & \nodata & \nodata & \nodata & \nodata\\
  77   SMM123712621211 &  4.04 (0.36) &  4.75 (0.88) & \nodata & \nodata &  2.59 (0.55) & \nodata\\
  78   SMM123658621451 &  3.99 (0.32) &  3.82 (0.89) & \nodata & \nodata & \nodata & \nodata\\
  80   SMM123656621205 &  3.97 (0.30) &  4.00 (0.84) & \nodata & \nodata & \nodata & \nodata\\
  82   SMM123605620836 &  3.87 (0.75) &  4.95 ( 1.4) & \nodata & \nodata & \nodata & \nodata\\
  86   SMM123634620940 &  3.65 (0.41) &  3.43 (0.96) &  4.23 ( 1.3) & \nodata & \nodata & \nodata\\
  91   SMM123609620800 &  3.57 (0.76) &  6.04 ( 1.4) & \nodata & \nodata & \nodata & \nodata\\
  95   SMM123719621021 &  3.50 (0.41) &  3.97 (0.95) & \nodata & \nodata & \nodata & \nodata\\
  98   SMM123608620852 &  3.39 (0.72) & \nodata & \nodata & \nodata &  2.42 (0.60) & \nodata\\
 102   SMM123713621156 &  3.24 (0.38) &  3.64 (0.89) & \nodata & \nodata & \nodata & \nodata\\
 105   SMM123717620801 &  3.16 (0.51) &  3.97 ( 1.1) & \nodata & \nodata &  2.74 (0.67) & \nodata\\
 110   SMM123634621239 &  2.92 (0.34) &  3.57 (0.90) & \nodata & \nodata & \nodata & \nodata\\
 112   SMM123700620910 &  2.85 (0.39) & \nodata & \nodata &  9.00 ( 2.1) & \nodata & \nodata\\
 115   SMM123646620833 &  2.76 (0.42) &  4.04 ( 1.0) & \nodata & \nodata & \nodata & \nodata\\
 126   SMM123608621251 &  2.50 (0.44) & \nodata & \nodata &  16.8 ( 4.0) & \nodata & \nodata\\
 128   SMM123623621629 &  2.45 (0.41) &  3.55 (0.99) & \nodata & \nodata & \nodata & \nodata\\
 131   SMM123619621003 &  2.39 (0.43) & \nodata &  6.01 ( 1.7) &  6.70 ( 1.6) & \nodata & \nodata\\
 134   SMM123702621302 &  2.36 (0.30) & \nodata & \nodata &  3.20 (0.60) & \nodata & \nodata\\
 141   SMM123652621856 &  2.23 (0.47) &  3.91 ( 1.0) & \nodata & \nodata & \nodata & \nodata\\
 143   SMM123611621033 &  2.19 (0.47) &  4.41 ( 1.0) & \nodata & \nodata & \nodata & \nodata\\
 158   SMM123612621221 &  1.94 (0.42) & \nodata &  3.75 ( 1.1) & \nodata & \nodata & \nodata\\
 166   SMM123720621102 &  1.83 (0.40) &  4.11 (0.92) & \nodata & \nodata & \nodata & \nodata\\
\enddata
\label{tab_previous}
\end{deluxetable*}

\end{document}